%% file: CompleteManuscriptV2_doublespace.tex
\documentclass{article}
\usepackage[utf8]{inputenc}
\usepackage{times}
\usepackage[margin=20mm]{geometry}
\usepackage{amsmath}
\usepackage{amssymb}
\usepackage{graphicx}
\usepackage[labelfont=bf]{caption}
\usepackage{booktabs}
\usepackage{refs}
\usepackage[hidelinks]{hyperref}
\usepackage{color}
\usepackage{xcolor}
\usepackage{scicite}

\usepackage{pdflscape}

\newcommand{\Msun}{\ensuremath{\,M_\odot}}
\newcommand{\Msunyr}{\ensuremath{\,M_\odot\, \mathrm{yr}^{-1}}}
\newcommand{\Lsun}{\ensuremath{\,L_\odot}}
\newcommand{\kms}{\ensuremath{\,\mathrm{km}\,\mathrm{s}^{-1}}}

\newcommand{\YG}[1]{{\color{blue}[YG: #1]}}

\newcommand{\HeI}{He\,{\sc i}}
\newcommand{\HeII}{He\,{\sc ii}}
\newcommand{\NIII}{N\,{\sc iii}}
\newcommand{\NIV}{N\,{\sc iv}}
\newcommand{\NV}{N\,{\sc v}}
\newcommand{\CIV}{C\,{\sc iv}}
\newcommand{\CIII}{C\,{\sc iii}}

\newenvironment{sciabstract}{%
\begin{quote} \bf}
{\end{quote}}

\title{Discovery of the missing \\ intermediate-mass helium stars stripped in binaries}
\author{M.\ R.\ Drout$^{1,2,*,\dagger}$, Y.\ G\"{o}tberg$^{2,*,\dagger}$, B.\ A.\ Ludwig$^{1}$, J.\ H.\ Groh$^{3}$, S.\ E.\ de Mink$^{4,5}$, \\ A.\ J.\ G.\ O'Grady$^{1,6}$, N.\ Smith$^{7}$ \\
}

\begin{document}


\maketitle

\noindent
\normalsize{$^{1}$ David A. Dunlap Department of Astronomy \& Astrophysics, University of Toronto, 50 St.\ George Street, Toronto, Ontario, M5S 3H4, Canada}\\
\normalsize{$^{2}$ The Observatories of the Carnegie Institution for Science, 813 Santa Barbara St., Pasadena, CA 91101, USA}\\ 
\normalsize{$^{3}$ Independent Researcher}\\ 
\normalsize{$^{4}$ Max-Planck-Institut f\"{u}r Astrophysik, Karl-Schwarzschild-Stra{\ss}e 1, 85741 Garching, Germany}\\
\normalsize{$^{5}$ Anton Pannekoek Institute for Astronomy, University of Amsterdam, Postbus 94249, 1090 GE Amsterdam, The Netherlands}\\
\normalsize{$^{6}$ Dunlap Institute for Astronomy \& Astrophysics, University of Toronto, 50 St. George Street, Toronto, Ontario, M5S 3H4, Canada}\\
\normalsize{$^{7}$ Steward Observatory, University of Arizona, 933 N. Cherry Ave., Tucson, AZ 85721, USA}\\

\noindent 
\normalsize{$^\ast$Corresponding Authors; E-mails:  maria.drout@utoronto.ca, ygoetberg@carnegiescience.edu }\\
\normalsize{$^{\dagger}$ These authors contributed equally.}

\begin{sciabstract}
The theory of binary evolution predicts that many massive stars should lose their hydrogen-rich envelopes via interaction with a companion---revealing hot helium stars with masses of $\sim$2--8M$_{\odot}$. However, only one candidate system had been identified, leaving a large discrepancy between theory and observation. Here, we present a new sample of stars---identified via excess ultraviolet emission---whose luminosities, colors, and spectral morphologies are consistent with predictions for the missing population. We detect radial velocity variations indicative of binary motion and measure high temperatures ($T_{\rm eff}\sim60-100$kK), high surface gravities ($\log(g)\sim5$) and depleted surface hydrogen mass fractions ($X_{\rm{H,surf}}\lesssim0.3$), which match expectations for stars with initial masses between 8--25 M$_{\odot}$ that have been stripped via binary interaction.
These systems fill the helium star mass gap between subdwarfs and Wolf-Rayet stars, and are thought to be of large astrophysical significance as ionizing sources, progenitors of stripped-envelope supernovae and merging double neutron stars.
\end{sciabstract}

The lifecycle of massive stars impacts every subfield in astrophysics---from galaxies to gravitational waves. Notably, recent studies indicate that $\sim$70\% of all massive stars should \emph{interact} with a binary companion during their lifetimes \cite{2012Sci...337..444S, 2017ApJS..230...15M}. One of the most common outcomes predicted are stars stripped of their hydrogen-rich envelopes via stable mass transfer or successful common envelope ejection, which leaves the hot and compact helium cores exposed. Binary-stripping can remove the hydrogen-rich envelope from lower mass stars than stellar winds (M$_{\rm{init}}$ $\lesssim$ 25 M$_{\odot}$) and the resulting ``stripped'' stars are relatively long-lived. As a result, they are predicted to be numerous. 

These binary-stripped massive stars are also expected to play important roles in multiple astrophysical processes. For example: (i) low ejecta masses, high rates, and a lack of direct progenitor detections all suggest that binary-stripped stars, not wind-stripped Wolf-Rayet (WR) stars, are the progenitors for most hydrogen-poor core-collapse supernovae \cite{Drout2011,Lyman2016,Smith2011,Eldridge2013}, (ii) binary neutron stars that merge in gravitational wave events are thought to have undergone two phases of envelope-stripping \cite{2017ApJ...846..170T}, and (iii) the extremely high surface temperatures of stripped stars make them promising candidates for the origin of the hardest ionizing photons observed in stellar populations as well as contributors to cosmic reionization 
\cite{Stanway2016, 2020A&A...636A..47S, 2020A&A...634A.134G}.

However, despite their importance and predicted ubiquity, a population of binary-stripped helium stars with masses between $\sim$2--8 M$_\odot$---descended from stars with initial masses of $\sim$8-25 M$_\odot$---has not been found. 
Of known helium stars, both low-mass subdwarfs ($\lesssim$ 1.5 M$_{\odot}$, \cite{2016PASP..128h2001H}) and high-mass WR stars ($\gtrsim$ 8 M$_{\odot}$, \cite{2007ARA&A..45..177C}) have been studied in detail, including in binary systems \cite{2021AJ....161..248W,2017MNRAS.464.2066S,1992Natur.355..703V}. However, neither are in the mass range expected to yield the majority of stripped-envelope supernovae or neutron-star mergers.
Only one ``intermediate mass'' stripped star has been published to date \cite{2008A&A...485..245G}.

If such systems are truly rare, significant modifications to models of binary evolution will be needed. Alternatively, this
may be an observational bias: the optical flux from intermediate mass stripped stars can be dominated by a bright main sequence (MS) companion, and they should also exhibit weaker wind features than luminous WR stars---potentially eluding detection in existing narrowband surveys.
However, it has recently been suggested that some stripped helium star binaries may be detectable via excess ultraviolet emission in their spectral energy distributions \cite{2018A&A...615A..78G}. To assess this, we create synthetic spectra for a large set of stripped star plus MS star binaries \cite{MM}. While many systems remain obscured by the MS, we find that
intermediate-mass helium stars paired with relatively low-mass MS companions (M$<$10 M$_\odot$) can appear in a unique region of UV-optical color-magnitude diagrams (CMDs): bluewards of the MS at intermediate luminosities ($-1$ mag $>$ M$_{\rm{UV}} > -4$ mag; see Figures~\ref{fig:Theoretical_CMD}-\ref{fig:Theoretical_CMD_V}).

\begin{figure}
\includegraphics[width=0.95\textwidth]{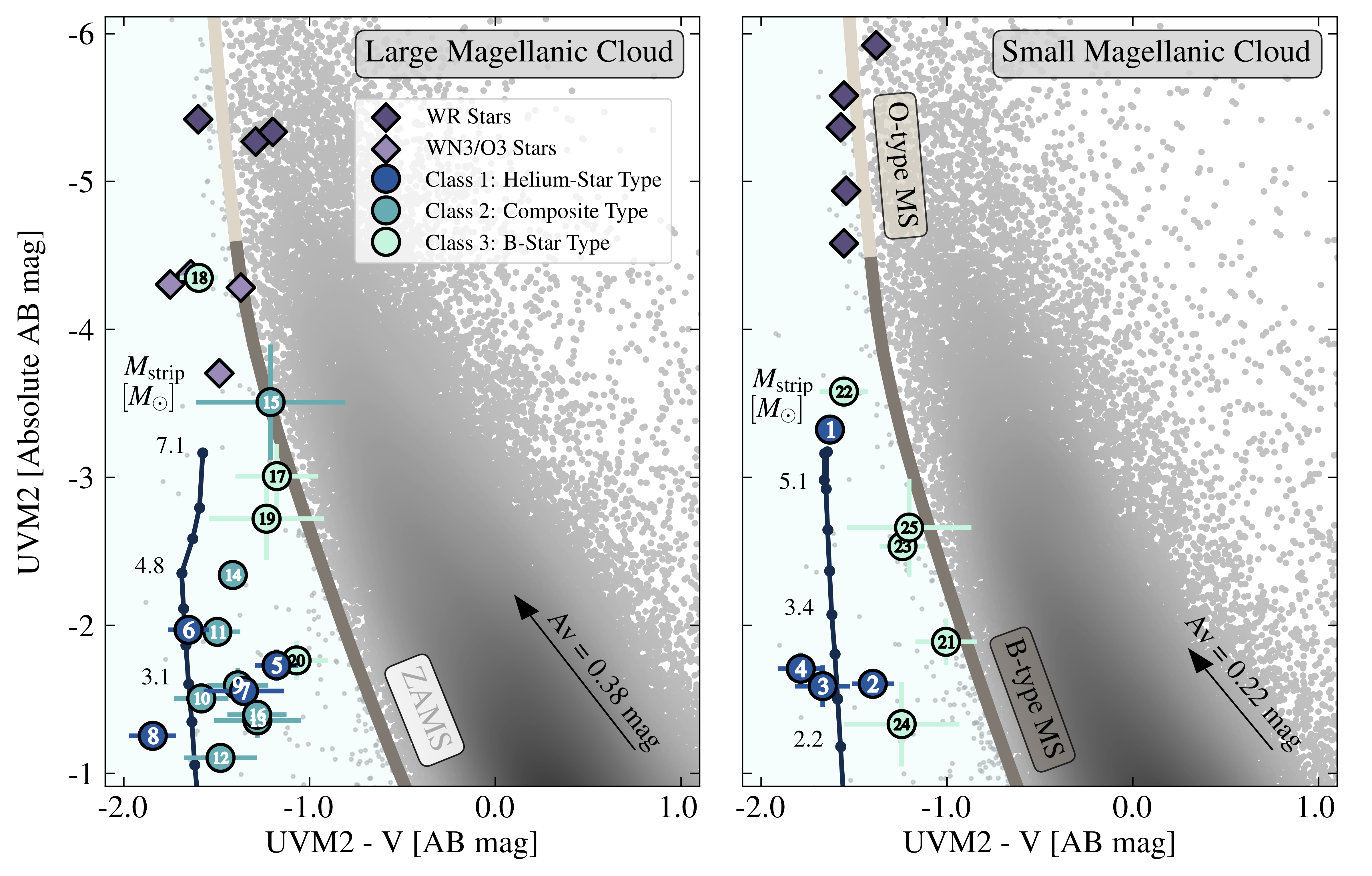}
\caption{
{\bf Identification of Candidate Stripped Helium Star Binaries in UV-optical Color-Magnitude Diagrams.} Gray dots represent the location of stars in the LMC (panel A) and SMC (panel B) based on new UV photometry performed on images from the \emph{Swift} satellite. Dozens of candidates stripped helium star binaries are visible bluewards of the ZAMS. The 25 stars presented in this work are shown as numbered circles, color-coded based on their observed spectral morphologies. These systems have similar UV-optical colors, but lower brightnesses than either traditional WR stars (dark purple diamonds) or the recently discovered weaker-wind WN3/O3 stars (light purple diamonds). For reference, we also show models for isolated helium-core burning stripped stars (black connected dots). All data has been de-reddened as indicated by the arrows.
}
\label{fig:CMD}
\end{figure}

With this as motivation, we carry out a search for massive stars with excess UV light. We target the Large and Small Magellanic Clouds (LMC/SMC) as they offer a view of a large number of massive stars at known distances and low reddening. We perform new UV photometry on images from the \emph{Swift-}UVOT Magellanic Cloud Survey \cite{Siegel2015,Hagen2017}. These images cover $\sim$3(9) deg$^2$ of the SMC(LMC) in three UV filters spanning 1928--2600~\AA\, with a resolution of 2.5$''$.
To partially mitigate effects of crowding, we use the forward modeling code \emph{The Tractor} \cite{Lang2016b} to perform forced PSF photometry at the location of known stars from the optical ground-based Magellanic Cloud Photometric Survey \cite{Zaritsky2002,Zaritsky2004}. We obtain new UV magnitudes for over 500,000 sources in the direction of the LMC/SMC \cite{MM}. These are shown in a UV-optical CMD in Figure~\ref{fig:CMD}. Both a dense band representing the MS and many candidate stripped helium star binaries bluewards of the MS are evident. 

Here, we present a sample of 25 of these blue systems, identified as having luminosities, colors, and spectral morphologies consistent with predictions for binaries containing intermediate mass helium stars \cite{MM}. We obtained between 1 and 30 spectra for each system with the MagE spectrograph mounted on the 6.5m Magellan-Baade telescope at Las Campanas Observatory.
All 16 systems with $>$1 epoch show radial velocity variations indicative of a binary nature (Table~\ref{tab:kinematics}). 

To select objects with brightnesses in excess of those expected for low-mass subdwarfs, we use kinematics to reject any likely foreground objects. 
The average radial velocities for all 25 systems are consistent with expectations for stars the Magellanic Clouds. When coupled with proper motions from \emph{Gaia}-DR3, 23 systems also show 3D motion consistent with known O/B-type stars in the LMC/SMC \cite{MM}. The remaining two objects (stars 5 and 6) show slight offsets in proper motion, but have highly significant excess-noise and poor goodness-of-fit statistics in \emph{Gaia}. We therefore include them as their radial velocities are consistent with LMC membership and their spectra closely resemble other stars in the sample.

The 25 stars are shown as numbered circles in Figure~\ref{fig:CMD}, where we have adopted distances of 50 and 61 kpc and reddening values of A$_{\rm{V}} = $ 0.38 and 0.22 mag for the LMC and SMC, respectively \cite{MM}. The stars are of similar brightness to late O/B-type MS stars (M$_{\rm{init}}$ $\lesssim$18 M$_{\odot}$) but---for these reddening values---appear bluewards of a theoretical Zero-Age MS (ZAMS) in nine distinct UV-optical CMDs \cite{MM}. They have similar UV-optical colors to WR stars but are intrinsically fainter.  For some systems, the observed colors and magnitudes approach predictions for \emph{isolated} helium stars with masses between $\sim$2--8 M$_\odot$.

\begin{figure}
\includegraphics[width=0.75\textwidth]{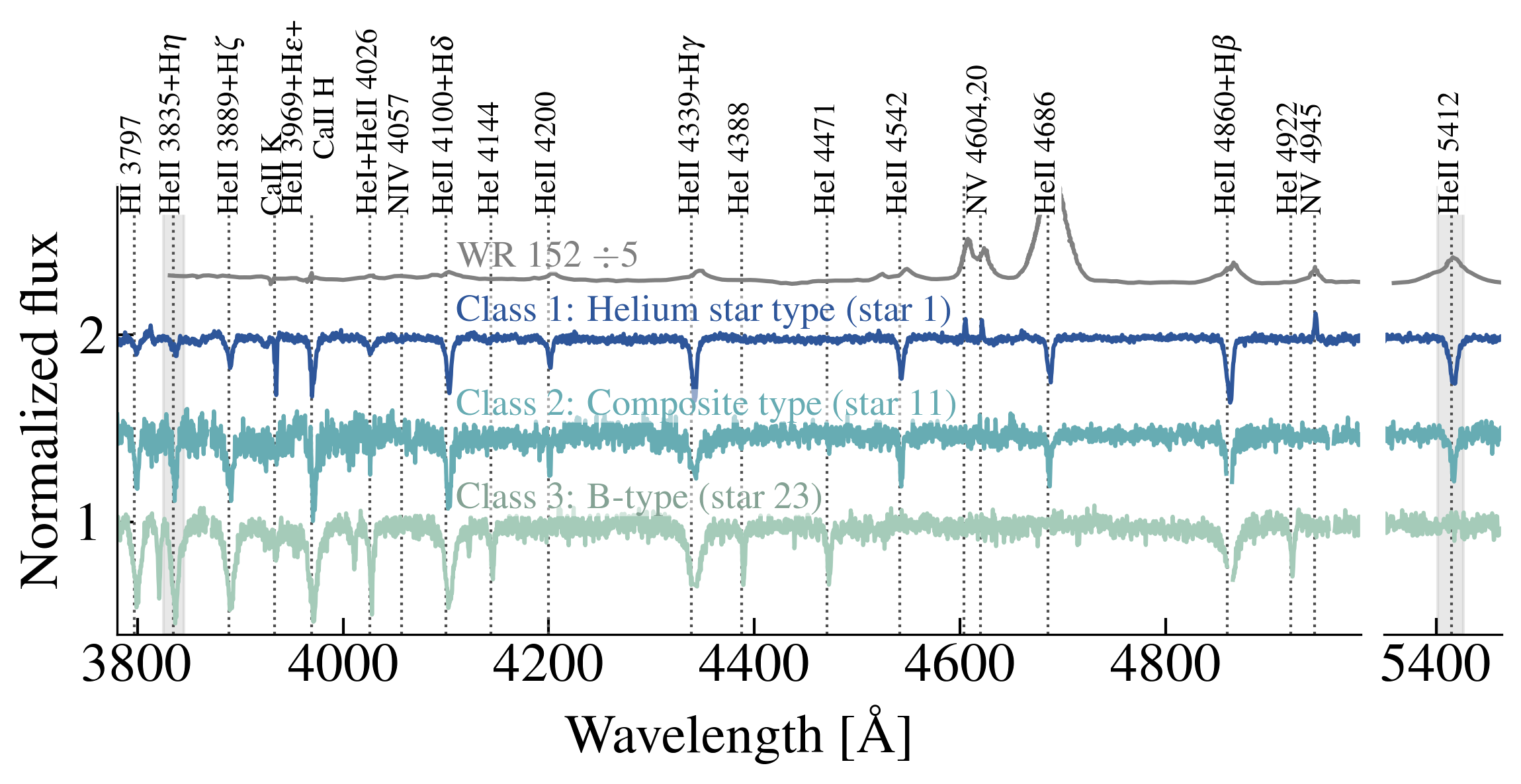}\\
\includegraphics[width=0.75\textwidth]{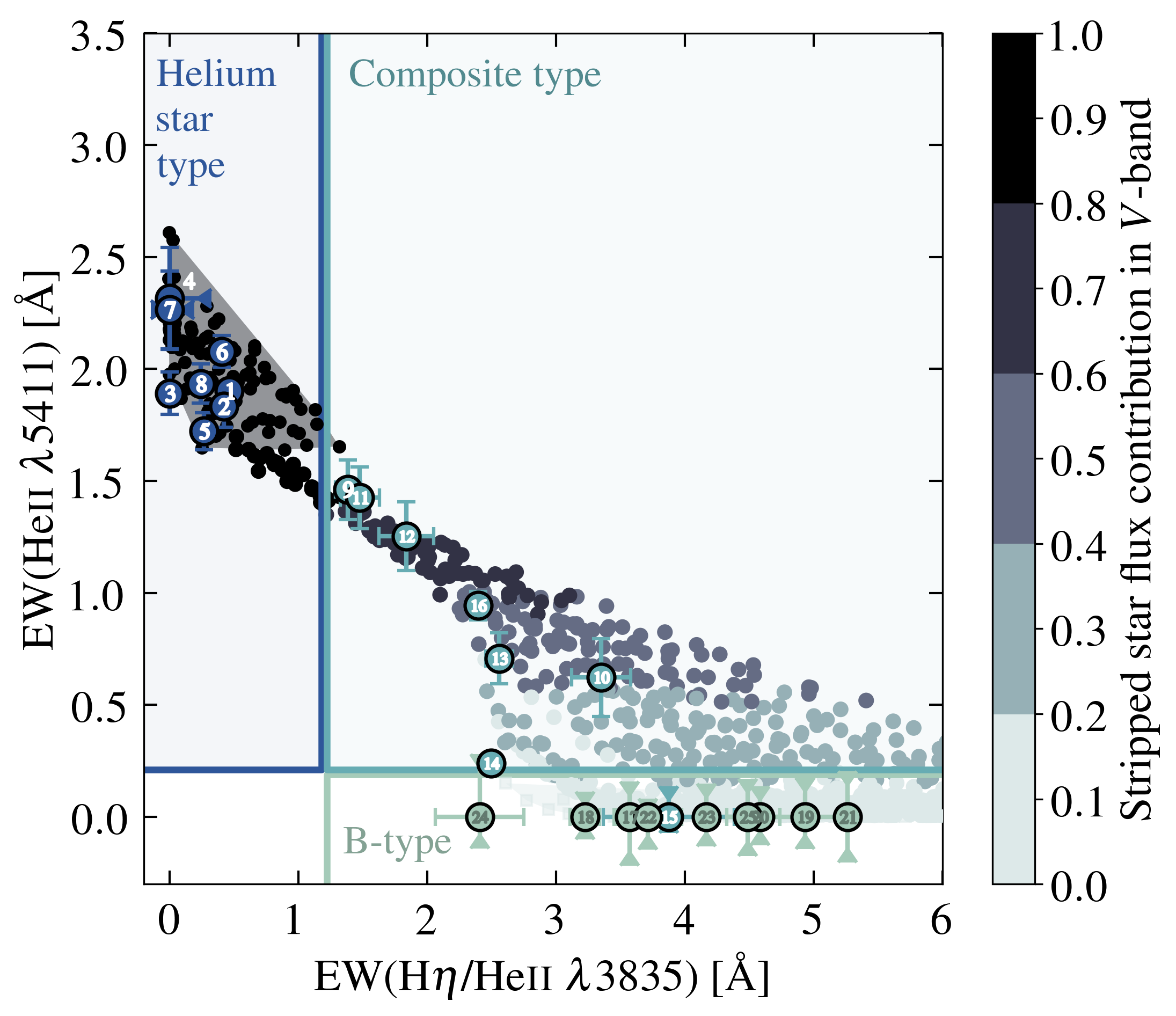}
\caption{{\bf Identification of Three Spectral Morphologies in Stars Bluewards of the ZAMS.} In panel A (top) we show three representative spectra of stars in Class 1, 2 and 3 which are characterized by a presence or lack of significant (i) \HeII\ absorption as expected for hot stars and (ii) short-wavelength Balmer lines characteristic of cooler B-type MS stars. For reference, we also show the optical spectrum of a WR star, which shows similar line transitions as the Class 1 stars in our sample, but in emission. In panel B (bottom) we plot the EWs of \HeII\ $\lambda 5411$ versus H$\eta$/\HeII\ $\lambda 3835$ for our full spectroscopic sample (large numbered circles) in comparison to models for single stripped stars, single B-type MS stars, and composites of the two (dots) \cite{MM}. Models are color-coded based on the fraction of $V-$band flux contributed by a stripped star. The observed sample forms a smooth sequence that overlaps with predictions from theoretical models of stripped helium-star binaries. Class 1 and 3 stars overlap with models dominated by the helium star and B-type star, respectively, while Class 2 overlaps with models that have non-negligible contributions from both.
}\label{fig:SpT}
\end{figure}

Optical spectra for these 25 stars all fall into three broad morphological classes (examples are shown in Figure~\ref{fig:SpT}a; for all spectra see Figures~\ref{fig:optical_spectra_isolated}--\ref{fig:optical_spectra_Btype2}):
\begin{itemize}
\vspace{-0.075in}
    \item Class 1 (8 stars): Spectra are dominated by He\textsc{ii} absorption lines and, in some cases, N\textsc{iv}/N\textsc{v} in emission and/or absorption. These are spectral transitions observed in only very hot stars, such as WR or the earliest O-type MS stars.
    \vspace{-0.1in}
    \item Class 2 (8 stars): Spectra show He\textsc{ii} absorption, but \emph{also} display significant short-wavelength Balmer lines. In two cases, a lack of accompanying He\textsc{i} lines firmly establish the presence of two stellar temperatures. 
    \vspace{-0.1in}
    \item Class 3 (9 stars): No He\textsc{ii} lines are observed in the optical. Spectra are dominated by strong Balmer and He\textsc{i} absorption, closely resembling those of B-type MS stars.  
    \vspace{-0.075in}
\end{itemize}
\noindent We emphasize that while some Class 3 stars may be single or binary B-type MS stars with over-corrected reddening, the location of the Class 1/2 stars on the CMD are inconsistent with O-type MS stars that show He\textsc{ii} for \emph{any} reddening value and their absorption-line morphology is distinct from WR stars (see Figure~\ref{fig:SpT}a). 

Instead, both the CMD locations and spectral morphologies of these stars are consistent with expectations for binary systems containing hot intermediate-mass helium stars. Synthetic spectra of such objects show similar line transitions to WR stars, but with significantly weaker emission or even absorption lines due to their lower luminosities and mass-loss rates \cite{MM}. As a result, we find the same three broad spectral classes in a set of helium star plus MS star composite spectral models, where the progression from Class 3 to 1 represents an increasing contribution from the helium star to the optical flux of the system \cite{MM}. To illustrate this, in Figure~\ref{fig:SpT}b we plot the equivalent width (EW) of He\textsc{ii} $\lambda$5411 vs.\ H$\eta$/He\textsc{ii} $\lambda$3835 for both our observed sample and the composite models. The former spectral line is chosen to probe the helium star, as it is not expected at the cooler temperatures of B-type MS stars, and the latter to probe the MS companion, as significant short-wavelength Balmer lines are not expected in hot hydrogen-depleted stars. 

The 25 observed stars all overlap in this parameter space with predictions from the composite models. For Class 3 stars, in which no He\textsc{ii} is detected, we find upper limits on He\textsc{ii} $\lambda$5411 EW of $\lesssim$0.2 \AA\ (Table~\ref{tab:ew}). This corresponds to portions of the model grid where the helium star contributes $<$20\% of the optical flux and hence appear as ``B-type'' MS spectra. While such stars can show a UV excess, they are only predicted relatively close to the ZAMS (as observed for the Class 3 stars in Figure~\ref{fig:CMD}).
In contrast, the Class 1 stars all have H$\eta$/He\textsc{ii} $\lambda$3835 EWs of $<$1.2 \AA, consistent with predictions for systems where a helium star contributes $>$80\% of the optical flux and appear as isolated ``helium-star-type'' spectra.  Such systems are only produced in the model grid when any MS companion has a mass $\lesssim$ 3.5 M$_\odot$. These properties are therefore indicative of systems that either (i) have extreme mass ratios and underwent non-conservative mass transfer or (ii) have compact object as opposed to MS companions. These systems should display the bluest colors and lie close to the models for isolated helium stars. While reddening to individual objects is uncertain, this is  broadly consistent with the location of the Class 1 objects in the CMD (Figure~\ref{fig:CMD}). Finally, Class 2 stars primarily overlap with the model grid where the helium star contributes 20--80\% of the V-band flux and display ``composite-type'' optical spectra where contributions from both stars are evident. 

\begin{figure}
\centering
\includegraphics[width=0.46\textwidth]{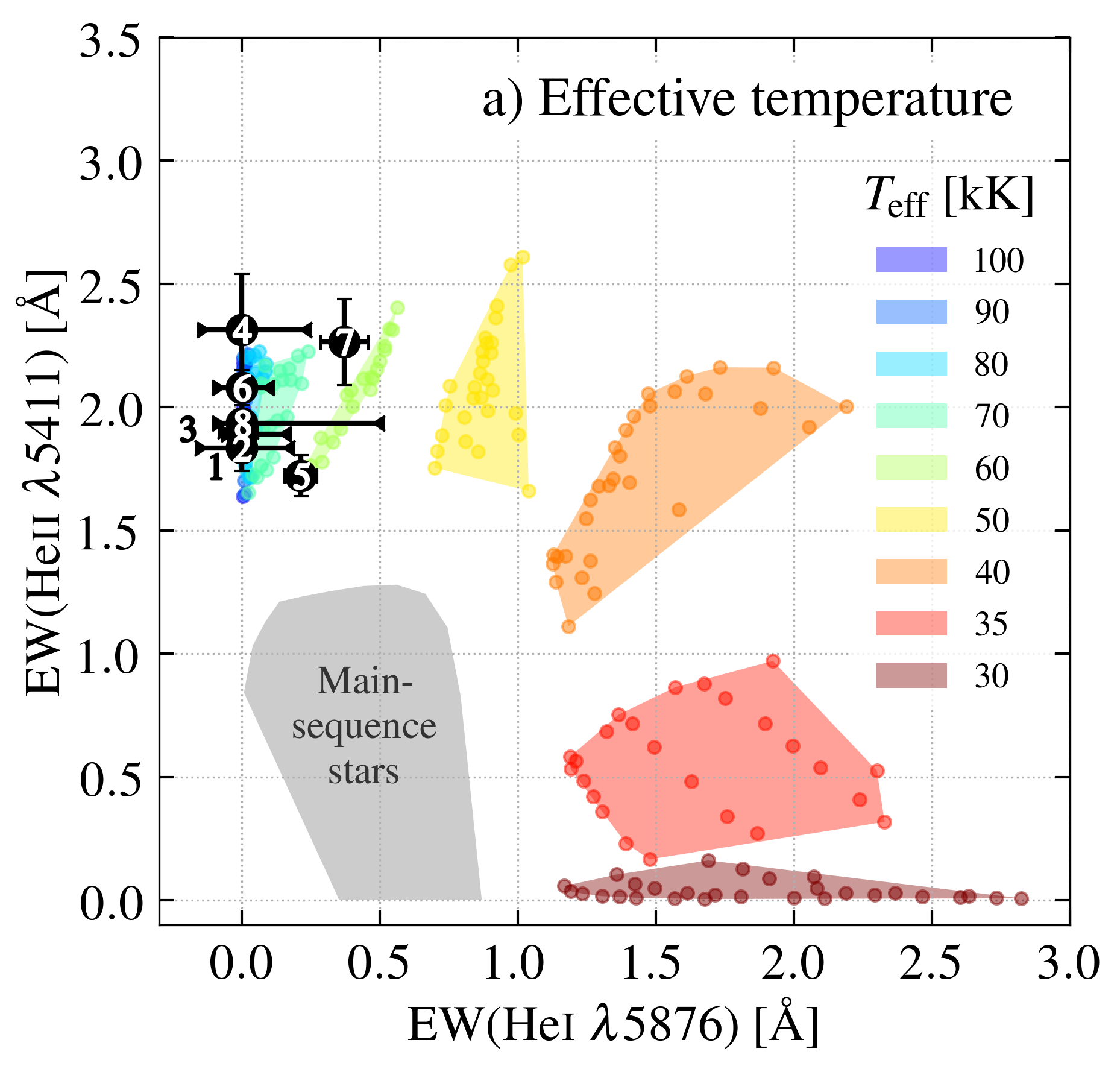}
\includegraphics[width=0.45\textwidth]{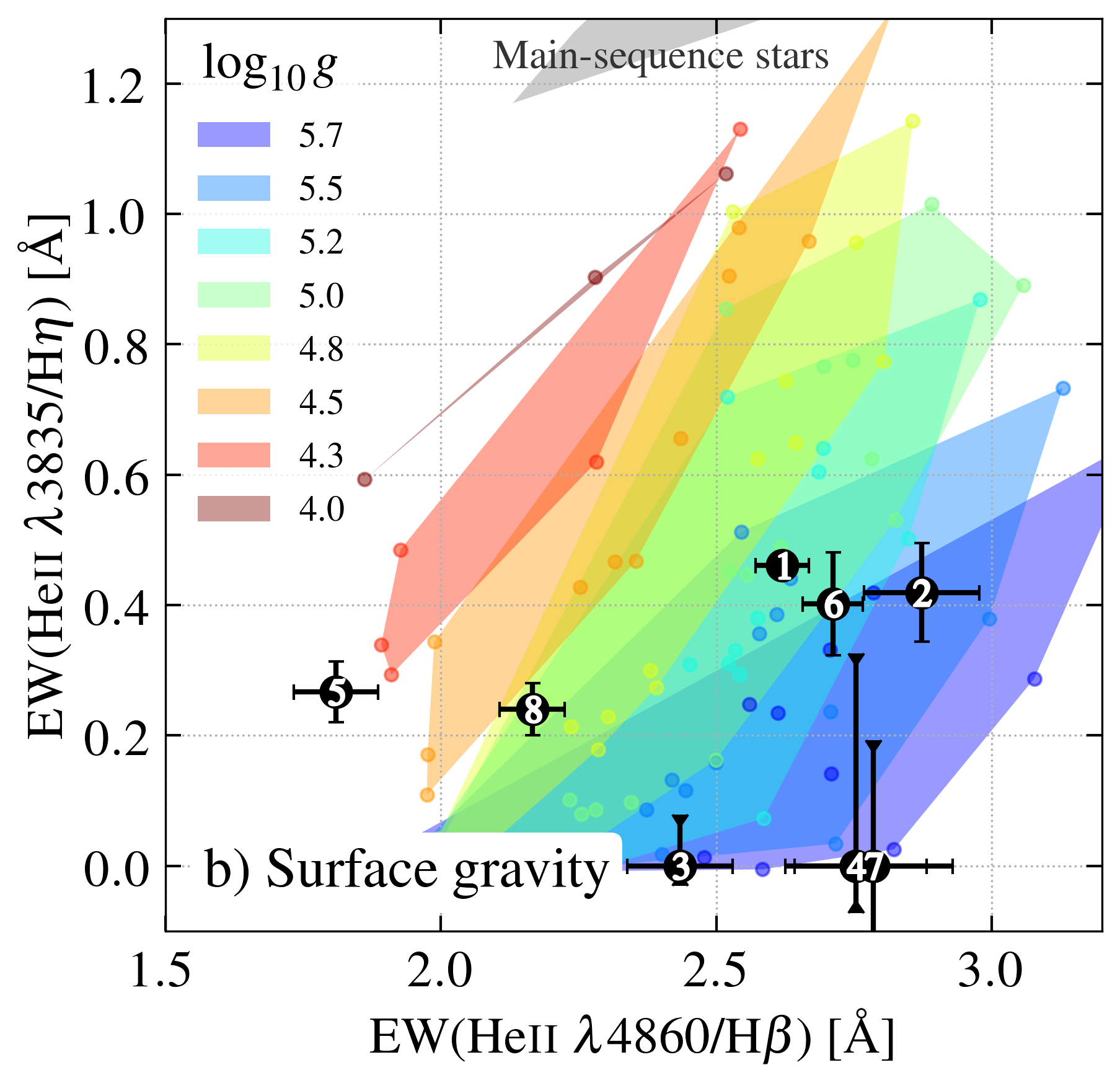}
\includegraphics[width=0.45\textwidth]{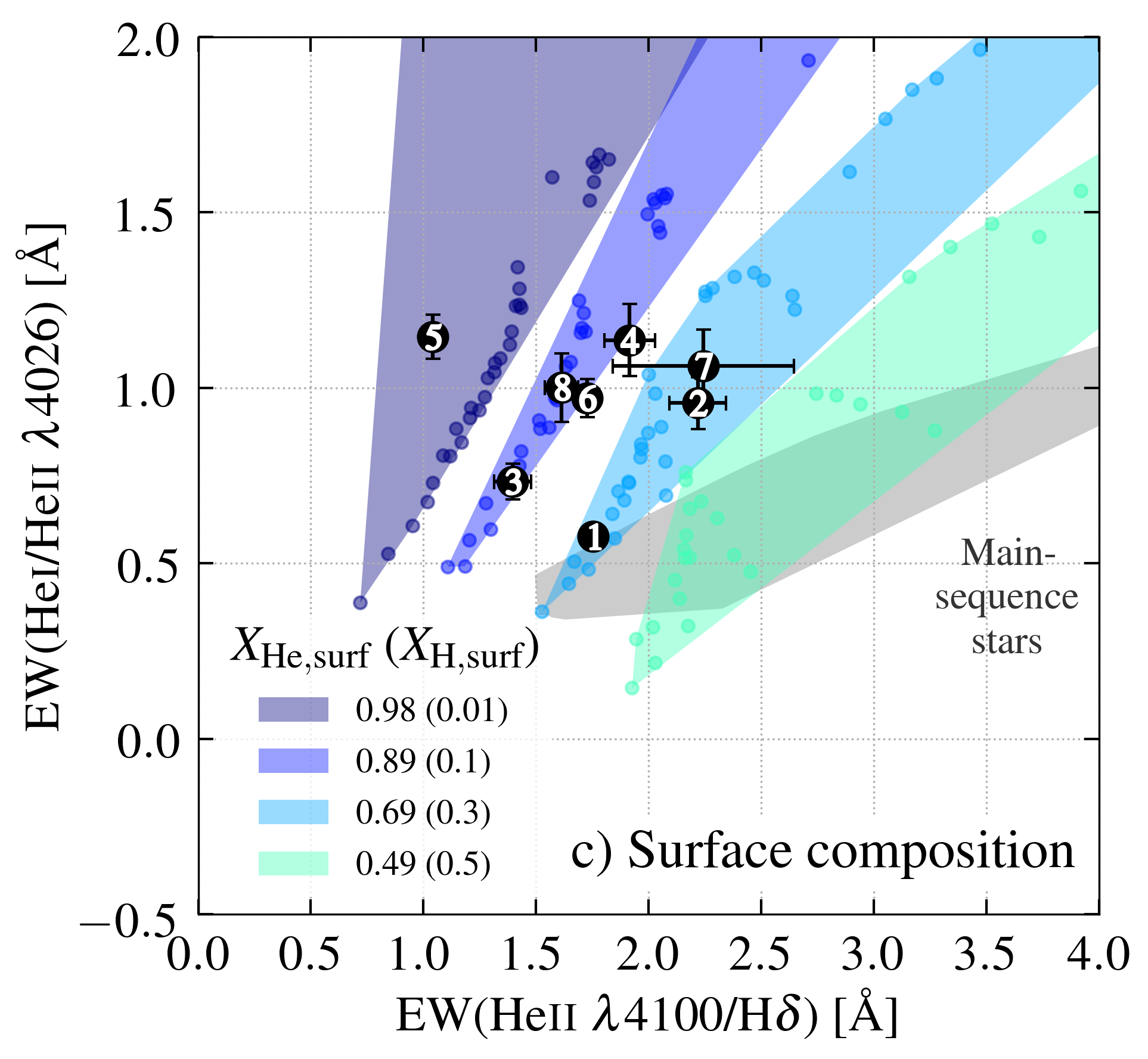}
\caption{{\bf Diagnostic Diagrams Used to Estimate Stellar Properties of Stripped Stars.} We show diagrams for three properties: effective temperature (panel A), surface gravity (panel B), and helium enrichment (panel C). In each panel, we compare measured equivalent widths for the observed stars in Class 1 (which exhibit ``helium-star-type'' spectra) with the corresponding values from a newly-computed spectral model grid \cite{MM}. Gray regions mark the location of MS star models from the TLUSTY O and B-star grids. In panel B we only show models (stripped stars and MS stars) with $T_{\rm eff} \geq 50$ kK, and in panel C we only show O-type MS star models. Based on these diagrams, we find that the Class 1 stars are hot (T$_{\rm{eff}} \gtrsim 60 \rm{kK}$), compact ($\log(g) \sim 5$), and hydrogen-poor ($X_{\rm{H,surf}} \lesssim 0.3$) and clearly distinct from MS stars.}
\label{fig:diagnostic_plots}
\end{figure}

To assess the nature of the hot stars in these systems, we develop a set of EW diagnostics that can be used to distinguish the optical spectra of stripped stars from MS stars and estimate their surface properties. We use the 1D non-LTE radiative transfer code CMFGEN \cite{1998ApJ...496..407H} to compute a new set of spectral models with a wide range of effective temperatures (30 kK $<$ T$_{\rm{eff}}$ $<$ 100 kK), surface gravities (4.0 $<$ $\log(g)$ $<$ 5.7) and depleted surface hydrogen mass fractions (0.01, 0.1, 0.3 and 0.5) \cite{MM}. These models are designed to cover a broad parameter space without making assumptions about the detailed evolutionary state of any system. In Figure~\ref{fig:diagnostic_plots} we display properties of these models compared to both the Class 1 stars and existing O and B-star MS model spectra \cite{2003ApJS..146..417L, 2007ApJS..169...83L} in three key parameter spaces. We focus on the Class 1 stars as they are expected to have minimal contamination from any companion star.

To constrain the temperatures of these stars, in Figure~\ref{fig:diagnostic_plots}A we plot the EW of the He\textsc{ii} $\lambda$5411 line vs.\ the He\textsc{i} $\lambda$5876 line, which provides a temperature diagnostic due to the shifting helium ionization balance. For all but two stars, we find \emph{lower limits} on the T$_{\rm{eff}}$ of 70kK due to the lack of detected He\textsc{i}---temperatures typical of WR stars and in excess of even the hottest O-stars \cite{2002AJ....123.2754W}. For some objects, the detection of lines of N\textsc{iv} and/or N\textsc{v} can provide a refined temperature estimate, which range from $\sim$70-80kK to $\gtrsim$90kK (see Table~\ref{tab:estimate_stellar_properties}).
In Figure~\ref{fig:diagnostic_plots}B, we assess surface gravity by plotting the EW of the He\textsc{ii} $\lambda$3835/H$\eta$ line vs.\ the He\textsc{ii} $\lambda$4860/H$\beta$ line. This provide rough diagnostic power for surface gravity due to the increase(decrease) in line strength that is observed for long(short) wavelength Balmer and Pickering lines with increasing $\log(g)$. 
The location of the observed sample is consistent with having surface gravities higher than observed in MS stars ($\log(g)\sim5$).  
Finally, in Figure~\ref{fig:diagnostic_plots}C, we probe the surface composition by plotting the pure helium blend He\textsc{i}/He\textsc{ii} $\lambda$4026 vs.\ the hydrogen/helium blend He\textsc{ii} $\lambda$4100/H$\delta$, which are impacted by the hydrogen and helium surface mass fractions.  The observed Class 1 stars are all consistent with having hydrogen-depleted surfaces, spanning the location of the the model grid from $X_{\rm{H,surf}} = 0.01$ (essentially hydrogen-free) to $X_{\rm{H,surf}} = 0.3$.

\begin{figure}
\centering
\includegraphics[width=0.6\textwidth]{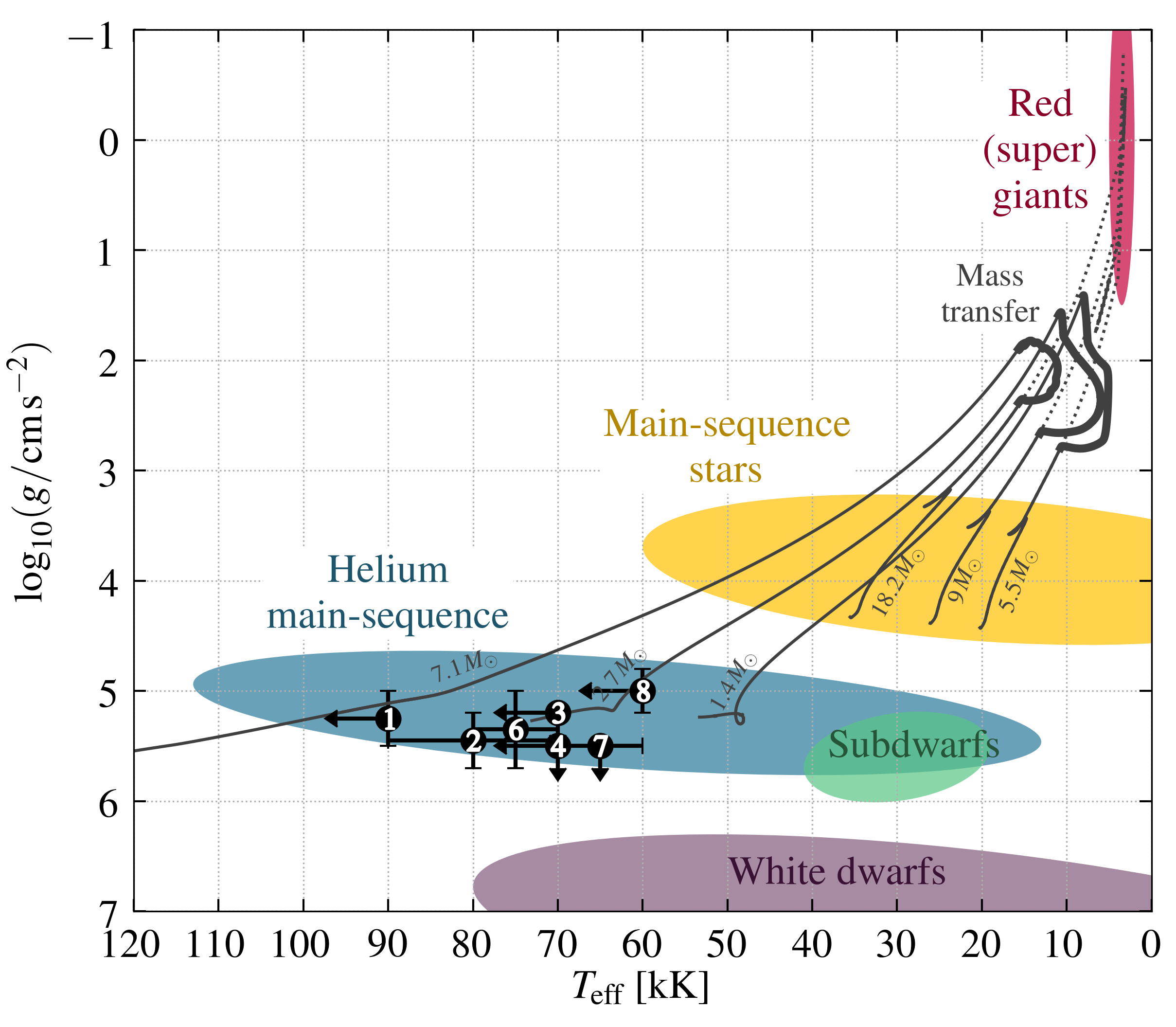}
\caption{{\bf Kiel Diagram Comparing the Physical Properties of the Class 1 Systems to Various Types of Stars.} Constraints on the effective temperatures and surface gravities of the Class 1 stars with ``helium-star-type'' spectra, are shown as numbered circles. They are consistent with expectations for stars evolving on the helium main-sequence (blue region), but distinct from either main-sequence stars (yellow region) or white dwarfs (purple region). They also show significantly hotter temperatures than observed for most subdwarfs (green region). For comparison, we also show evolutionary tracks for stars with ZAMS masses of 5.5\Msun, 9.0\Msun, and 18.2\Msun\ through the end of central helium burning \cite{MM}.  Single stars of these masses (dotted lines) evolve to burn helium as cool and extended red (super)giants (red region). In contrast, binary stars of these initial masses that undergo envelope-stripping via mass transfer (thick solid lines) evolve to burn helium as hot and compact helium stars with masses of 1.4\Msun, 2.7\Msun, and 7.1\Msun. Our observed population is consistent with predictions for binary-stripped intermediate-mass (M$\sim$2--8M$_\odot$) helium stars originating from progenitors with masses of $\sim$8--25 \Msun.
}
\label{fig:kiel}
\end{figure}

The diagnostics in Figure~\ref{fig:diagnostic_plots} demonstrate that the Class 1 stars are hot, compact, and hydrogen-poor. At the same time, Figure~\ref{fig:CMD} shows that their brightnesses complete a sequence, connecting traditional WR stars and the slightly lower luminosity WN3/O3 stars \cite{Neugent2017} to subdwarfs---a progression that is mirrored in the strength of wind features in their optical spectra (see Figure~\ref{fig:comparison_spectra}).
In Figure~\ref{fig:kiel} we plot the constraints for T$_{\rm{eff}}$ and $\log(g)$ for the Class 1 stars to further assess their consistency with theoretical predictions for intermediate mass helium stars, independent of distance and reddening. The observed stars have surface gravities between those expected for MS stars and white dwarfs---consistent with the helium MS---and temperatures hotter than most subdwarf stars \cite{2013A&A...551A..31D}. Also shown are a set of evolutionary tracks \cite{2018A&A...615A..78G}. The observed stars are consistent with expectations for $\sim$2.5--8 M$_\odot$ stripped stars, which originate from progenitors with initial masses between $\sim$9 and 25 M$_\odot$. These are massive enough to reach core-collapse \cite{2020ApJ...890...51E} and should therefore eventually explode as stripped-envelope supernovae \cite{2017ApJ...840...10Y}. As the winds from stars with initial masses $<$25 M$_\odot$ are too weak to remove the hydrogen-rich envelope \cite{2021ApJ...922...55B}, binary interaction is the primary means by which such stars can be stripped \cite{MM}.

Other possible interpretations of the nature of the stars in the observed sample all present challenges or inconsistencies \cite{MM}. Central-stars of planetary nebulae, extremely young post-AGB stars, and accreting white dwarfs may reach very high temperatures and similar brightnesses, but they are characterized by surrounding material leading to emission features or infrared excess \cite{2003ARA&A..41..391V,2003A&A...409..969H}, which we do not observe. Extremely rapidly rotating stars could potentially undergo chemically homogeneous evolution---resulting in hot and compact helium stars---but this is only expected at higher masses and luminosities \cite{2011A&A...530A.115B}. Finally, while some hot low-mass objects (e.g.\ evolved subdwarfs, white dwarf merger products) could pollute our sample if located in the Galactic halo, analysis of a control field indicates that we expect $<$1 foreground object with colors, magnitudes and kinematics similar to our spectroscopic sample along the line-of-site to the Magellanic Clouds \cite{MM}.

Thus, while the current state, binary companions, and evolutionary history of the individual systems in the observed sample are likely diverse---they clearly demonstrate that a population of massive stars stripped via binary interaction \emph{does} exist. Furthermore, as only a subset of stripped star binaries are expected to show a UV excess, this detected population represents only a small fraction of intermediate mass helium stars that are predicted. Many other systems may remain hidden by the light of their companion stars.
The stripped stars identified here are therefore valuable for constraining the physical properties of this important but elusive population. 
With estimated masses of $\sim$2--8 M$_\odot$, they fill the helium-star mass gap, connecting subdwarfs with WR stars, and represent the first sample of the most likely progenitors for stripped-envelope core-collapse supernovae that can be studied in detail.

\clearpage 

\clearpage

\def\aap{{A\&A}}
\def\aapr{{A\&AR}}
\def\aj{{AJ}}
\def\apj{{ApJ}}
\def\apjs{{ApJS}}
\def\apjl{{ApJL}}
\def\araa{{ARA\&A}}
\def\mnras{{MNRAS}}
\def\nat{{Nature}}
\def\pasp{{PASP}}
\def\physrep{{Physics Reports}}
\def\ssr{{Space Sci.\ Rev.}}
\def\zap{{ZAP}}
\def\aaps{{A\&AS}}

\section*{Acknowledgements}

The authors would like to thank Katie Auchettl, Katie Breivik, Chris Burns, Ashley Carpenter, Jim Fuller, Beryl Hovis-Afflerbach, Alexander Ji, Cole Johnston, Dan Kelson, Marten van Kerkwijk, Dustin Lang, Colin Norman, Tony Piro, Mathieu Renzo, Anaelle Roc, Peter Senchyna, Santiago Torres, and Manos Zapartas for fruitful discussions and support. 
This paper includes data gathered with the 6.5 meter Magellan Telescopes located at Las Campanas Observatory, Chile. We pass our extreme gratitude to John Mulchaey (Carnegie Observatories director), Leopoldo Infante (Las Campanas Observatory director), and the entire Las Campanas staff for their hard work and dedication to keep the observatory operating through the COVID-19 pandemic, when a significant fraction of this data was obtained.

M.R.D.\ acknowledges support from the NSERC through grant RGPIN-2019-06186, the Canada Research Chairs Program, the Canadian Institute for Advanced Research (CIFAR), and the Dunlap Institute at the University of Toronto. Y.G.\ by NASA through the NASA Hubble Fellowship Program grant \#HST-HF2-51457.001-A awarded by the Space Telescope Science Institute, which is operated by the Association of Universities for Research in Astronomy, Inc., for NASA, under contract NAS5-26555. A.J.G.O.\ acknowledges support from the Lachlan Gilchrist Fellowship Fund.  Support was provided to N.S.\ by the National Aeronautics and Space
Administration (NASA) through HST grant GO-15824.
Computing resources used for this work were made possible by a grant from the Ahmanson Foundation.

Final reduced and combined spectra for the 25 stars presented in this work, as well as the new stripped helium star and main sequence spectral models will be available for download from zenodo upon publication. In addition, we provide $UVW2, UVM2, UVW1, U, B, V, I-$band photometry for the observed stars and models in Tables~\ref{tab:absolute_magnitudes_OB},~\ref{tab:absolute_magnitudes_stripped_stars}, and ~\ref{tab:photom} and in machine readable form at the same repository.

\clearpage

\noindent {{\bf Supplemetary Materials:}\\
{\tt www.sciencemag.org}\\
Materials and Methods \\
Figures~S1--S18\\
Tables~S1--S8\\
References $36$--$127$}

\clearpage
\setcounter{page}{1}
\setcounter{figure}{0}
\renewcommand{\thefigure}{S\arabic{figure}}
\renewcommand{\thetable}{S\arabic{table}}
\renewcommand{\thesection}{S\arabic{section}}
\renewcommand{\theequation}{S\arabic{equation}}

\begin{center}

\title{{\LARGE Supplementary Materials for}\\[0.5cm] {\bf\large{} Discovery of the missing intermediate-mass helium stars stripped in binaries}}

\author{M.\ R.\ Drout$^{1,2,*,\dagger}$, Y.\ G\"{o}tberg$^{2,*,\dagger}$, B.\ A.\ Ludwig$^{1}$, J.\ H.\ Groh$^{3}$, S.\ E.\ de Mink$^{4,5,6}$, A.\ J.\ G.\ O'Grady$^{1,7}$, N.\ Smith$^{8}$ \\
}

\normalsize{$^{*}$Correspondence to: maria.drout@utoronto.ca, ygoetberg@carnegiescience.edu}\\
\normalsize{$^{\dagger}$These authors contributed equally.}

\end{center}

\baselineskip 12pt

\noindent {{\bf This PDF file includes:}\\
\indent \indent  Materials and Methods \\
\indent \indent Figures~S1--S18\\
\indent \indent Tables~S1--S8\\
\indent \indent References $36-127$}

\renewcommand*\contentsname{Detailed Contents of Materials and Methods}
\setcounter{tocdepth}{2}
\tableofcontents

\clearpage

\noindent {\bf \LARGE Materials and Methods}

\section{Theoretical Models of Main Sequence and Stripped Helium Stars}\label{sec:models}

Throughout this work, we use a combination of binary stellar evolution and spectral models to both (i) predict the characteristics of binaries containing stripped helium stars and (ii) to estimate the stellar properties of an observed sample of stars based on their optical spectra. Here we summarize these models and their primary uses in our analysis. We convert all spectral models to air wavelengths before use.

\subsection{MESA Models of Main Sequence and Stripped Binary Stars}\label{sec:evolutionary_models}

We use stellar evolution models computed with MESA to determine realistic surface properties of both main sequence (MS) and stripped stars. These are required for three distinct reasons: (i) as inputs for a spectral model grid (\S~\ref{sec:MESAspec} and \ref{sec:MESAMSspec}), (ii) to determine the appropriate flux scaling for each component when combining B-type MS and stripped helium stars in a new composite model grid (section \ref{sec:prediction}), and (iii) to compare with the physical properties estimated for our observed sample.
To achieve this, we use the binary stellar evolutionary models of \cite{2018A&A...615A..78G}, which simulate envelope-stripping via Roche-lobe overflow. Here, we summarize key inputs, predictions, and caveats for these models. We refer to  \cite{2018A&A...615A..78G} and references therein for a full description.

Models were computed using the stellar evolution code MESA (version 8118 \cite{2011ApJS..192....3P, 2013ApJS..208....4P, 2015ApJS..220...15P}). We use the grid with metallicity $Z = 0.006$, appropriate for the LMC. There are 23 models in this grid and they are separated evenly in the logarithm of the initial mass of the most massive star in the system, ranging from 2 up to 18.2\Msun. We chose the initial orbital period and mass of the secondary such that the systems experience stable mass transfer early during the Hertzsprung gap evolution of the donor star. 
These models therefore provide predictions for the properties of both MS stars (pre-interaction) and stripped stars (post-interaction). After envelope stripping, the donor stars have remaining masses between 0.27 and 7.14\Msun.

The evolutionary models predict that, during central helium burning, intermediate mass stripped stars ($\sim 2-7\Msun$) 
are very hot ($\sim 50-100$ kK), 
have high surface gravity ($\log_{10} (g/\text{cm\, s}^2) \sim 4.8-5.2$), 
and are helium-rich and hydrogen-poor ($X_{\text{He,surf}} \sim 0.5-1$, $X_{\text{H, surf}} \sim 0-0.5$). 
Because they are the exposed cores of stars that fuse hydrogen to helium via the CNO cycle, 
their surfaces are also expected to be nitrogen enriched ($X_{\text{N, surf}} \sim 0.004$). 
The models predict that stripped stars with masses $\lesssim 7 \Msun$ 
reach bolometric luminosities up to $10^5 \Lsun$.

While these models are sufficient for our present purposes, some assumptions can affect the detailed properties predicted for individual systems. The assumed wind mass loss rate for stripped stars has recently been debated and considered too high \cite{2017A&A...607L...8V}, which can affect the surface temperature and composition \cite{2019MNRAS.486.4451G}. The assumed convective overshoot is that of \cite{2011A&A...530A.115B}, which was adapted for $16\Msun$ MS stars and may be too high at the lower mass end (e.g.\ \cite{2015A&A...580A..27M}). This can impact what initial mass is inferred for a given stripped star. Finally, we note that stars stripped via common envelope ejections (CEE) remain poorly studied due to the computational challenges associated with the evolution of such systems. It is possible that stars stripped via CEE have somewhat different stellar properties than predicted by these stable mass transfer models.

\subsection{Spectral Models of Stripped Helium Stars}\label{sec:general_strip_spectral_models}

We compute two distinct grids of model spectra for intermediate mass helium stars. The first uses the physical properties from the binary evolution models described above as input, and will be used to predict the appearance (both photometric and spectroscopic) of binaries that contain stripped stars in \S~\ref{sec:prediction}. The second is instead computed for a wide range of effective temperatures, surface gravities, and surface hydrogen (and helium) mass fractions, and will be used to estimate the physical properties of our observed sample, agnostic to the evolutionary history of the system.

\subsubsection{CMFGEN Spectra Based on Binary Evolution Models}\label{sec:MESAspec}

In order to predict both the photometric and spectroscopic appearance of binaries that contain stripped helium stars it is necessary to use models that have been created with realistic stellar radii--and thus bolometric luminosities--in addition to accurate temperatures, surface gravities, surface compositions, and wind mass loss parameters. 
We therefore compute model spectra for a set of stripped helium stars based on the evolutionary models described above. We closely follow the methodology of \cite{2018A&A...615A..78G}, but with a few minor changes to the wind. Here, we briefly summarize settings used in \cite{2018A&A...615A..78G} that remain the same for the updated models followed by a description of how we adapt the treatment of stellar winds.

As in \cite{2018A&A...615A..78G}, atmosphere models are computed for stripped stars halfway through central helium burning (defined as when $X_{\rm He, center} = 0.5$) with the non-LTE radiative transfer code CMFGEN \cite{1990A&A...231..116H, 1998ApJ...496..407H}\footnote{The models presented in \cite{2018A&A...615A..78G} were computed using the CMFGEN version 30th of June 2014, while the updated models were computed using the newer version of CMFGEN of the 5th of May 2017.}. We input surface properties predicted by the binary stellar evolution models at the base of the wind and account for the presence of the elements H, He, C, N, O, Si, and Fe. To simulate the stellar wind, a standard $\beta$-law is assumed, $v(r) = v_{\infty}(1-R_{\star}/r)^{\beta}$ (the velocity law number 7 in CMFGEN is used without employing the parameter BETA2). 
The wind velocity structure parameter $\beta$ is set to 1 and the terminal wind speed is set to be $v_{\infty} = 1.5\times v_{\text{esc}}$, where $v_{\text{esc}}$ is the surface escape speed (following the relation for WR stars, \cite{1999isw..book.....L,2000A&A...360..227N}). This results in terminal wind speeds in the range $\sim 1300-2200 \kms$, with increasing speeds for increasing stellar mass. 

The models of \cite{2018A&A...615A..78G} were computed using an extrapolation of the empirical wind mass loss rate prescription for Wolf-Rayet stars from \cite{2000A&A...360..227N}, resulting in mass loss rates between $10^{-7.8}$ and $10^{-6} \Msunyr$ for stripped stars more massive than 2\Msun. However, recent theoretical work has suggested that intermediate mass stripped stars likely have lower mass loss rates \cite{2017A&A...607L...8V, 2020A&A...634A..79S}. This is because, unlike Wolf-Rayet stars, intermediate mass stripped stars have luminosities far from the Eddington limit and, therefore, radiation pressure should not boost their wind mass loss rates (cf.\ \cite{2014A&A...570A..38B}). Mass loss rates have a large impact on the predicted spectral morphology of hot stars, as various features shift from emission to absorption and, if the mass loss rate is sufficiently strong, also the effective temperature is affected. Thus, to obtain more realistic spectral morphologies, we compute new model atmospheres where we decreased the wind mass loss rates by a factor of $\sim$50-100 for models with initial masses $>6\Msun$, reaching mass loss rates of $10^{-10}$-$10^{-8} \Msunyr$. A comparison between the wind mass loss rates assumed in \cite{2018A&A...615A..78G} and the updated ones can be seen in \tabref{tab:updated_stripped_star_models}. 

In the previous models, the higher mass loss rates allowed us to compute the stellar atmospheres out to about 500 times the stellar radius (see RMAX, \tabref{tab:updated_stripped_star_models}). But, with the new weaker wind mass loss rates, the atmosphere is very dilute at such large distances. Thus, to reach numerical convergence we decreased the extent of the atmospheres, as displayed in \tabref{tab:updated_stripped_star_models}. 
For most of the models, we also increased the number of mesh points, from around 40-80 to 80 or above (see ND, \tabref{tab:updated_stripped_star_models}).  
Finally, we account for wind clumping by assuming a volume filling factor. In the original models we assumed the volume filling factor to be 0.5 for stars with initial masses $<14\Msun$ and 0.1 for stars with initial masses $>14\Msun$. In accordance with the assumed weaker winds, we also update the volume filling factor to 0.5 for all stars. 

\begin{figure}
\centering
\includegraphics[width=0.9\textwidth]{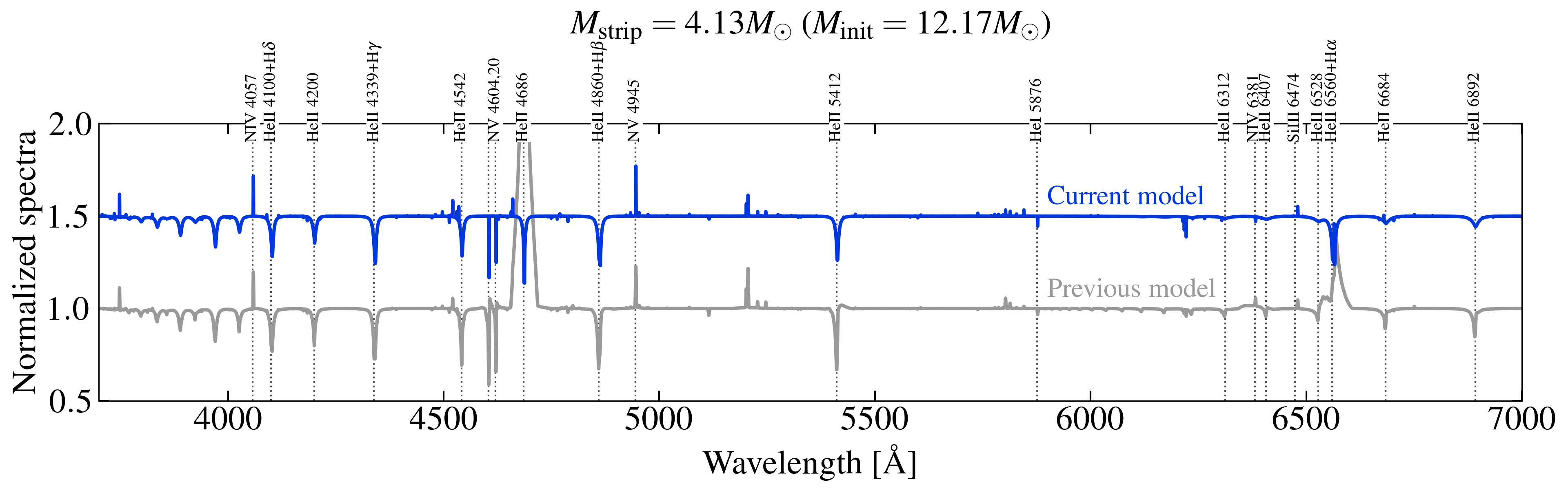}
\caption{Comparison of the current (blue) and previous (gray) models of the optical spectra for the 4.13\Msun\ stripped star in the $Z =0.006$ grid. The previous model was published in \cite{2018A&A...615A..78G}, while the current is a version with decreased wind mass loss rate (see \tabref{tab:updated_stripped_star_models}).}
\label{fig:example_update}
\end{figure}

As expected, the most noticeable difference between the original models from \cite{2018A&A...615A..78G} and the updated ones is the spectral morphology. In the updated spectral models, all stars have almost pure absorption line spectra, while in the previous models they had both pure emission line spectra and spectra with both absorption and emission lines. As an example, we compare the optical spectra of the models for the 4.13\Msun\ stripped star, originating from a 12.17\Msun\ progenitor, in \figref{fig:example_update}. The figure shows that the strong emission features and P~Cygni profiles created in the stellar wind in the previous model have disappeared and turned into absorption in the current model. There remain, however, a few weak and narrow emission features of highly ionized nitrogen that originate from photospheric processes unrelated to the stellar wind (e.g., \cite{2011A&A...536A..58R,2012A&A...537A..79R}).  

We show the normalized optical spectra for the full spectral model grid in \figref{fig:optical_spectra_006}, highlighting the updated models in color. As shown in \figreftwo{fig:example_update}{fig:optical_spectra_006}, the characteristic spectral lines remain the same as in \cite{2018A&A...615A..78G}: the Balmer series blended with every second \HeII\ line in the Pickering series, the \HeII\ Pickering series, \HeII\ $\lambda 4686$, \HeI\ lines, \NIV\ and \NV\ lines.
The absolute magnitudes of the models in the grid for ultraviolet Swift-bands and optical UBVI bands are shown in \tabref{tab:absolute_magnitudes_stripped_stars}. We present the absolute magnitudes first in the AB magnitude system and second in the Vega magnitude system. 
With values between $1.0$ AB mag $>$ M$_V$ $>$ $-1.5$ AB mag, stripped stars with masses above 2\Msun\ should have absolute V-band magnitudes similar to 2--8\Msun\ main-sequence stars, while their hot temperature lead to notably bluer colors. In the ultraviolet, their brightnesses ($-0.5$ AB mag $>$ M$_{UVM2}$ $>$ $-3$ AB mag) are more comparable to 5--11\Msun\ main-sequence stars (see \S~\ref{sec:MESAMSspec}).

\subsubsection{CMFGEN Spectra Covering a Large Range of Stripped Star Physical Properties}\label{sec:agnostic_grid}

In order to estimate physical properties for an observed sample of stripped stars, a spectral model grid that covers large parts of the parameter space is necessary. In this way, such analysis remains as neutral as possible to assumptions about the current evolutionary state and previous history of the system. For the spectral analysis, we therefore computed a new spectral model grid using CMFGEN. The considered parameter space covers: effective temperatures of 30, 35, 40, 50, 60, 70, 80, 90, and 100kK, surface gravity $\log_{10} (g/\mathrm{cm\, s}^{-2}) = $ 4.0, 4.3, 4.5, 4.8, 5.0, 5.2, 5.5, and 5.7, and surface hydrogen(helium) mass fraction $X_{\text{H,surf}}$($X_{\text{He,surf}}$) $=$ 0.01(0.98), 0.1(0.89), 0.3(0.69), and 0.5(0.49). 
Models with the combination of low surface gravity and high effective temperature did not converge, but it is also unlikely to find a very hot, but fluffy stripped star. In total, the grid is composed of 208 converged spectral models\footnote{\YG{Link to where these models are publicly available.}}. 

Because the shape and strength of the spectral lines of a stripped star primarily are determined by the effective temperature, surface gravity, and hydrogen/helium mass fraction, we choose to keep the remaining input parameters fixed in this exploratory grid. Informed by the properties predicted by the evolutionary models (\S~\ref{sec:evolutionary_models}), we set the stellar radius to $0.5 R_\odot$ and choose the surface mass fractions of carbon, nitrogen, oxygen, silicon and iron to be $X_{\text{C,surf}} = 3\times 10^{-5}$, $X_{\text{N,surf}} = 4\times 10^{-3}$, $X_{\text{O,surf}} = 10^{-4}$, $X_{\text{Si,surf}} = 1.5\times 10^{-4}$, and $X_{\text{Fe,surf}} = 2.5\times 10^{-4}$, respectively.

We produce absorption line spectra (as found in our observed sample; \S~\ref{sec:samplesum}) by setting the terminal wind speed to $v_{\infty} = 2500\kms$, the volume filling factor that accounts for wind clumping to 0.5, and the wind mass loss rate to the lowest possible in the range $10^{-8} - 10^{-10}\Msunyr$, along with assuming a standard $\beta$-law for the wind velocity profile and setting $\beta = 1$.
The extent of the atmosphere (RMAX) is typically 10-100 times the stellar radius and the number of mesh points (ND) is 40 or above. 

\subsection{Spectral Models of Main Sequence Stars}\label{sec:general_MS_spectral_models}

We require spectral models of main sequence stars both (i) to predict the photometric and spectroscopic appearance of stripped star plus MS binaries, and (ii) as a control sample against which the physical properties of our observed sample can be compared. We use a combination of both newly computed models and models available in the literature. 

\subsubsection{CMFGEN Spectra Based on MESA Main Sequence Models}\label{sec:MESAMSspec}

Because the analysis of spectral morphology requires high-resolution spectra, we compute new CMFGEN models for MS stars that will be used to develop predictions for stripped star plus MS binaries (\S~\ref{sec:prediction}). Models are computed for MS stars that are hotter than 10,000K, which is the limit down to which spectra can be computed with CMFGEN. This temperature is roughly coincident with the boundary between A and B type stars, and we also find that this boundary lies somewhere between 2 and 3\Msun\ depending on how evolved the main sequence star is. 

Similar to the models of \cite{2014A&A...564A..30G, 2018A&A...615A..78G}, we use the surface properties predicted by evolutionary models as input at the base of the atmosphere. We use the evolutionary models described in \S~\ref{sec:evolutionary_models}, which all have unperturbed main-sequences as interaction is initiated during the Hertzsprung gap. Because main-sequence stars expand, we compute three spectral models for each evolutionary model, which correspond to when the star has reached 20\%, 60\%, and 90\% of the duration of its main-sequence evolution. We compute models for stars with initial masses between 2.21 and 18.17\Msun. 

We base the computations on the OB-star grid available on CMFGEN's website\footnote{\url{http://kookaburra.phyast.pitt.edu/hillier/web/CMFGEN.htm}}. In addition to hydrogen, helium, carbon, nitrogen, oxygen, silicon and iron, the base grid also contains phosphorus and sulfur. We keep the surface mass fractions of these elements fixed at $X_{\text{P,surf}} = 5.679 \times 10^{-6}$ and $X_{\text{S,surf}} = 5.493 \times 10^{-4}$, respectively. For the other elements, we use the output from the evolutionary models. The late B-stars are affected by gravitational settling, causing a mild helium deficiency. 

The wind mass loss rate during the main sequence evolution is computed using the \cite{2001A&A...369..574V} theoretical algorithm, which for low-mass stars can reach very low values. For computational reasons, we place a floor on the wind mass loss rate to $10^{-12} \Msunyr$. Following the findings of \cite{1995ApJ...455..269L} for O- and early B-type stars, we set the terminal wind speed to 2.6 times the surface escape speed, which typically results in wind speeds of 2000-3000\kms. We use a standard $\beta$-law for the velocity profile and set $\beta = 1$. We assume the wind is smooth and un-clumped. These models are made publicly available\footnote{\YG{Put link here for where the B-star models are.}}.

We present the surface temperatures, surface gravities and bolometric luminosities predicted by the evolutionary models for the times we compute spectral models in \tabref{tab:OB_properties}.
The optical range of the normalized spectra of the main sequence stars 20\% through their main sequence evolution are shown in \figref{fig:optical_spectra_OB}. The figure shows that for the full main-sequence star grid, the Balmer series is dominating the optical spectrum with strong absorption lines that decrease in separation with decreasing wavelength. For stars with temperature $>15$kK, \HeI\ lines become clearly visible in the spectra (e.g., \HeI\ 4026, 4144, 4388, 4471, 4713, 4922, 5876) and remain strong throughout the rest of the grid. \HeII\ lines (e.g., \HeII\ 4200, 4542, 4686, 5411) become visible for stars with temperature $>30$kK. 

We also compute the absolute magnitudes for the spectral models computed for main sequence companions and present these in \tabref{tab:absolute_magnitudes_OB} for the models evolved 20\%, 60\%, and 90\% through the main-sequence duration. The table shows that the stars increase in optical brightness with evolution, but decrease in UV brightness. This is an effect of the star is becoming brighter and cooler during the main sequence evolution.

\subsubsection{TLUSTY and ATLAS Models}\label{sec:lit}

Finally, throughout the analysis below, we also utilize two previously released spectral atmosphere model grids: 

\emph{TLUSTY:} To compare physical properties estimated for our observed sample \secref{sec:Physical_properites} with hot main-sequence stars, we use the $Z = Z_{\odot}/2$ TLUSTY O-star and B-star model grids \cite{2003ApJS..146..417L, 2007ApJS..169...83L}. These cover effective temperatures between 15,000~K and 55,000~K, and surface gravities ($\log_{10} g$) between 1.75 and 4.75. These models are not ideal to use to represent the radiative contribution from companion MS stars when constructing a composite grid (\S~\ref{sec:prediction}) since the effective temperatures are hotter than many of the expected companions. 

\emph{ATLAS:} In order to determine unique regions in observed color-magnitude diagrams where stripped binaries, but not single MS stars, are expected to appear, we require models for the zero-age MS (ZAMS) over a wide range of masses. We create these using spectral models from the Kurucz ATLAS grids \cite{1992IAUS..149..225K} that match the properties of stars at the zero-age main sequence (ZAMS) according to the evolutionary models described in \S~\ref{sec:evolutionary_models} and originating from \cite{2018A&A...615A..78G, 2019A&A...629A.134G}. To account for the metallicity of the Large and Small Magellanic Clouds, we use the evolutionary (spectral) model grids with $Z = 0.006$ ($Z/Z_{\odot} = 10^{-0.5}$) and $Z=0.002$ ($Z/Z_{\odot} = 10^{-1}$), respectively. We note that the choice of metallicity for the spectral models minimally affects the colors of the resulting ZAMS, since the determining quantities for the spectral slope is the effective temperature and the models are shifted to match the bolometric luminosity expected from the evolutionary models.

\section{Predictions for the Photometric and Spectroscopic Appearance of Binary Systems Containing Stripped Helium Stars}\label{sec:prediction}

To represent the emission from binaries containing stripped stars, we create a composite model grid by combining the spectral models for stripped stars and MS stars that were computed based on evolutionary models and described in \S~\ref{sec:MESAspec} and \ref{sec:general_MS_spectral_models}, respectively. We combine each stripped star model with each main-sequence star model. While theory predicts that the majority of stripped stars are accompanied by MS stars (e.g., \cite{2021ApJ...908...67S,2017ApJ...842..125Z}), they can also orbit compact objects \cite{2017ApJ...846..170T} or perhaps be single, runaway stars \cite{1994A&A...290..119P, 2019A&A...624A..66R}. In the latter two cases, the observable optical spectrum is likely completely dominated by the stripped star and we therefore also include the stripped star models alone. This way, we attempt to create a diversity of binary configurations. With this composite model grid, we (1) estimate the shape of the spectral energy distributions and UV excess introduced by the stripped star, and (2) identify broad groups of spectral morphologies.

Most stripped star binaries are likely well-represented by the described composite model grid, and it should be sufficient for making the theoretical predictions we require. 
However, there are three caveats that are worth mentioning.  
First, the rapid contraction phase after detachment, when the stripped star is still inflated, is not well-represented by our models of helium-core burning stripped stars. While they are rare, because the contraction phase is short, their cooler temperatures lead to brighter optical emission.
Second, rotation in the main-sequence star, induced by mass accretion, is not accounted for. Rotation primarily broadens the spectral features, and/or sheds material leading to the Be-phenomenon \cite{1998ApJ...493..440G}, which would make it easier to distinguish the two stars in a binary (see e.g., \cite{2021AJ....161..248W}).
Third, the binary evolutionary model grid is created for stars stripped via stable mass transfer and may not completely represent stars stripped via common envelope ejection, which could, for example, be hotter. However, we expect the difference to be small. 
Overall, while each of these effects could increase the diversity of observed properties of stripped star binaries that are possible, they do not impact the main conclusions we draw from this model grid. Namely, (i) that there are unique regions of observed color-magnitude diagrams where binaries containing stripped helium stars are expected (\S~\ref{sec:UVexcess_prediction}), and (ii) a range of spectral morphologies are possible based on the relative flux contribution from the stripped star and its companion (\S~\ref{sec:SpTs}).

\subsection{The Spectral Energy Distribution of Stripped Star Plus Main Sequence Binaries and The Expectation of a UV Excess}\label{sec:UVexcess_prediction}

\begin{figure}
\centering
\includegraphics[width=0.32\textwidth]{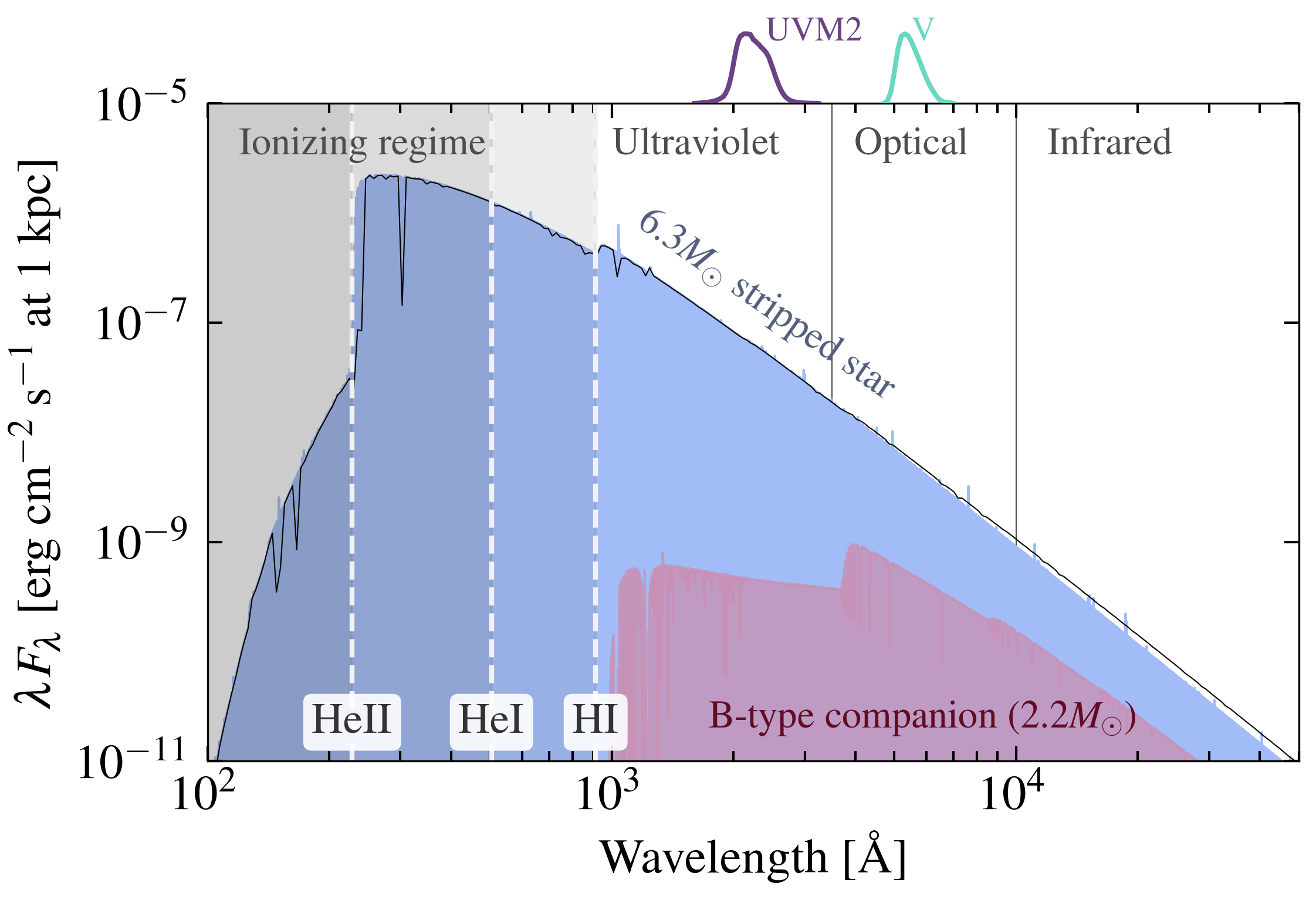}
\includegraphics[width=0.32\textwidth]{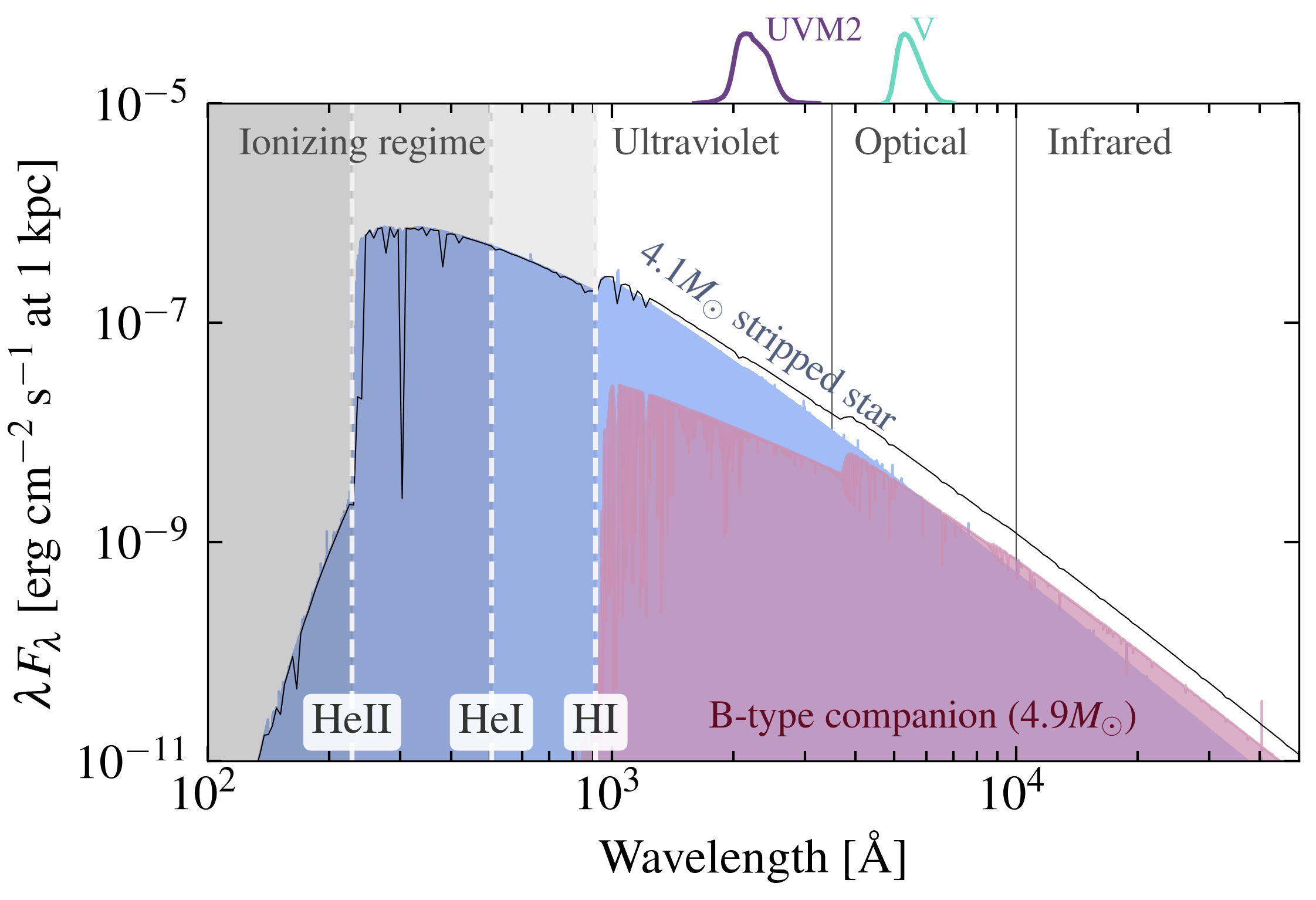}
\includegraphics[width=0.32\textwidth]{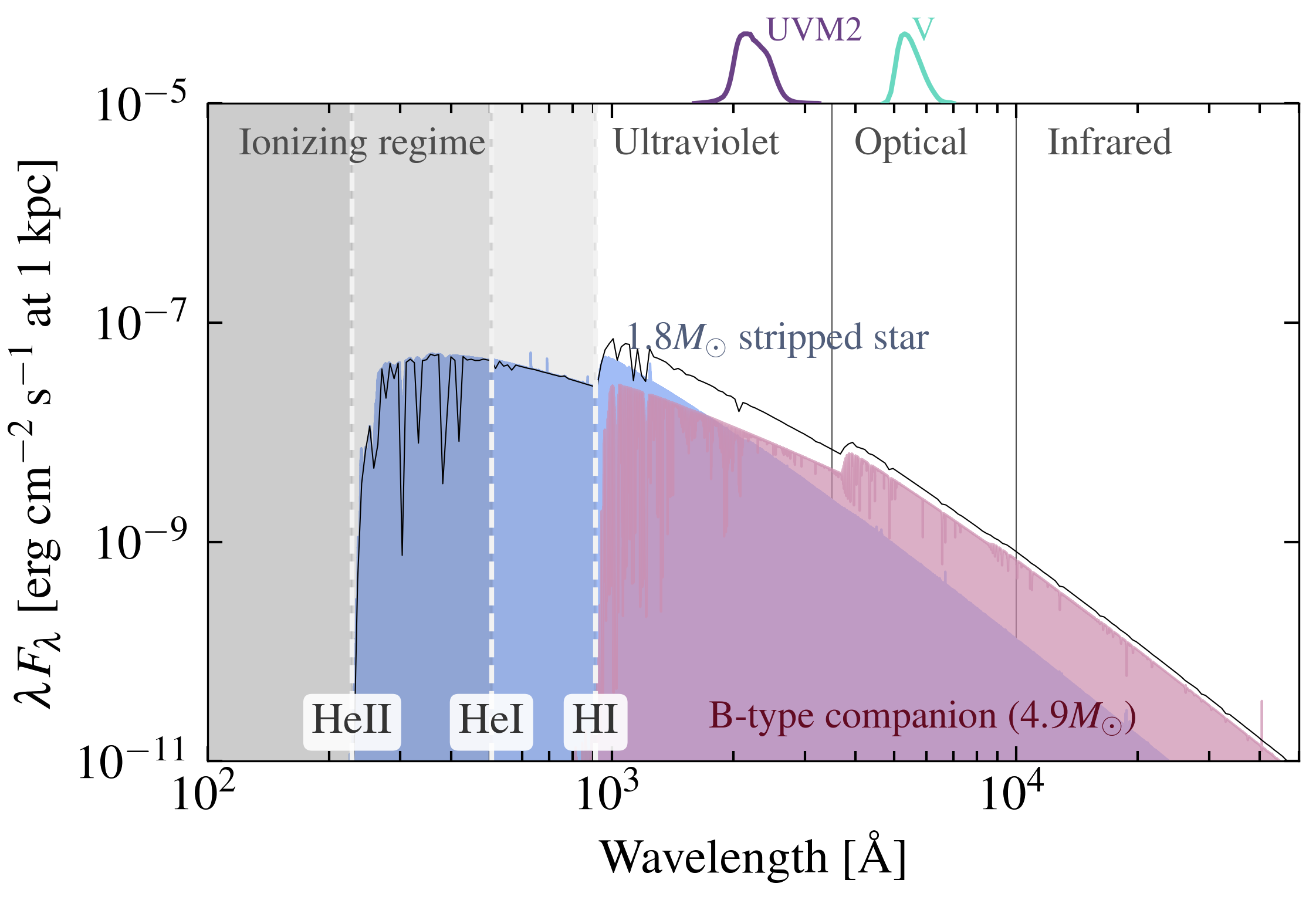}
\caption{Modeled spectral energy distributions of three binary combinations with a 6.3\Msun\ stripped star combined with a 2.2\Msun\ late B-type companion (left), a 4.1\Msun\ stripped star combined with a 4.9\Msun\ B-type companion (middle) and a 1.8\Msun\ stripped star combined with a 4.9\Msun\ B-type companion (right). The radiation from the stripped stars are shown in blue and the B-type companions in pink, while the total composite flux is indicated with a thin black line (using degraded resolution for visibility). We show the transmission curves for the ultraviolet filter Swift/UVM2 and the optical Bessell V-filter above the panels. In all three combinations, the stripped star causes the system to have UV excess.}
\label{fig:SED_MScompanions}
\end{figure}

Due to their high temperatures, helium core-burning stripped stars emit most of their radiation in the ionizing regime. As a result, they can be outshone by a bolometrically fainter B-type companion in optical and IR wavelengths but are somewhat harder to obscure in the ultraviolet (UV) \cite{2017ApJ...843...60W,2018A&A...615A..78G,2018ApJ...853..156W, 2021AJ....161..248W}. We illustrate this in \figref{fig:SED_MScompanions}, where we show three examples of binary combinations consisting of a 6.3\Msun\ stripped star combined with a 2.2\Msun\ B-type star (left), a 4.1\Msun\ stripped star with a 4.9\Msun\ B-type companion (middle), and a 1.8\Msun\ stripped star with a 4.9\Msun companion (right). In this example, we use spectral models for the companion stars that have evolved 20\% of their MS duration. The figure shows that the companion star is the bolometrically fainter of the two stars in each system, but in the optical it contributes significantly and can even be brighter than the stripped star (right panel). In the UV, on the other hand, all of the stripped stars dominate the total emission when coupled with their B-star companions. Thus, these systems are all expected to have \textit{excess} UV radiation.

\begin{figure}
\centering
\includegraphics[width=\textwidth]{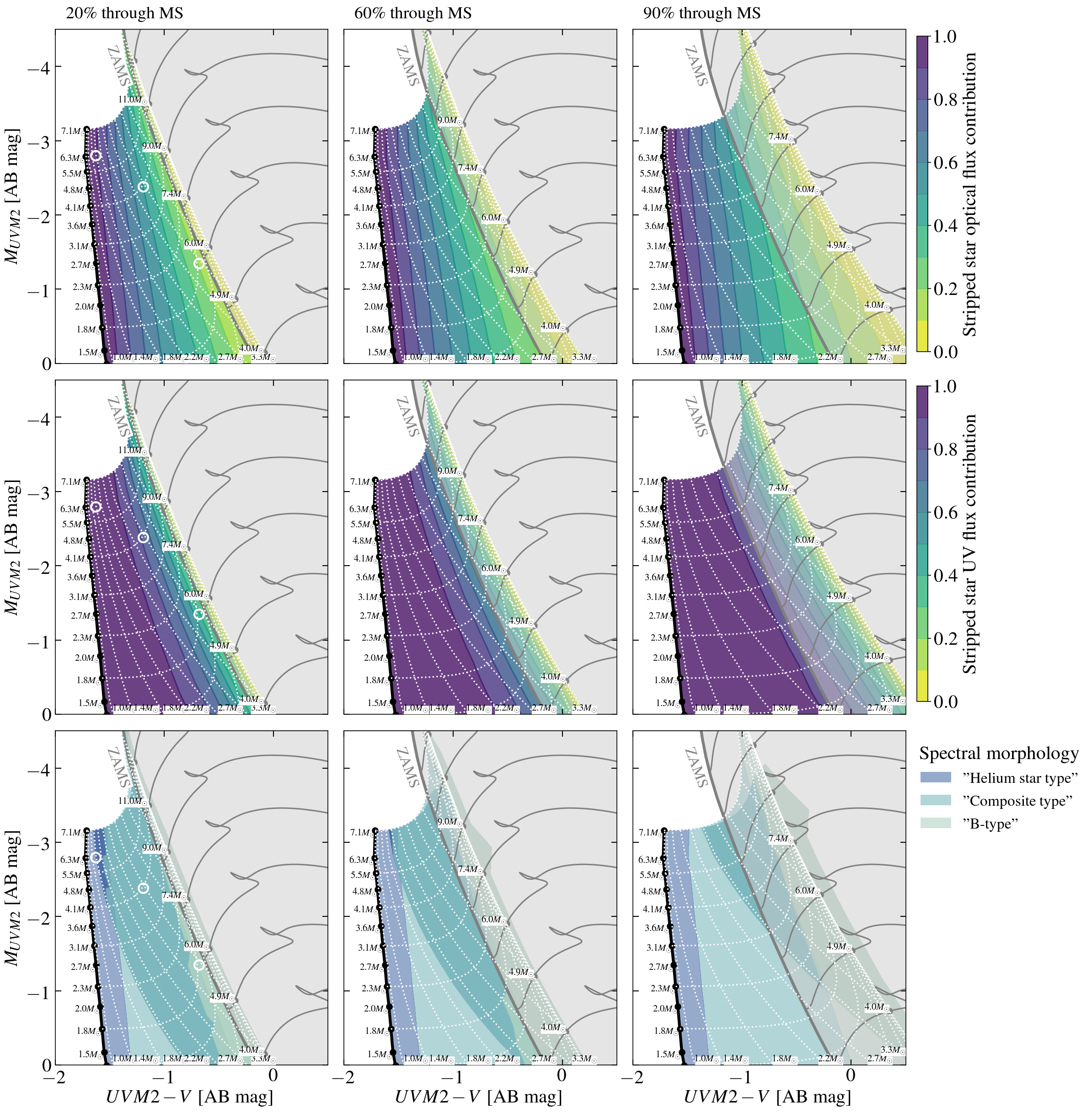}
\caption{Theoretical color-magnitude diagram using AB magnitudes in the optical Bessell/V-filter and the ultraviolet Swift/UVM2-filter. The locations of helium core burning stripped stars are shown in black dots connected with a line. Single star evolutionary tracks from the MIST database are shown in gray and their zero-age main sequence (ZAMS) is marked with a thick gray line (see \S~\ref{sec:lit}). From left to right, we show the location of binary stars composed of helium-core burning stripped stars combined with main-sequence stars that are evolved 20, 60, and 90\% through the duration of their main-sequence evolution (marked with gray dots on the evolutionary tracks). Each white, dotted line follows a stripped star or a main-sequence star with the labeled mass. The three example systems shown in \figref{fig:SED_MScompanions} are marked with white circles. \textit{Top:} Colored shading displays the flux contribution in the Bessell/V-band of the stripped star to each binary combination. \textit{Middle:} Colored shading displays the flux contribution in the Swift/UVM2-band of the stripped star to each binary combination. \textit{Bottom:} Vivid colors indicate the spectral morphology of each composite model spectrum: blue is for ``Helium star type'', teal for ``Composite type'', and light green for ``B-type'' (see \S~\ref{sec:SpTs}). When no high-resolution spectral model exists for the companion star, that is when $T_{\mathrm{eff}} < 10\,000$K, we approximate the spectral morphology using the flux contribution. These approximations are shown with shaded colors.}
\label{fig:Theoretical_CMD}
\end{figure}

\begin{figure}
\centering
\includegraphics[width=\textwidth]{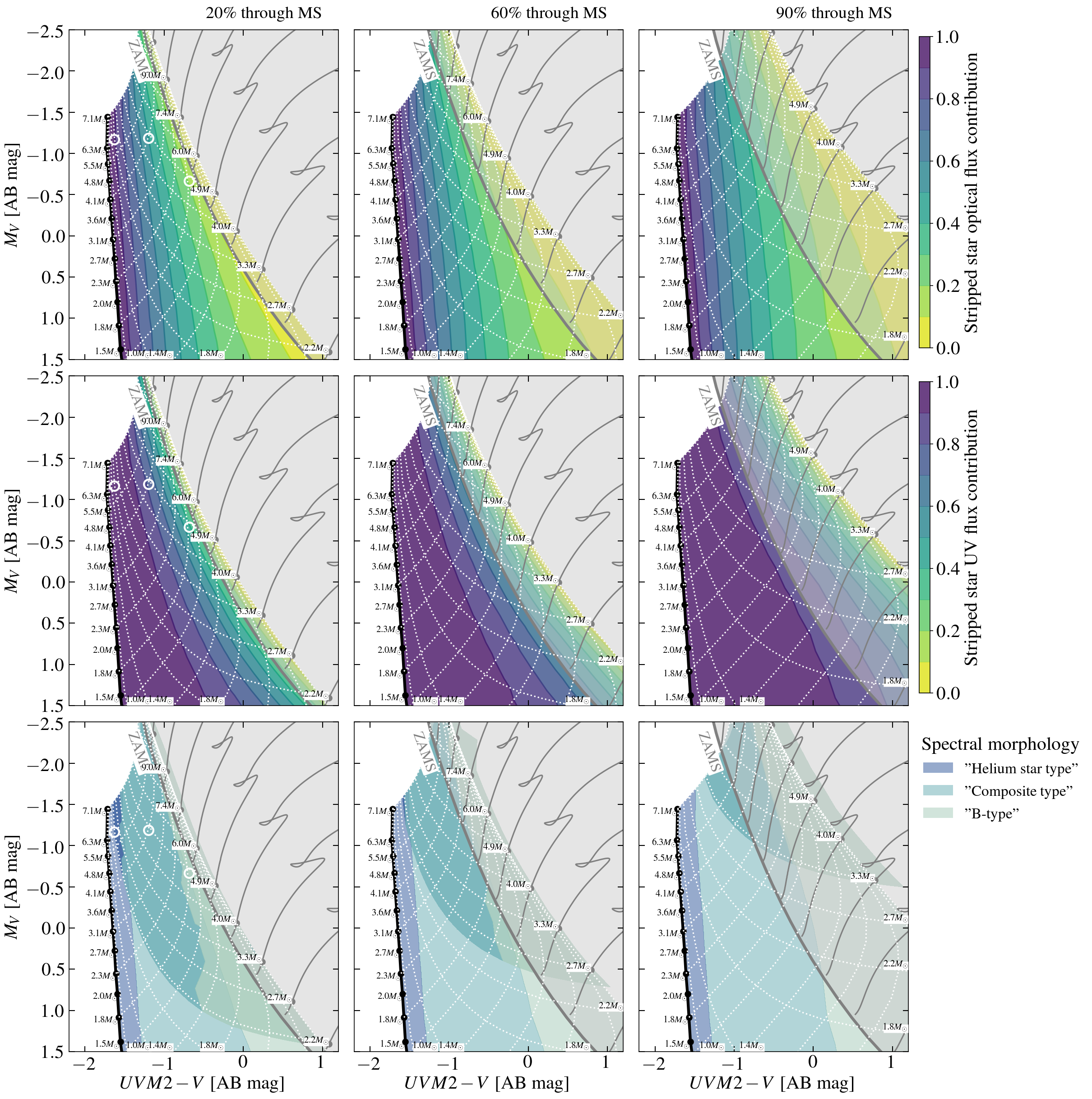}
\caption{Same as \figref{fig:Theoretical_CMD}, but showing the absolute V-band magnitude on the vertical axes. }
\label{fig:Theoretical_CMD_V}
\end{figure}

Using the full composite model grid we find that---due to this excess UV emission---a subset of binaries containing stripped stars will lie in a unique region of UV-optical color-magnitude-diagrams (CMDs). This is demonstrated in \figreftwo{fig:Theoretical_CMD}{fig:Theoretical_CMD_V}, where each panel shows a UV-optical CMD constructed using the optical Bessel/V-filter and the ultraviolet Swift/UVM2-filter (see also \figref{fig:SED_MScompanions}).
We draw the zero-age main-sequence (ZAMS) using the evolutionary models in combination with ATLAS 
models as described in \S~\ref{sec:lit} and mark the part of the diagram populated by single stars using gray shading. For reference, we display the evolutionary tracks of a set of single stars using the MIST models \cite{2016ApJ...823..102C} with initial masses of 4-13.5\Msun. 
It is clear that at these brightnesses, stars bluewards of the ZAMS are not expected from single star evolution (see also \S~\ref{sec:nature}). 

However, based on our current suite of models, stripped helium binaries with certain companions \emph{will} populate the region bluewards of the ZAMS. In \figreftwo{fig:Theoretical_CMD}{fig:Theoretical_CMD_V} we use colors to show where binaries compose of a stripped star and a main sequence star will appear. In the left, middle and right panels, we present results assuming young, intermediate, and evolved main-sequence companions by using the models 20\%, 60\%, and 90\% through the MS duration. The white dotted lines follow either one stripped star model coupled with a range of main-sequence star models, or one main-sequence star model coupled with a range of stripped star models. Because we only have spectral models for main-sequence stars with $T_{\rm eff}>10,000$K (roughly corresponding to masses $>2 \Msun$), we use the photometry from MIST models to draw these dotted lines and cover the full UV excess regime. The MIST photometry matches well with our expectations in the regime where we have available spectral models. We mark the location of the three binary combinations displayed in \figref{fig:SED_MScompanions} using white circles. 

Predictions from our composite model grid smoothly fill the entire parameter space between the MS companion and the models for stripped stars described in \S~\ref{sec:MESAspec}, which are shown as black dots and lie approximately 1 magnitude bluewards of the ZAMS in this CMD. The location of a given system is dictated by the relative flux contributions for each binary component. We use color shading to show optical (Bessell/V) the ultraviolet (Swift/UVM2) flux contributions from the stripped star to the emission of the different binary systems in the top and middle panels of \figreftwo{fig:Theoretical_CMD}{fig:Theoretical_CMD_V}, respectively. The top panels show that, only when coupled with the very lowest mass companions ($\lesssim 2-3\Msun$), the stripped star is responsible for more than 80\% of the optical emission from the binary. There is also a large spread in the optical flux contribution from the stripped star for systems that lie bluewards of the ZAMS, spanning from 10-100\%. This is in stark contrast with the UV flux contribution. As shown in the middle panels, for most binaries that lie bluewards of the ZAMS, the stripped star is responsible for more than $\sim$70-80\% of the ultraviolet radiation. 
It is also interesting to note that some systems in which the stripped star contributes less than 20\% of the optical emission can still be detectable with UV excess. 

For each stripped star mass, we can find a maximum MS companion mass for which the system will lie bluewards of the ZAMS. For stripped stars with mass between 2 and 7\Msun, this maximum MS companion mass is between $\sim$5 and 10\Msun\ (see also \figref{fig:M1M2}). 
Considering that 2-7\Msun stripped stars originate from progenitors with at least 7.5-18\Msun, this means that the stripped star binaries that are detectable with UV excess either (i) cannot have had initial mass ratios close to one or (ii) do not have luminous companions, but instead orbit compact objects or are single, runaway, stars. In the former case, this means that mass transfer cannot have been fully conservative. Therefore, we expect that the UV excess detection technique favors the detection of stripped stars created via common envelope ejection, since common envelopes are thought to develop when the mass ratio is large at the time of interaction and negligible mass accretion is expected. However, stable, non-conservative mass transfer could also result in systems that are observable via UV excess.

\subsection{Expected Spectral Morphology of Stripped Star Plus Main Sequence Binaries}\label{sec:SpTs}

Even if a star shows excess UV radiation as described in \S~\ref{sec:UVexcess_prediction}, it can only be confirmed as a stripped star when its stellar properties have been corroborated. The spectral features contain the necessary information for determining the stellar properties. The optical wavelength range is accessible by ground-based observations, but can also be significantly affected by the presence of the companion star, as seen by the optical flux contribution from the stripped companion in a binary in the top row of panels in \figreftwo{fig:Theoretical_CMD}{fig:Theoretical_CMD_V}. 
To better understand the spectral morphology of stripped star binaries, we scrutinize the spectra in our composite grid.  
Because we only have high-resolution spectra for stars hotter than 10,000K, we are, therefore, limited to exploring spectral morphologies of binaries containing main sequence stars more massive than $\sim2\Msun$ (see \tabref{tab:OB_properties}).

We identify three broad spectral classes for stripped star binaries \emph{that are identifiable with UV excess}: 
\begin{itemize}
\item[\textbf{(1)}] \textbf{``Helium-star-type''}\newline
The composite spectrum appears as an isolated stripped star. The main characteristic features are: (i) \HeII\ lines are present, (ii) \NIV\ and \NV\ lines may be present, and (iii) the short-wavelength Balmer lines are weak or not present.
\item[\textbf{(2)}] \textbf{``Composite-type''}\newline
Characteristic features of both a stripped star and a B-type companion star are present in the composite spectrum. A combination of \HeII\ lines and prominent short-wavelength Balmer lines is expected.
\item[\textbf{(3)}] \textbf{``B-type''}\newline
The composite spectrum appears as an isolated B-type main-sequence star. The main characteristic features are: (i) the short-wavelength Balmer lines are prominent, and (ii) no \HeII\ lines are present.
\end{itemize}

\begin{figure}
\centering
\includegraphics[width=\textwidth]{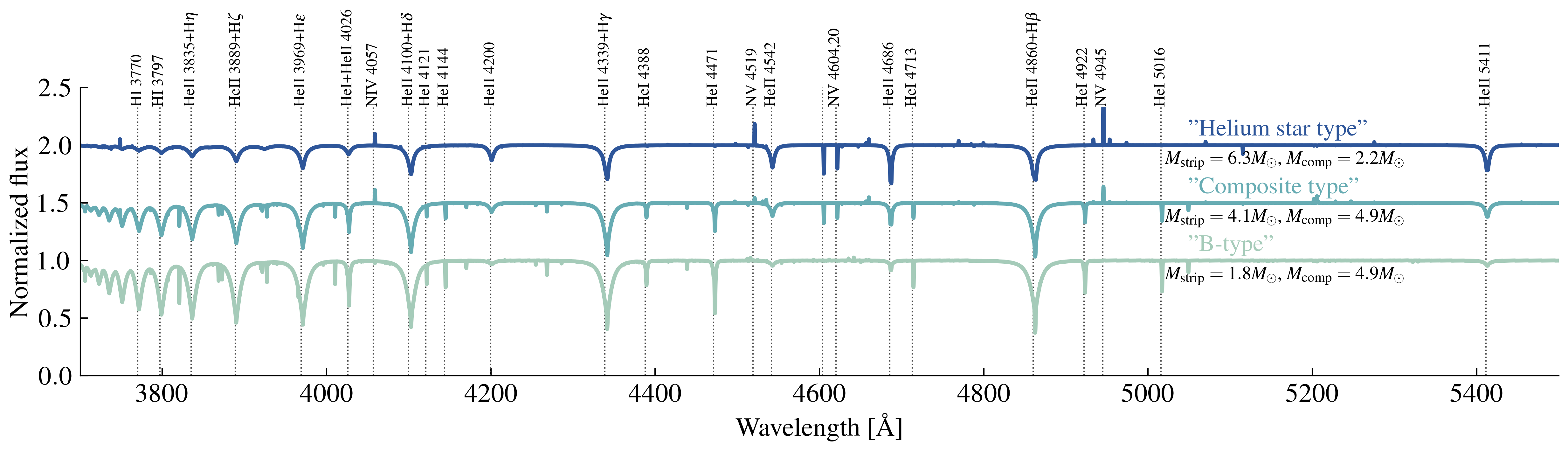}
\caption{Examples of the three types of spectral morphology expected for stripped stars orbiting main-sequence stars. From top to bottom, we show the modeled composite optical spectrum of a 6.3\Msun\ stripped star in a binary with a 2.2\Msun\ late B-type star (dark blue), a 4.1\Msun\ stripped star orbiting a 4.9\Msun\ B-type companion (teal), and a 1.8\Msun\ stripped star orbiting a 4.9\Msun\ B-type companion (light green). These are examples of the type of spectra expected for the ``Helium-star'', ``Composite'', and ``B-type'' groups described in \S~\ref{sec:SpTs}, respectively. These are the same example systems as shown in \figref{fig:SED_MScompanions} and marked with white circles in \figreftwo{fig:Theoretical_CMD}{fig:Theoretical_CMD_V}. We label several spectral features that are characteristic to stripped stars and B-type stars.}
\label{fig:example_SpT}
\end{figure}

Examples for these three types of spectra are shown in \figref{fig:example_SpT}. In the figure, we display the optical morphology of the same three binary combinations shown in \figref{fig:SED_MScompanions}: a 6.3\Msun\ stripped star in a binary with a 2.2\Msun\ late B-type star (top, dark blue), a 4.1\Msun\ stripped star orbiting a 4.9\Msun\ B-type companion (middle, teal), and a 1.8\Msun\ stripped star orbiting a 4.9\Msun\ B-type companion (bottom, light green). The figure clearly shows the presence of the pure \HeII\ lines in the ``Helium-star'' and ``Composite'' spectra and prominent short-wavelength Balmer features for the ``B-type'' and ``Composite'' spectra. 
This is consistent with the expected temperatures of the two components -- stripped stars are very hot ($\sim 30-100$kK) and helium-rich, meaning that their spectra should contain strong lines of ionized helium. B-type stars are so cool that lines of ionized helium cannot form in the stellar atmosphere ($< 30$kK), but the lower temperature, in combination with that they are hydrogen rich, is beneficial for the creation of strong Balmer features \cite{2009ssc..book.....G}.

Assigning individual spectra to each of these classes can be complex, since the composite model grid shows a range of spectral morphologies. We therefore take a simplified approach where we compare the equivalent widths (EW) of two spectral lines. 
We choose to use H$\eta$, which is blended with \HeII\ $\lambda 3835$, to represent the short-wavelength Balmer lines, and \HeII\ $\lambda 5411$ to represent the pure \HeII\ lines. Because the spectral model grid for isolated stripped stars described in \secref{sec:agnostic_grid} predicts that H$\eta$/\HeII\ $\lambda 3835$ should have EWs that are less than roughly $1.2$\AA, we choose to use EW(H$\eta$/\HeII\ $\lambda 3835$) $=1.2$\AA\ as a dividing line between the ``Helium-star'' and ``Composite'' groups. For the presence of \HeII\ $\lambda 5411$, the signal-to-noise ratio of an observed spectrum constrains how weak spectral lines can be while still being detectable. To determine the corresponding threshold in EW of \HeII\ $\lambda 5411$, we add noise to our models such that the resulting signal-to-noise ratio is 35 and identify that \HeII\ $\lambda 5411$ can be detected if its EW is more than 0.2\AA. Thus, for illustrative purposes, we choose to use EW(\HeII\ $\lambda 5411$) $=0.2$\AA\ as a dividing line between the ``Composite'' and ``B-type'' groups.

\begin{figure}
\centering
\includegraphics[width=\textwidth]{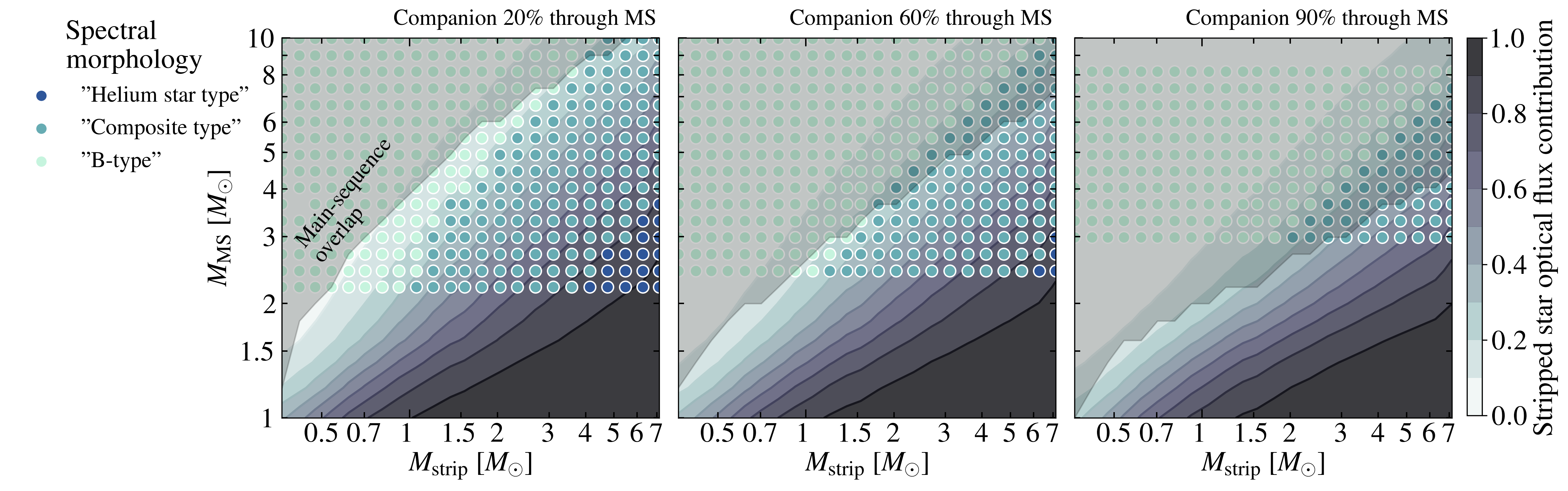}
\caption{Colored circles show the available high-resolution composite spectra and their corresponding spectral morphology in the mass plane for the two components in a binary composed of a stripped star and a main-sequence star. Gray shading shows systems that overlap with the main-sequence in the color-magnitude diagrams in \figref{fig:Theoretical_CMD_V}, while the other systems should show UV excess. The figure shows that systems with all three spectral morphologies may be detectable with UV excess. The background shading shows the stripped star flux contribution in the V-band. Comparing to the color of the circles, it can be seen that the boundary between the three types of spectra occurs roughly at $80\%$ and $20\%$ flux contributions. The three panels correspond to binary combinations with unevolved, intermediate, and evolved main-sequence companions.}
\label{fig:M1M2}
\end{figure}

Using these criteria, we find that the predicted optical spectral morphology is closely linked to the flux contribution of the two components in the binary star. To have an optical spectrum of type ``Helium-star'', the stripped star needs to contribute at least 80\% of the optical flux. For the ``Composite'' type, the stripped star contributes between 20 and 80\% of the optical flux, while for the ``B-type'' morphology, the stripped star contributes less than 20\% of the optical flux. This is illustrated in \figref{fig:M1M2}, where we mark the spectral morphology of a range of different binaries from the high-resolution spectra using colored circles in the mass plane of the two components. As background shading, we show the flux contribution from the stripped star in the V-band. The transitions between the spectral morphologies can be seen to occur roughly at 20 and 80\% flux contribution. 

With gray shading, we mark the location of binary combinations that overlap with the main sequence. This shading suggests that all stars with ``Helium-star'' type spectra should be detectable via UV excess. For binaries with the spectral type ``Composite'', most stripped stars with un-evolved main-sequence companions should be detectable via UV excess, while a smaller fraction should be detectable if the companion is more evolved. Finally, most of the systems in the ``B-type'' group have such bright companions that they will overlap with the main sequence, with the exception of lower-mass stripped stars ($\lesssim 2\Msun$) orbiting un-evolved companions. 

Using both the determined spectral classes from high-resolution spectra and their relation with the optical flux contribution, we can mark the locations of the different spectral types in the UV-optical color-magnitude diagram, as shown in the bottom three panels of \figreftwo{fig:Theoretical_CMD}{fig:Theoretical_CMD_V}. In dark blue, teal and light green, we show the regions expected to contain stripped star binaries with the optical spectral types ``Helium-star'', ``Composite'', and ``B-type'', respectively. The more vivid colors show the region where high-resolution composite spectra could be used, while the lighter semi-transparent colors show the regions where the spectral morphology was inferred from the optical flux contribution. The figure shows that stars with UVM2-V $\lesssim-1.2$ ABmag are expected to appear as the type ``Helium-star''. The ``Composite'' type is expected for UVM2-V colors between $\sim -1.2$ and $\sim 0.5$ ABmag. The figure also shows that only low-mass main-sequence companions ($\lesssim 3\Msun$) are expected to give rise to the ``B-type'' spectra and still show a detectable UV excess. The higher mass combinations for the ``B-type'' that are visible in \figref{fig:M1M2} are located very close to the zero-age main-sequence ($\lesssim0.1-0.3$ ABmag) and may be difficult to identify as having a UV excess.

\section{Observations and Sample Selection}

Motivated by the theoretical predictions outlined above, we carried out a search for stripped helium stars. We specifically identify a set of 25 stars that: 
\begin{enumerate}
    \item have photometric colors that show an ultraviolet excess relative to expectations for the Zero Age Main Sequence.
    \item have constraints on their distances that place their absolute magnitudes in the range expected for intermediate mass stripped helium stars. As described below, for most objects, this consists of having 3-D kinematics consistent with membership in the Magellanic Clouds. 
      \item have an overall spectral morphology that falls in one of the three main categories expected for stripped helium star binary systems, as outlined above.
\end{enumerate}
Here, we describe the main observations and details of the selection process used to identify this sample.

\subsection{Preliminary UV Photometry of Stars in the Magellanic Clouds}\label{sec:prelimphotom}

We targeted the Magellanic Clouds for our initial search for stripped helium star binaries for two primary reasons. First, the LMC/SMC offer us a nearly complete view of their stellar populations, with the advantage of known distances and low reddening when compared to the Milky Way. Second, with distances of $\sim$ 50 kpc and 61 kpc, respectively, intermediate mass stripped helium stars (2.0 M$_\odot$ $<$ M$_{\rm{strip}}$ $<$ 7.0 M$_\odot$) in the LMC/SMC are expected to have $V-$band apparent magnitudes between 17.0 mag and 19.5 mag (see Figure~\ref{fig:Theoretical_CMD_V})---still within reach of ground-based spectroscopic follow-up. 

While HST has imaged portions of the Magellanic Clouds at high resolution in the UV \cite{Brosch1999} and GALEX has imaged the full extent of the galaxies at 5$''$ resolution \cite{Simons2014}, we found that there was no existing UV point source catalog of the Clouds with sufficient footprint, resolution, and depth for our purposes. We therefore chose to compute new UV photometry for stars in the Magellanic Clouds using images that were recently obtained by the \emph{Swift-}UVOT satellite. Full photometric catalogs of sources identified in these images will be presented in a future paper. Here, we provide details of the photometry process that was used to perform the initial selection of the spectroscopic sample that is the focus of this manuscript.

\subsubsection{The Swift UVOT Magellanic Clouds Survey}\label{sec:SUmaC}
Between 2011 and 2013, the \emph{Swift} satellite carried out the Swift-UVOT Magellanic Clouds Survey, or SUMaC \cite{Siegel2015,Hagen2017}. Over this time, \emph{Swift-}UVOT imaged 165 fields in the LMC and 50 fields in the SMC in all three UV filters (UVW2: $\lambda_{\rm mid} =$ 1928 \AA, UVM2: $\lambda_{\rm mid} =$ 2246 \AA, and UVW1: $\lambda_{\rm mid} =$ 2600 \AA) at a resolution of 2.5$''$. This corresponds to a total area of approximately 3 and 9 deg$^2$ in the LMC and SMC, respectively. While this data has been used to study the extinction law and star formation history in the Magellanic Clouds \cite{Hagen2017}, no point source catalog is currently available.

We downloaded Swift-UVOT Level 2 image files for the SUMaC data from the Swift swiftmastr catalogue
\footnote{https:\/\/www.swift.ac.uk\/swift\_live\/index.php\#advanced}. This data has already been processed by the UVOT reduction pipeline, which includes bad pixel, flatfield, and boresight distortion corrections. Level 2 products also come with RA and DEC coordinates attached. However, we found that in the crowded regions of the Magellanic Clouds the standard Swift-UVOT astrometric solutions---which are based on matching sources detected in the images to those found in the USNO-B1 catalog \cite{Monet2003}---often failed. We, therefore, compute new astrometric solutions for all images by running \texttt{astrometry.net} \cite{Lang2010} on the Level 2 sky images. When running \texttt{astrometry.net}, we use custom index files constructed from all sources in the ground-based Magellanic Clouds Photometric Survey (MCPS)  
\cite{Zaritsky2002,Zaritsky2004} 
with U-band magnitudes brighter than 19th mag, in order to ensure and improved solution. 

Each field targeted by the SUMaC survey, was visited between 1 and 5 times between 2011 and 2013 (termed an ``observation segment''). During each of the these visits, between 2 and 8 individual exposures (or ``snapshots'') were obtained per UV filter. We make no attempt to combine any of these images, but rather treat them independently.

\subsubsection{Forward-Modeled Photometry with \emph{The Tractor}}

While the 2.5$''$ resolution of Swift offer a substantial improvement over GALEX, the crowding in the Magellanic Clouds still presents a challenge for the standard Swift photometry packages provided by HEASARC, which rely on aperture photometry (e.g., \texttt{uvotsource}). We, therefore, chose to compute photometry with \emph{The Tractor} \cite{Lang2016}, a forward-modeling photometry code that previously has been used to produce the NEOWISE photometry based on data from the entire WISE mission \cite{Lang2016b}.  We provide the following components as input to \emph{The Tractor}: 
\begin{enumerate}
    \item A Swift UVOT image, divided by its exposure time such that it is in units of counts/s.
    \item A model for the background in the Swift-UVOT image. We compute this using the \texttt{Background2D} routine distributed as part of \texttt{photutils} package of \texttt{Astropy} \cite{photutils,astropy2018}, after sigma-clipping bright sources. 
    \item A model of the Swift-UVOT point spread function (PSF) in a given filter. We construct this from the curve of growth distributed with CALDB (see also the Swift-UVOT CALDB release note number 104\footnote{https:\/\/heasarc.gsfc.nasa.gov\/docs\/heasarc\/caldb\/swift\/docs\/uvot\/uvot\_caldb\_psf\_02.pdf}).
    \item A set of coordinates where sources are located in the image. We utilize coordinates of sources from the MCPS \cite{Zaritsky2002,Zaritsky2004}, which imaged the LMC/SMC in the UBVI filters. This is beneficial for two primary reasons: (i) these same sources were used to calibrate the astrometry in the Swift-UVOT images (see \S~\ref{sec:SUmaC}) and (ii) this guarantees ease of matching our UV magnitudes to an associated optical source. Both UV and optical magnitudes will be necessary to assess the presence of a UV excess (see below).
\end{enumerate}
With these as input, \emph{The Tractor} adjusts the count rate associated with each source in the image, continuing until a log-likelihood condition ($<0.001$) is met. As a final product, it produces both a forward-modeled image and the associated count rate for each input source---calibrated within the standard 5$''$ radius aperture utilized by Swift-UVOT. We supplement these with a measurement of the background count rate for each source, calculated within the same 5$''$ apertures, using \texttt{photutils}. 

With these two values (background count rate and source count rate) we compute UV magnitudes using standard HEASARC UVOT calibration tools. Namely we: (i) correct total count rate for coincidence loss with \texttt{uvotcoincidence}, (ii) correct the background count rate for coincidence loss and subtract this from the corrected total count rate to yield a coincidence loss corrected source count rate (iii) apply the recommended 5\% systematic error in count rate to account for variations in PSF shape (iv) exclude sources located in small regions of the UVOT detector with reduced sensitivity based on the small scale sensitivity file provided in CALDB (v) apply a large-scale sensitivity correction using \texttt{uvotlss}, (vi) apply a correction for the temporal degradation of Swift-UVOT sensitivity based on the sensitivity correction file provided in CALDB, and (vii) convert the final source count rates to magnitudes and fluxes using \texttt{uvotflux}. Finally, we convert the resulting magnitudes to AB scale using standard zeropoint offsets.

For the preliminary photometry described here, which was used to perform initial target selection for spectroscopic follow-up (see \S~\ref{sec:spectroscopic_observations}), we freeze the source positions within \emph{The Tractor} and consider only 1 exposure per field in the LMC and SMC. As there exist both small variations in the Swift PSF as a function of source brightness and small uncertainties in the astrometric solutions, we also compute and save a the residual between \emph{The Tractor} forward-modeled image and the original observation with the 5$''$ aperture utilized for each source. This is used as one metric for the reliability of the resulting photometry. Overall we obtained preliminary UV photometry for $\sim$300,000 sources in the LMC and and $\sim$200,000 sources in the SMC, down to UV magnitudes of approximately 19 ABmag.

\subsection{Spectroscopic Observations with Magellan/MagE}\label{sec:spectroscopic_observations}

We carried out spectroscopic observations for a number of stars that we identified as targets of interest based on the combination of their UV photometry from SUMaC images and optical photometry from the MCPS. We began by selecting blue sources by requiring that the reddening free index, Q $=$ (U-B) - 0.72*(B-V) $<$ -0.5, B-V $<$ 0.2, and U-B $<$ 0.4. We then exclude sources with unphysical UV photometry by considering only sources with (UVM2-U)$_{\rm{Vega}}$ $>$ -2 mag. At this point we prioritize targets that (i) appear bluewards of the ZAMS in \emph{multiple} UV-optical color-magnitude diagrams and (ii) exhibit low residuals between the observed UVOT images and forward-modeled images used to compute the UV photometry. At this stage we do not correct the photometry for extinction and compare colors to a theoretical ZAMS, but rather select targets that appear bluewards of the large over-density of stars (which represents the MS) in the ``raw'' observed CMDs (see \S~\ref{sec:UVexcess} for a discussion of reddening). Highest priority for spectroscopic follow-up was given to stars with the bluest colors and V-band magnitudes between 16.5 mag $<$ V$_{\rm{Vega}}$ $<$ 19 mag. We emphasize that the initial sample of stars targeted for spectroscopy was not designed to be systematic or complete, but rather exploratory. Our goal was to understand the properties and nature of targets that appeared to exhibit a UV excess in our data. 

Spectroscopic observations were carried out over 18 nights between December 2018 and January 2022. All spectra were obtained with the MagE spectrograph mounted on the 6.5m Magellan-Baade telescope \cite{2008SPIE.7014E..54M}.  MagE is a moderate-resolution optical echellette, offering continuous wavelength coverage between $\sim$3500 \AA\ and 1 micron. For all our observations we utilize the 0.85$''$ slit, which provides a resolution of $R \sim 4100$. By default, observations were obtained at the parallactic angle. However, in some particularly crowded cases, a rotation was applied to avoid having multiple nearby objects within the slit. Thorium-Argon (ThAr) lamps were obtained prior to every science observation to reduce the impact of instrument flexure on the wavelength calibration.  While observing nights in Dec 2018 and Dec 2019 were devoted to obtaining spectra of a large number of candidate systems, nights between Jan 2020 and Jan 2022 were primarily used to monitor the radial velocities of targets of particular interest. In total, we obtained spectra of 42 unique systems with 1--30 observations per star.

\subsection{Spectroscopic Data Reduction}\label{sec:specreduc}

Initial spectroscopic reduction of all Magellan/MagE observations were performed with the \texttt{CarPy} python-based pipeline\footnote{\url{https://code.obs.carnegiescience.edu/mage-pipeline}} \cite{Kelson2000,Kelson2003}. The main steps carried out by the pipeline include bias subtraction, flat-field correction, echelle order identification, 2D background subtraction, extraction of individual echelle orders, and wavelength calibration based on ThAr lamp exposures taken immediately prior to each science observations. Multiple exposures of the same object taken sequentially can also be automatically combined. We manually inspect the 2D spectra to ensure that the pipeline selects the proper target in any case where multiple stars fell within the 10$''$ slit length of MagE. 

The final output of the \texttt{CarPy} pipeline is a multispec FITS file with extensions for each echelle order. For each order, the pipeline provides both the 1D sum of the object over the extraction aperture and the combined noise spectrum due to the object, sky, and readout. We use these to create final combined and normalized spectra for each object in our sample. First, using the \texttt{pyraf} interface for IRAF \cite{pyraf2012}, we flatten and combine the multiple echelle orders for each individual observation of every target. We select portions of every echelle order that are above a chosen signal-to-noise threshold, use the \texttt{pyraf} task \texttt{continuum} to normalize each order with a cubic spline fit, and subsequently stitch multiple orders together into a single 1D spectrum spanning the full wavelength range of MagE. Error spectra are propagated through the same process. We begin with a signal-to-noise (S/N) threshold of 25 to select portions of each echelle order. This threshold is lowered if necessary to maintain continuous wavelength coverage between 3800-7000 \AA, with a minimum S/N threshold of 10. Finally, each individual observation is corrected to a heliocentric reference frame using the \texttt{pyraf} task \texttt{rvcor}. 

Second, we combine multiple observations taken of the same target, when available, in order to produce the highest possible signal-to-noise spectrum of each star. Multiple observations were acquired between hours and years apart, depending on the target. As our scientific goal in selecting our targets was to identify stripped helium binaries, we must account for possible binary motion when combining spectra acquired at different times. This is done via an iterative approach: 

We begin by cross correlating our highest signal-to-noise observation of a given target against the full grid of CMFGEN helium star and main-sequence star model spectra spectra described in \S~\ref{sec:agnostic_grid} and \ref{sec:general_MS_spectral_models} using the \texttt{pyraf} task \texttt{fxcor}. We select the model spectrum for which the cross-correlation lead to the highest Tonry \& Davis \cite{Tonry1979} $r-$parameter as our initial ``template''. All individual spectra of the target in question are then cross-correlated against this same model template, shifted by the relative velocity output by \texttt{fxcor}, and then averaged together. As the models described in \S~\ref{sec:general_strip_spectral_models} and \ref{sec:general_MS_spectral_models} are not perfect templates for our spectra, we then iterate this process. We adopt the combined data spectrum as our new template, cross correlate each individual spectrum against this combined template, shift each individual spectrum by the new relative velocity, and average the spectra together. This process is repeated 10-15 times until the velocity shifts for each spectrum have converged. Finally, we perform one additional cross-correlation between this final combined data spectrum and the ``best-fit'' model spectrum in order to shift it to the rest-frame. Throughout this process, we identify signatures of binary motion in many of our targets for which we possess $>3$ epochs of observations. These will be analyzed in detail in subsequent publications (see also \S~\ref{sec:radvel}). Finally, error spectra were added in quadrature to produce a corresponding error spectrum for the final combined spectra for each target. Typical signal-to-noise for these final combined spectra range from $\sim$ 25 for targets with only single observations to $\sim$150 for targets with many.  

\subsection{Final UV Photometry of Spectroscopic Sample}

While the preliminary UV photometry described in \S~\ref{sec:prelimphotom}, was sufficient for initial target selection, we now compute improved photometry for our final spectroscopic sample. We again run the \emph{The Tractor} on the \emph{Swift-UVOT} SUMaC images but with two modifications: 
\begin{enumerate}
    \item we allow the positions of the input sources to be varied slightly by \emph{The Tractor}, in addition to their count rates. This is accomplished by setting a gaussian prior on their position centered at the MCPS coordinates and with a sigma of 0.05$''$. This yields lower overall residuals between the \emph{Tractor} models and the observed images, and accounts for slight errors in the astrometric solutions across the chip.
    \item we run \emph{The Tractor} on a 30$''$ $\times$ 30$''$ square around our spectroscopic targets in \emph{every} SUMaC image that contains them. Between the multiple observing segments and snapshots obtained for each field, as well as small overlaps between different SUMaC fields, this amounts to 2--15 independent measurements of each source in each filter.
\end{enumerate} 
The final photometry presented for our spectroscopic sample is the weighted average of each of the independent measurements. The standard deviation of these measurements is then formally incorporated into the error in our photometric measurements, along with the statistical uncertainties provided by \emph{The Tractor}. Prior to averaging these independent observations, we remove any images where either the statistical uncertainty on our target is greater than 0.25 mag or where the residual between the \emph{Tractor} model and the observed image within the 5$''$ aperture for the source of interest is greater than 30\% of the flux assigned to the source itself. These restrictions work to exclude images where the source is not detected at high signal-to-noise as well as cases where tracking errors cause source in the Swift images to be poorly described by the input PSF.

Finally, in addition to formal photometric errors, we compute a number of flags that provide context for how crowded the region surrounding a given source is in the \emph{Swift-}UVOT images. In particular, we provide (i) the distance to the closest star to each spectroscopic target, (ii) the number of stars within a 5$''$ radius of each spectroscopic target, and (iii) the fraction of the flux within a 5$''$ radius of each spectroscopic target that was actually ``assigned'' to those object by \emph{The Tractor}. While the methodology used by \emph{The Tractor} is a significant improvement over aperture photometry, because the Swift-UVOT images are somewhat under-sampled (1$''$ pixels coupled with a 2.5$''$ PSF) some issues persist in robustly disentangling flux from very close sources of similar brightnesses. These flags therefore help gauge when additional caution is warranted in interpreting the robustness of the calculated UV magnitudes. Final photometry for our spectroscopic sample is given in Table~\ref{tab:photom}.

\subsection{Presence of an Ultraviolet Excess}\label{sec:UVexcess}

Using the final UV photometry described above, we perform a more detailed investigation of the evidence for a UV excess in the spectral energy distributions of our spectroscopic sample. As above, we consider objects to have a ``UV excess'' if they are located bluewards of the Zero Age Main Sequence (ZAMS) in a range of color-magnitude diagrams. To describe the location of the ZAMS, we rely on the Kurucz spectral models described in \S~\ref{sec:lit} and used in \S~\ref{sec:UVexcess_prediction} to illustrate that some stripped star binaries are predicted to show an observable UV excess. However, comparing our observed photometry to this theoretical ZAMS requires that we correct for both distance and reddening. 

For the former, we adopt distances of 50 Mpc for stars in the direction of the LMC \cite{Pietrzynski.G.2013.LMCDistance} and 61 Mpc for stars in the direction of the SMC \cite{Hilditch.R.2005.SMCDistance}. While membership of the stars in our spectroscopic sample in the Magellanic Clouds will be assessed in \S~\ref{sec:Member}, we note that this is a conservative approach. If any object is actually a Milky Way foreground star, then its true absolute magnitude will be fainter than assumed here. In this case, any observed UV excess will become more significant as the ZAMS moves to cooler temperatures.

We do not model the reddening for each star in our spectroscopic sample independently, but rather correct for a mean extinction value in both the LMC and SMC. We adopt the average LMC/SMC reddening curves from \cite{Gordon2003}. To determine the appropriate correction coefficients for the \emph{Swift-UVOT} bands for our purposes, we perform synthetic photometry on the stripped star models described in \S~\ref{sec:MESAspec} both before and after reddening. 

To determine an appropriate A$_{V}$ value for both the LMC and SMC, we turn to the MCPS, which provides distributions of A$_{V}$ values for the ``hot'' and ``cool'' stars in the LMC and SMC \cite{Zaritsky2002,Zaritsky2004}. Mean A$_{V}$ values for the ``hot'' star distributions in the LMC/SMC are 0.55 and 0.46 mag, respectively, however, we find that if we adopt these mean extinction values when constructing our UV-optical CMDs, the theoretical ZAMS falls in the \emph{middle} of the large overdensity of observed stars (see Figure~\ref{fig:CMD}). As it is unlikely that tens of thousands of stars exhibit a UV excess we consider instead that the population of stars we obtain UV photometry of are not well described by the \emph{mean} of the hot star A$_{V}$ distribution from \cite{Zaritsky2002,Zaritsky2004}. In particular, we note that these distributions show large tails to high values of A$_{V}$ and while the \emph{Tractor} method of performing photometry on the Swift UV images is an improvement, the 2.5$''$ resolution of Swift still prohibits useful photometry from being obtained in the most heavily clustered regions, where extinction is also enhanced. Instead, we find that adopting mean A$_{V}$ values of 0.38 mag and 0.22 mag in the LMC/SMC, respectively, leads to the theroetical ZAMS being located near the edge of the overdensity of observed stars. These values lie close to both the the mean values found for the cool star distributions (0.43 and 0.18 mag), as well as the \emph{peaks} of the distributions of the hot stars ($\sim$0.2-0.4 mag) in \cite{Zaritsky2002,Zaritsky2004}. We, therefore, adopt these values when assessing the potential UV excess of our spectroscopic sample. Throughout, we emphasize that extinction of individual stars remains an uncertainty in our analysis. Implications of this for the nature of the objects in our spectroscopic sample will be discussed in \S~\ref{sec:nature}.

After correcting for distance and reddening, we compare our spectroscopic sample to the ZAMS in nine different color-magnitude spaces: (UVW2,UVM2,UVW1) versus (UVW2,UVM2,UVW1) $-$ (B,V,I). To be considered to display a robust UV excess, we require that a given star is consistent with being bluewards of the ZAMS (within errors) in \emph{all nine} color-magnitude spaces, or, if any are lacking that this could be explained by a single (possibly errant) photometric band. Upon completing this analysis, several objects in our original spectroscopic sample were deemed to \emph{not} show robust UV excess. The two most common reasons for this shift were: (i) the original identification of a UV excess was based on an errant optical photometric band in the MCPS or (ii) for some objects located within 2$''$ of another source, the process of averaging multiple independent measurements of the UV flux removed the appearance of a UV excess. In general, the objects that lacked a robust UV excess upon further inspection also displayed spectra consistent with single B-type stars. In the end, out of our original sample of 42 objects we are left with 33 stars with robust UV excess (23 in the LMC and 10 in the SMC).

\subsection{Membership in the Magellanic Clouds}\label{sec:Member}
We do not place an explicit luminosity cut on stars in our sample. However, we are primarily interested in stars located within the Magellanic Clouds, as stripped stars with intermediate masses ($\sim$2-7 M$_\odot$) are expected to have similar apparent magnitudes as our spectroscopic sample at these distances. We therefore use information on the kinematics of our spectroscopic sample to eliminate any likely foreground stars.

\subsubsection{Radial Velocity Determination}\label{sec:radvel}

To assess the radial velocities of the stars in our sample, we rely on cross-correlation with the grid of helium star and main-sequence star model spectra, as described in \S~\ref{sec:specreduc}. In particular, for a given target, each individual observation was cross-correlated against the model spectrum that yielded the highest Tonry \& Davis \cite{Tonry1979} $r-$parameter. For the purposes of kinematic assessment, we simply report the \emph{average} of the radial velocities for each individual observation of a given target. 

We emphasize that these velocities will likely vary from the true systemic velocity for binaries in our sample: for objects with only one observation they could be obtained at any point in the binary orbit, while for objects with multiple observations the quoted average will be affected by the cadence of our observations relative to the orbital period of the system. This should provide a rough estimate for the radial velocity of the center of mass for the system, but for this reason some scatter in the kinematics may be expected.
In Table~\ref{tab:kinematics} we provide the average, minimum, and maximum radial velocity measured for each target, as well as the number of spectra these measurements were based upon. The average velocities for stars in sample range from $\sim$ 215--310 km s$^{-1}$ and $\sim$130--200 km s$^{-1}$ for the LMC and SMC, respectively.

\begin{figure}
\centering
\includegraphics[width=0.95\textwidth]{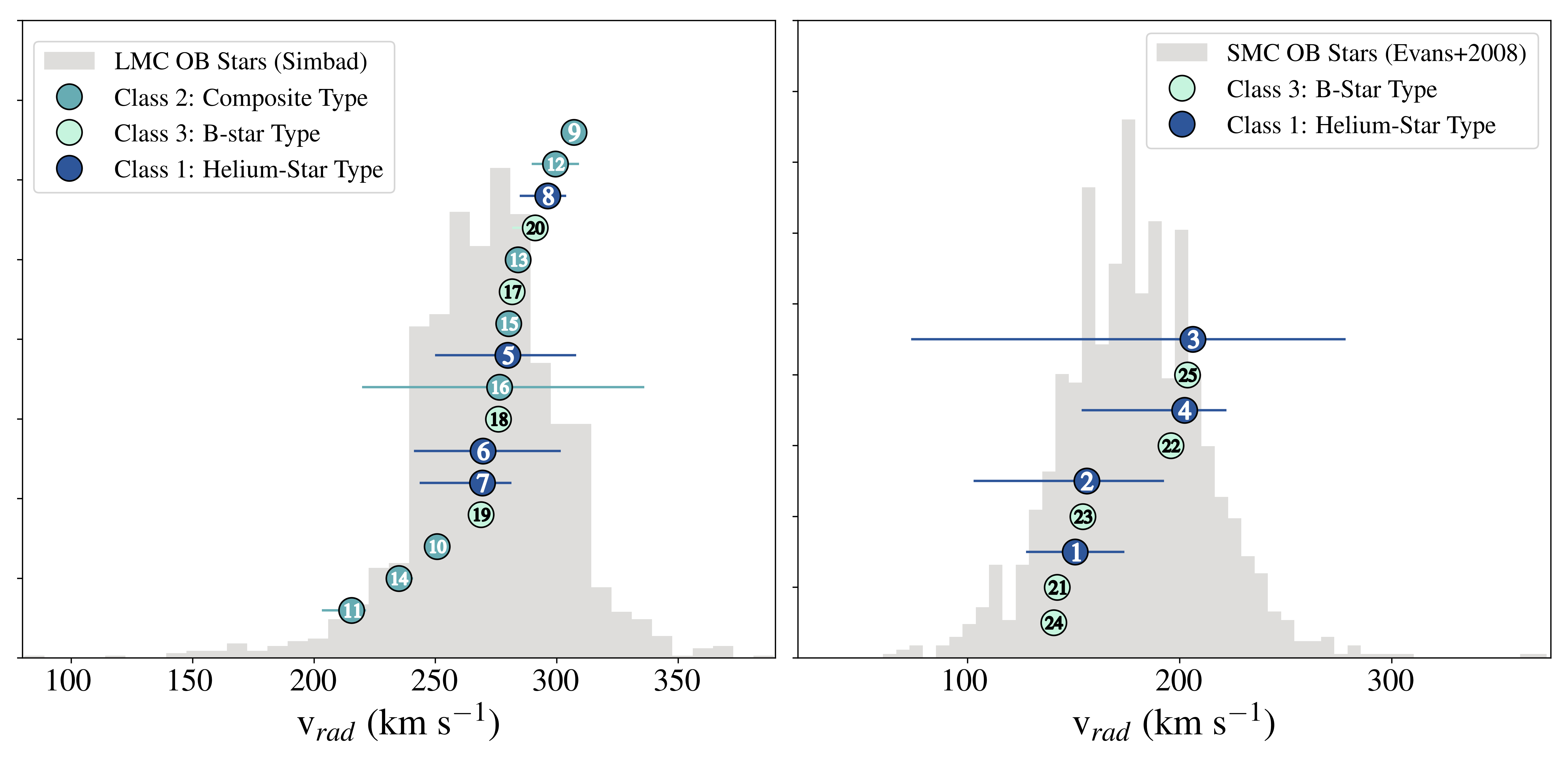}
\caption{The average radial velocities for the stars in our spectroscopic sample in the LMC (left) and the SMC (right) are here plotted using large colored and numbered circles and at different y-value just for clarity. The extents of the errorbars indicate the variation we detect for objects with multiple spectroscopic epochs. The gray histogram in the backgrounds demonstrate the distributions of the radial velocities of OB-type stars in the Magellanic Clouds.}
\label{fig:radvel}
\end{figure}

In Fig.~\ref{fig:radvel} we plot these average radial velocities of our final sample versus comparison samples of OB stars in the Magellanic Clouds. For the SMC, we compare to a histogram of all stars classified as O- or B-type in \cite{Evans2008}. For the LMC, we compare to a sample of all stars listed in the Simbad database that (i) overlap with the LMC, (ii) have an O or B-type spectral class, and (iii) have a listed radial velocity. The majority of this sample comes from \cite{Evans2015,Evans2015b}, but comparison to the full set within Simbad ensures that our comparison kinematics are not restricted to the 30 Dor/northern region of the LMC (which were the main targets for the \cite{Evans2015,Evans2015b} surveys). In all cases, the average radial velocities observed for our stars overlap with the bulk of the LMC/SMC OB star samples. 
The typical precision of our radial velocity measurements are $sim$5 km s$^{-1}$. In \figref{fig:radvel}, we use errorbars to indicate the \emph{range} of radial velocity we detect for objects with multi-epoch spectra. The individual radial velocity measurements for these stars will be reported in a future paper.

\subsubsection{Proper Motions and Parallaxes from \emph{Gaia EDR3}}\label{sec:gaia}

\begin{figure}
\centering
\includegraphics[width=0.95\textwidth]{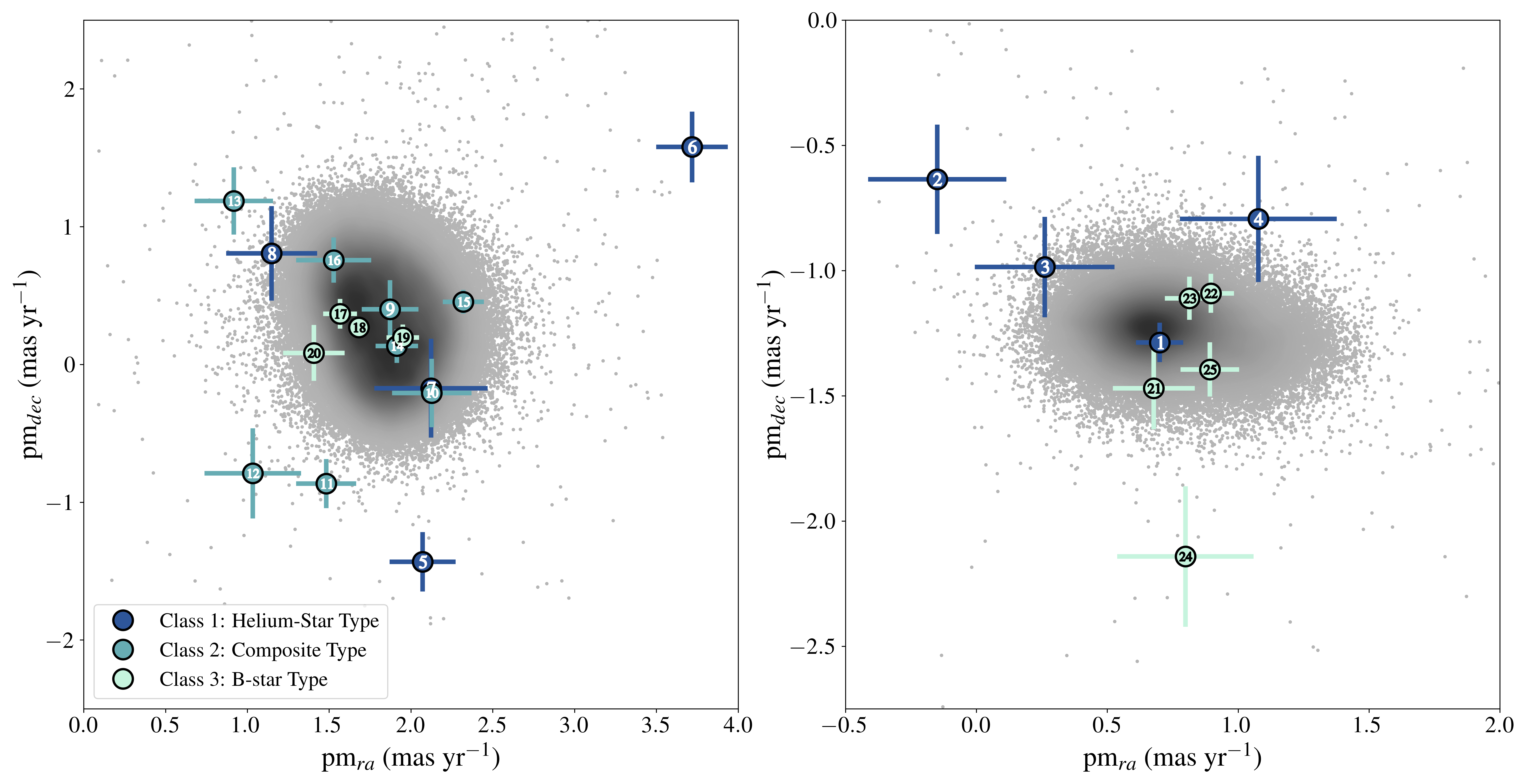}
\caption{Proper motions measured by \emph{Gaia} for our spectroscopic sample in the LMC (left) and the SMC (right) and marked with large colored and numbered circles. Although our sample has moderately large errors in proper motion, because of their faint magnitudes, the measurements agree well with the expectations for stars within the Magellanic Clouds as shown with the gray density plot in the background.}
\label{fig:gaia}
\end{figure}

We cross match our spectroscopic sample with \emph{Gaia} Early Data Release 3 \cite{Gaia.Collaboration.2020.EDR3} to obtain proper motion and parallax measurements. In all cases, we examine \emph{Gaia} magnitudes versus those from the MCPS as a check that the sources are likely the same. In Fig.~\ref{fig:gaia} we compare the \emph{Gaia} proper motions for our final sample to a distribution of $\sim$1,000,000 ``highly likely'' LMC/SMC members. To construct this comparison sample, we first take all stars brighter than 19 mag within 4 and 2.1 degrees of the center of the LMC/SMC, respectively. We then exclude stars with parallax\_over\_ error $>$ 4, stars in the yellow portion of the CMD (0.7 mag $\lesssim$ BP-RP $\lesssim$1.1 mag) where foreground contamination has been shown to be significant, and stars $>$ 3$\sigma$ from the mean in proper motion in RA and DEC (pmra and pmdec). Finally, we restrict the sample to stars with zero excess astrometric noise. Consistency with these distributions are quantitatively assessed below.

\subsubsection{Kinematic Assessment}\label{sec:kinematics}

To remove foreground objects from our sample, we first reject stars with \emph{Gaia} parallaxes detected at greater than 3 sigma. This excluded two objects, whose distances in \cite{Bailer-Jones2021} are $\sim$ 2 kpc. Both objects also had measured radial velocities inconsistent with the Clouds ($<$ 30 km s$^{-1}$) and broad spectral lines indicative of high surface gravities and consistent with expectations for subdwarfs and white dwarfs.

Second, for the remaining objects in our sample, we utilize a technique based on that of \cite{Gaia2018} and subsequently applied in other works \cite{Ogrady2020} to assess the consistency of our spectroscopic sample with membership in the Magellanic Clouds. Broadly, we determine the location of our objects within the kinematic distribution formed by a sample of LMC/SMC members. For our analysis we consider two separate comparison samples: (i) the large set of ``likely'' LMC/SMC members described in \S~\ref{sec:gaia} and shown in Figure~\ref{fig:gaia} and (ii) the sample of OB stars with radial velocities described in \S~\ref{sec:radvel} and shown in Figure~\ref{fig:radvel}. Despite being smaller in size we consider the second sample for two reasons. First, it allows us to perform 3D assessment of the motion of the star in radial velocity ($v_{\rm{rad}}$), pmra ($\mu_{\alpha}$), and pmdec ($\mu_{\delta}$) in addition to a 2D plane-of-sky comparison. Second, due to a combination of systematic (e.g., faint optical magnitudes; crowding) and physical (e.g., runaway velocity) effects it is possible for some hot stars in the Magellanic Clouds to have measured \emph{Gaia} proper motions slightly offset from the bulk distribution. For example, \cite{Aadland2018} found that some spectroscopic O-stars with radial velocities consistent with the clouds had slightly discrepant proper motions. This uncertainty can be somewhat mitigated by comparing our sample directly to spectroscopically confirmed OB stars.   

To describe the distribution formed by each of the comparison samples, we construct two dimensional covariance matrices (C) based on their kinematic properties: $\overrightarrow{\mu}$ = ($\mu_{\alpha}$, $\mu_{\delta}$) for comparison sample (i) and $\overrightarrow{\mu}$ = ($\mu_{\alpha}$, $\mu_{\delta}$, $v_{\rm{rad}}$) for comparison sample (ii). Following \cite{Ogrady2022} we take measurement uncertainties into account by minimizing the total negative log-likelihood for a set of matrices, each weighted by the measurement uncertainties of a single object. This yields an ``optimal'' covariance matrix, C$_{*}$, which is used in the rest of our analysis.  

For each object in our spectroscopic sample, we then calculate a chi-square statistic between its kinematic properties and the comparison samples as: $\chi^{2} =$ $( \overrightarrow{\mu}-\overrightarrow{X} )^{T}\mathrm{C}_{*}^{-1}(\overrightarrow{\mu}-\overrightarrow{X})$, where $\overrightarrow{X}$ is the median kinematic properties of the comparison sample. To consider a star broadly consistent with the kinematics of the Magellanic Clouds, we require that they overlap with the region containing 99.7\% of the comparison sample (i.e. a 3$\sigma$ threshold). For the 2D proper motion comparison sample, this corresponds to a requirement that  $\chi^{2} < 11.6$, while for the 3D OB star comparison sample (including radial velocities) this requires that $\chi^{2} < 13.9$. These requirements eliminated one star, which displayed proper motions discrepant from the Clouds. $\chi^{2}$ values for each object in our final sample are listed in Table~\ref{tab:kinematics}.

We note two exceptions to the requirements stated above. Stars 5 and 6 both have radial velocities consistent with membership in LMC, but proper motions slightly offset (see Figures~\ref{fig:radvel} \& \ref{fig:gaia}), resulting in moderately high $\chi^{2}$ values relative to both comparison samples. However, both stars have highly significant excess noise and poor goodness of fit statistics within \emph{Gaia} EDR3, indicating that the astrometric fit to these stars was poor. We therefore maintain these stars in our sample, which exhibit very similar spectra and physical parameters to other stars in our sample (see below). 

In total, we find that 29 stars in our initial sample show proper motions and/or radial velocities consistent with membership in the Magellanic Clouds. However, we emphasize that the statistics here demonstrate broad consistency between our sample and the kinematics of the Magellanic Clouds. They do not definitively rule out the possibility of any individual star being a foreground (halo) member of the Milky Way with distances greater than $\sim$2 kpc. This possibility, in conjunction with the derived physical properties for our sample, will be discussed in \S~\ref{sec:nature}, below.

\subsection{Spectral Classification}

A handful (4) of the stars in our spectroscopic sample appear to belong to other classes of stars than those indicated by our theoretical predictions. This includes objects that show a wealth of metal lines not predicted by our stripped star grid above as well as objects with strong emission lines similar to planetary nebulae or indicative of the presence of a disk. While these systems are interesting in their own right (and some may indeed contain stripped helium stars in different evolutionary stages) analysis of their properties and likely origin are beyond the scope of this manuscript. For the remaining 25 objects in our sample---whose spectral features are dominated by combinations of H, He\textsc{i}, and/or He\textsc{ii} absorption lines---we assess their spectral morphology in the context of the three main spectral types predicted in \S~\ref{sec:SpTs} for binary systems with stripped helium stars. This is done via a quantitative comparison of the equivalent widths of their absorption lines to those from the theoretical models described in Section~\ref{sec:models}.

\subsubsection{Equivalent Width Measurements: Observed Sample}
 
We begin by determining the appropriate continuum level for each spectral line of interest in our final combined and flattened spectra (see \S~\ref{sec:specreduc}), as the normalization and averaging process sometimes produced continua slightly offset from a value of 1.0.  For spectral lines with wavelengths below 4500 \AA, we calculate the continuum within predefined ``clean'' segments of the spectrum that are free from significant Balmer absorption. For spectral lines at redder wavelengths, we compute the continuum based on portions of the spectrum between 15 and 60 \AA\ from the line of interest. In both cases, the continuum is determined as the average of the portion of the spectrum under consideration, after sigma clipping (with high and low reject thresholds of 3.0$\sigma$ and 2.0$\sigma$, respectively) and iterating 30 times. Results were manually inspected for our observed spectroscopic sample to ensure that this process yielded reasonable continuum values.

Subsequently, we determine the wavelength range to calculate the equivalent width over on a line-by-line basis. We first select the point within a 3 \AA\ window around the rest wavelength of the line of interest that deviates the most from the calculated continuum level (either above or below the continuum, designating emission and absorption lines, respectively). From this point we then move to both bluer and redder wavelengths until the normalized flux comes to within 1\% of the continuum level. After adding an additional padding of $\pm$1 \AA, we compute the equivalent width over this wavelength range. Overall, the window over which we calculated equivalent widths ranged from $\sim$3$-$4 \AA\ for some nitrogen emission lines to $\sim$50$-$60\AA\ for some strong Balmer absorption features. 

We use a Monte Carlo approach to calculate errors on our equivalent widths. We create 200 versions of our observed spectra with each point is drawn from a gaussian distribution defined by its nominal value and 1$\sigma$ errors. Each of the 200 versions of the data is processed in the same manner described above. Our final quoted equivalent widths and associated errors are the mean and standard deviation of the 200 resulting equivalent widths measurements. For cases where this final mean equivalent width is less than three times standard deviation, we consider that the line is not robustly detected and instead compute upper limits on the equivalent width. This is done by adding a gaussian to the data at the rest wavelength of the line under consideration and with the typical Magellan/MagE resolution of $R = 4100$. The strength of this injected gaussian is increased until the Monte Carlo approach described above yields an equivalent width that is $\geq$3 times its error. This entire process is repeated twice: once injecting absorption lines and once injecting emission lines to the observed data. Thus, for non-detected lines, we quote a three sigma upper limit on the equivalent width of both an absorption feature (a positive value) and an emission feature (a negative value).

In total, we compute equivalent widths for 10 features, which we found to be the most useful in classifying and determining the physical properties for our sample of stars: He\textsc{ii}~$\lambda$3835/H$\eta$, He\textsc{i}/\textsc{ii}~$\lambda$4026, N\textsc{iv}~$\lambda$4057, He\textsc{ii}~$\lambda$4100/H$\delta$, He\textsc{ii}~$\lambda$4339/H$\gamma$, N\textsc{v}~$\lambda$4604, N\textsc{iii}~$\lambda$4634, He\textsc{ii}~$\lambda$4860/H$\beta$, He\textsc{ii}~$\lambda$5411, and He\textsc{i}~$\lambda$5876. Results are given in Table~\ref{tab:ew}.  

We note that some of the stars in our sample may be double-line spectroscopic binaries. In this case, our process of producing a final combined spectrum of each star by shifting each individual epoch based on a certain set of lines may slight blur other lines if they are dominated by different members of the binary system. We test the impact of this on our equivalent width measurements by repeating the process above on two combined templates for our objects: one aligned based on the isolated HeI/II lines in the spectrum of each star and one based on the Balmer lines. While individual equivalent width measurements vary slightly, they are always within the errors.

\subsubsection{Equivalent Width Measurements: Theoretical Models}

Equivalent widths for absorption/emission lines in the model spectra described in \S~\ref{sec:models} were measured by the same code developed for our observed sample. However, two adjustments were required to ensure that the equivalent widths measured for the theoretical models were truly ``equivalent'' to those measured for the observed data. 

First, the theoretical CMFGEN models presented in \S~\ref{sec:models} were normalized by the \emph{true} stellar continuum. In many cases, this results in a flat spectrum centered at a normalized flux level of 1.0 over the full wavelength range of interest (3700 \AA\ $<$ $\lambda$ $<$ 6500 \AA). However, for certain stellar parameters, Balmer absorption increases significantly---such that the entire spectrum has a normalized flux level of $<$ 0.9 at blue wavelengths ($\lambda$ $<$ 4300 \AA). While calculating equivalent widths on these models directly would provide the most accurate description of the total absorption provided by a given line, this would \emph{not} be directly comparable to values calculated from our observed spectroscopic sample. As described in \S~\ref{sec:specreduc}, we effectively normalize our Magellan/MagE spectra by fitting a low order spline to the local ``pseudocontinuum''. Therefore, prior to measuring the equivalent widths for the theoretical models, we run the \texttt{pyraf} routine \texttt{continuum} on the portion of the models with $\lambda < 4300$ \AA, in order to renormalize them to their local pseudocontinua as well.

Second, after initially measuring the equivalent widths on these renormalized models, we noted what appeared to be a systematic offset to larger equivalent width values compared to our observed sample. Upon investigation, this was due to the ``threshold'' described above wherein we determine the wavelength range to compute the equivalent width over which is based on when a line comes within 1\% of the calculated continuum level. As the CMFGEN models contain no noise, it takes longer for a given line to reach this threshold compared to a very similar observed spectrum that---depending on its signal-to-noise---will ``scatter'' over this threshold sooner. Thus, we found we were effectively integrating the theoretical models over a larger wavelength range than the observed data. 

We chose to account for this second issue by adding gaussian random noise to the theoretical CMFGEN models prior to calculating their equivalent widths. We added noise at multiple levels, based on the typical signal-to-noise of our observed spectroscopic sample. The main equivalent widths used throughout this manuscript correspond to those measured from the theoretical models when gaussian random noise was added with a signal-to-noise of 25 (e.g.\ Figure~\ref{fig:SpT}) or 100 (e.g\ Figure~\ref{fig:diagnostic_plots}). In each case, we calculate the equivalent widths for each model 10 times and average the results. We measure equivalent widths in this manner for the CMFGEN stripped helium star grid, the CMFGEN OB main sequence stars, the helium star plus main sequence star grid, and the TLUSTY OB main sequence star models, all described above.

\subsubsection{Model Comparisons and Resulting Classifications}

As shown in \figref{fig:SpT} and described in the main text, all 25 stars in our final sample have spectral morphologies consistent with the predictions of our composite model grid. Using the specific equivalent width criteria defined in \S~\ref{sec:SpTs} we would classify eight stars (four from the LMC and four from the SMC) as Class 1 -- ``Helium-star-type'', seven stars (all from the LMC) as Class 2 -- ``Composite-type'', and 10 stars (five in the LMC and five in the SMC) as Class 3 -- ``B-type''. However, one of the stars that lacks \HeII\ $\lambda 5411$ (placing it in our fiducial ``B-type'' region) does have a detected \HeII\ line (\HeII\ $\lambda 4686$, star 15 in the LMC) indicating the presence of a hot star. We therefore re-classify it to be in the ``Composite'' group\footnote{\HeII\ $\lambda 4686$ is a stronger line than \HeII\ $\lambda 5411$, but since \HeII\ $\lambda 4686$ is very sensitive to wind mass loss, we chose to use the next strongest \HeII\ line, which is \HeII\ $\lambda 5411$ for classification.}.

\subsection{Summary of Final Observed Sample}\label{sec:samplesum}

Our final sample contains 25 stars, 8 with ``Helium-star-type'' spectra (Class 1), 8 with ``Composite-type'' spectra (Class 2), and 9 with ``B-type'' spectra (Class 3). For simplicity, we number the stars from 1-8, 9-16, and 17-25 for the three spectral groups. Tables \ref{tab:photom}, \ref{tab:kinematics} and \ref{tab:ew} provide summaries of the photometric, kinematic, and spectroscopic properties, respectively. Figure \ref{fig:CMD} shows the location of these stars on a UV-optical CMD. None of these objects had previously been classified on Simbad.
The final combined optical spectra for all 25 objects are shown in Figs.~\ref{fig:optical_spectra_isolated}-\ref{fig:optical_spectra_Btype2}. We label each star and mark relevant spectral lines. In some cases we perform cosmetic clipping to remove contamination from poorly subtracted Earth sky lines and lines of surrounding ionized nebulae. The interstellar calcium lines Ca\textsc{ii} H \& K are present in many of the stars. Here, we provide a more detailed description of the spectral morphology observed in each class: 

\subsubsection{Class 1: ``Helium-star'' group}

\figreftwo{fig:optical_spectra_isolated}{fig:optical_spectra_isolated2} show the normalized optical spectra for the stars in the ``Helium-star'' group, that is, stars 1-8. 
To begin, we highlight that all the spectra are absorption line spectra, with the exception of a few weak emission features from metal ions created via photospheric effects. This suggests that the wind mass loss is weak, explaining why these stars have not been identified in previous narrow-band surveys, such as \cite{2014ApJ...788...83M}. 

The most dominant type of line in the spectra is \HeII, in particular the Pickering series (the $n \rightarrow 4$ line series, including 4026, 4100, 4200, 4339, 4542, 4860, 5412, 6560) and \HeII\ $\lambda 4686$ (the $\alpha$ line of the $n \rightarrow 3$ series). All of the stars in Class 1 exhibit these mentioned \HeII\ lines, some show also shorter-wavelength Pickering lines. In some cases, the signal is sufficient to notice \HeII\ lines from the $n \rightarrow 5$ series, for example \HeII\ $\lambda 6684$. 
\HeI\ lines are either weak or not present at all in the spectra of this class. If we were to follow the MK or MKK system, which implemented the O-type classification based on \HeII\ $\lambda 4542$/\HeI\ $\lambda 4471$ line ratios, this would mean that these stars should have earlier spectral type than O4 \cite{2002AJ....123.2754W, 2009ssc..book.....G}. However, we note (and also describe in more detail in \secref{sec:Teff_diagnostic}) that \HeI\ $\lambda 5876$ is somewhat more temperature sensitive than \HeI\ $\lambda 4471$, and is present in the spectra of stars 5 and 7.

It is somewhat complicated to identify lines of hydrogen in the ``Helium-star'' group, because each of the Balmer lines, which are the only hydrogen lines present in this wavelength range, overlap with every second Pickering line. 
However, it is possible to identify the presence of hydrogen lines ``by eye'' if these blends are stronger than the surrounding pure Pickering lines. This is the case for stars 1, 2, 4, and 6. 

Stars 1, 2, 3, 4, 5, and 6 show lines of highly ionized nitrogen, while the spectra of stars 7 and 8 are likely too noisy for these weak features to be robustly identified. Quadruply ionized nitrogen is easiest visible from \NV\ $\lambda 4945$ (stars 1, 2, 3, 4, 6), but also the \NV\ $\lambda\lambda 4604/20$ doublet (stars 1, 2, 5, 6) and \NV\ $\lambda 4519$ (star 1). Triply ionized nitrogen is easiest identified from \NIV\ $\lambda 4057$, present in the spectra of stars 5, and 6. We do not identify lines of doubly ionized nitrogen.

\begin{landscape}
\begin{figure}
\centering
\includegraphics[width=24cm]{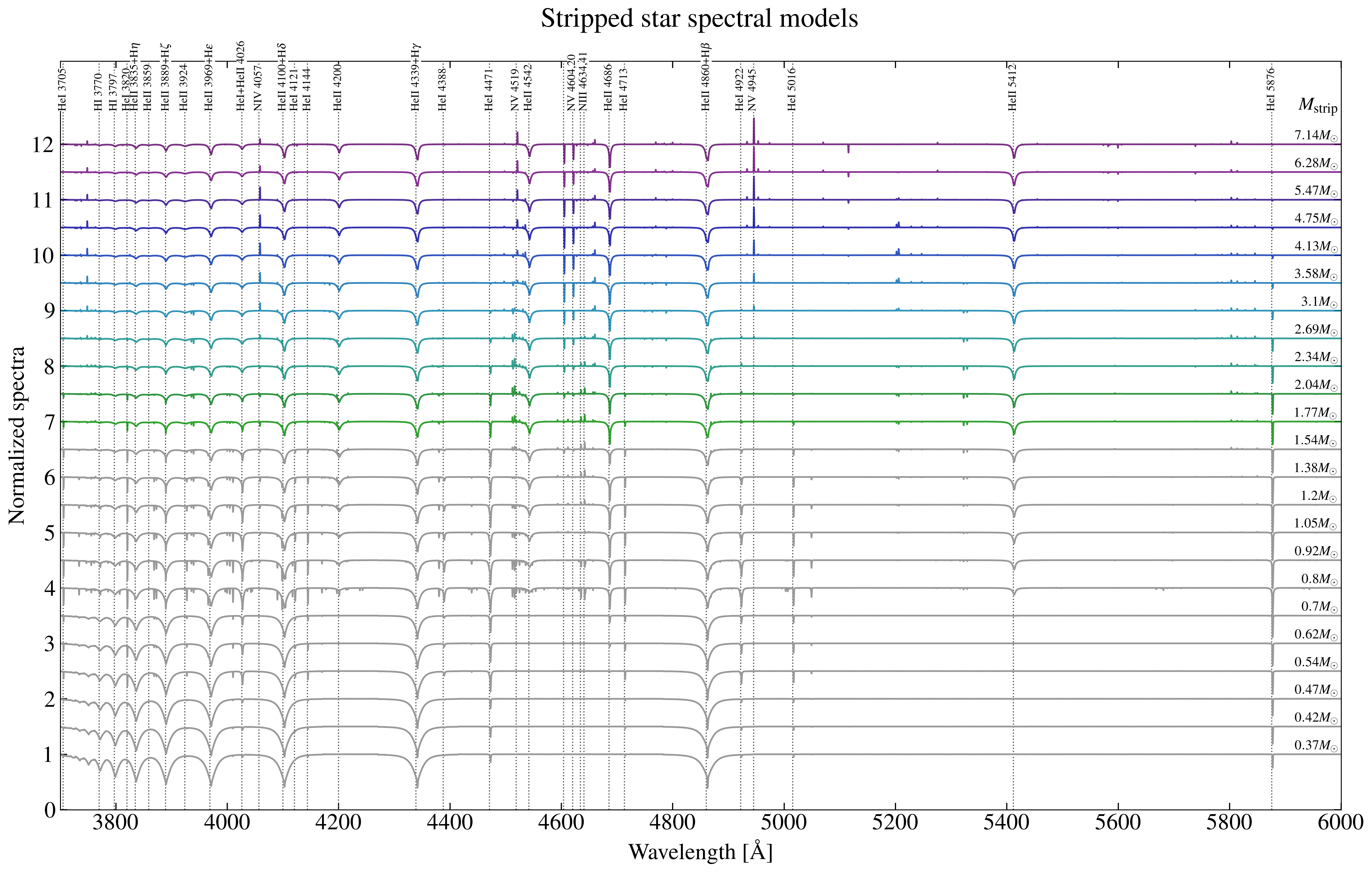}
\caption{Models for the normalized optical spectra of stars stripped in binaries, computed based on the surface properties predicted by the evolutionary models presented in \S~\ref{sec:evolutionary_models} and described in \S~\ref{sec:MESAspec}. The models shown in gray are unchanged since their publication in \cite{2018A&A...615A..78G}, while we have updated the winds of the models shown in color (see \tabref{tab:updated_stripped_star_models}). We label characteristic spectral lines, which include lines of neutral and ionized helium, line blends of ionized helium and hydrogen, and lines of doubly, triply and quadruply ionized nitrogen. The stripped star masses are displayed to the right.}
\label{fig:optical_spectra_006}
\end{figure}
\end{landscape}

There are lines of ionized carbon present in the spectra of stars 2, 5, 7, and 8. Triply ionized carbon can be seen from the doublet \CIV\ $\lambda\lambda 5801/12$ (stars 2, 5, and 7) and \CIV\ $\lambda 4658$ (star 7 and 8), while doubly ionized carbon is identified from a \CIII\ triplet at 4647, 4650, and 4651 (stars 5, 7, and 8) (cf., \cite{2012A&A...545A..95M}). While helium and nitrogen lines are expected for stripped stars, carbon lines are not expected (cf.\ \figref{fig:optical_spectra_006}). This is because stripped stars are the exposed cores of massive stars, meaning that the material in their surfaces have once been processed via the CNO-cycle during which nitrogen is enhanced at the expense of carbon and oxygen. 
Carbon enrichment has been identified in He-sdO stars \cite{2009PhDT.......273H, 2013A&A...551A..31D}. Its origin has been debated, for example, carbon enrichment occurs in the ``late hot flasher'' scenario for low-mass star evolution and in the evolution of a star produced by the fast merger of two white dwarfs \cite{2016PASP..128h2001H}. Given the brightness of the stars in Class 1, it is unlikely that they are the exposed cores of low-mass stars or merger products of white dwarfs (see \S~\ref{sec:nature} for a discussion of possible foreground contamination).

\subsubsection{Class 2: ``Composite'' group}

The optical spectra for the stars in the ``Composite'' group are shown in \figreftwo{fig:optical_spectra_composite}{fig:optical_spectra_composite2}. These include stars 9-16. As in the case of Class 1, these are absorption line spectra.
The most dominating features in the spectra of the ``Composite'' type is the hydrogen Balmer series lines. Their presence is particularly prominent in the wavelength range between 3750 and 4100 \AA, within which six Balmer lines are present. Comparing to the stars in the ``Helium-star'' group, which do not show any signs of strong short-wavelength Balmer lines (see \figref{fig:optical_spectra_isolated}), this part of the spectrum is the most different. 

\begin{landscape}
\begin{figure}
\centering
\includegraphics[width=24cm]{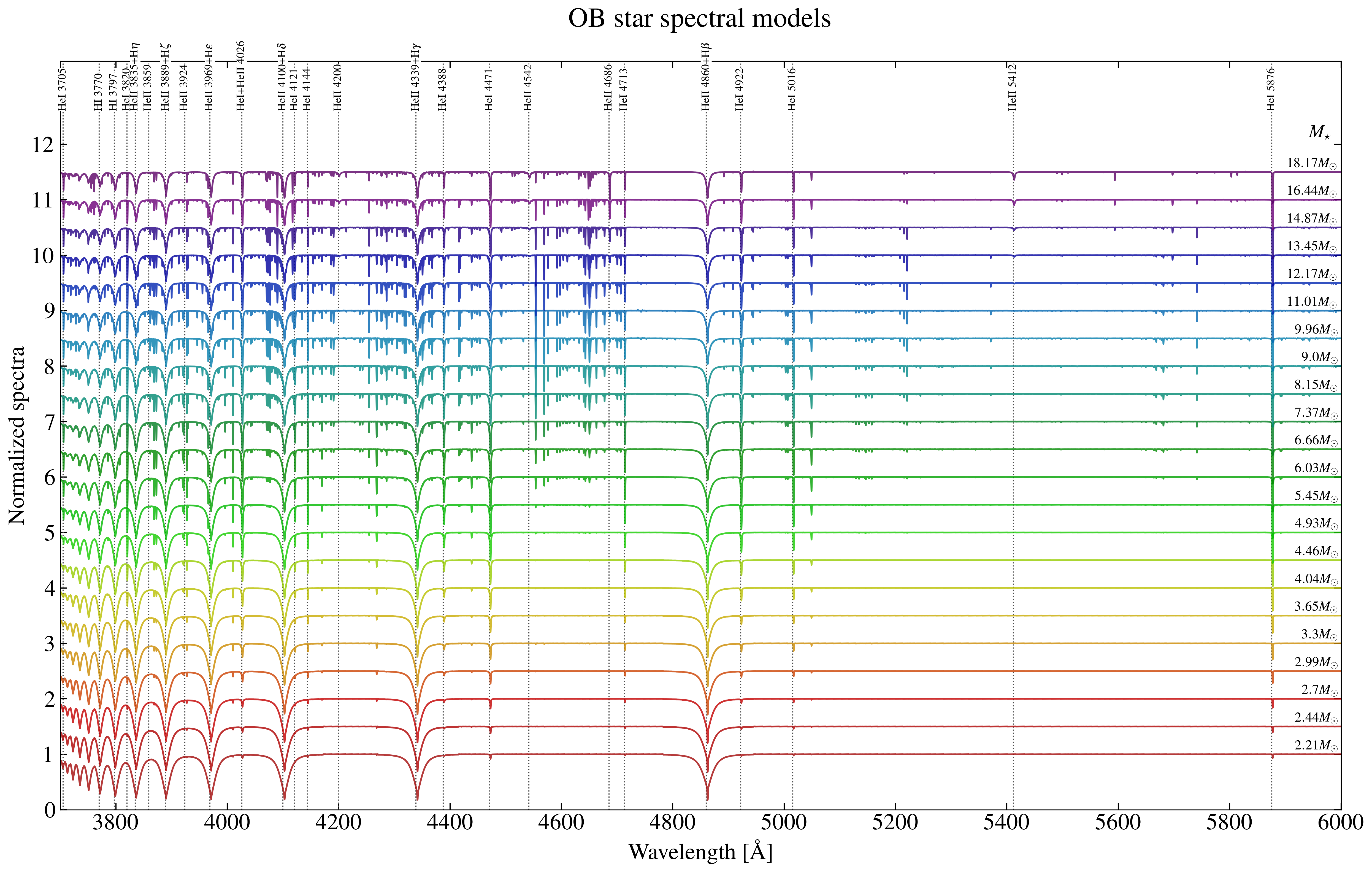}
\caption{Models for the normalized optical spectra of main-sequence stars early on their main-sequence evolution (20\% of the MS duration has passed), and described in \S~\ref{sec:MESAMSspec}. The spectra are organized, from top to bottom, by decreasing mass, and therefore also decreasing temperature. Relevant spectral features are marked, which includes primarily hydrogen Balmer lines (all stars), lines of neutral helium ($M_\star \gtrsim 3\Msun$), and lines of ionized helium in the most massive MS stars ($M_\star \gtrsim 15\Msun$). }
\label{fig:optical_spectra_OB}
\end{figure}
\end{landscape}

For these stars to be grouped into the ``Composite'' group, we required that they show lines of ionized helium. All stars show \HeII\ $\lambda 4686$, and all but star 15 also show \HeII\ $\lambda 4542$ and \HeII\ $\lambda 5412$. \HeII\ $\lambda 4200$ is present in stars 10, 12, 13, 14, and 16. 
Lines of neutral helium is present in stars 9, 10, 13, 14, 15, and 16, but cannot be identified in stars 11 and 12. The most prominent lines are \HeI\ $\lambda 6678$, \HeI\ $\lambda 5876$, \HeI\ $\lambda 4922$, \HeI\ $\lambda 4713$, \HeI\ $\lambda 4471$, \HeI\ $\lambda 4388$, \HeI\ $\lambda 4144$, \HeI(+\textsc{ii}) $\lambda 4026$, and \HeI\ $\lambda 3820$. 
Lines of neutral helium is characteristic to the B-type classification and disappear for stars cooler than $\sim 10,000$~K (very late B or A-type), corresponding roughly to $\sim 2-3\Msun$ main-sequence stars (see \figref{fig:optical_spectra_OB} and \tabref{tab:OB_properties}). The lack of \HeI\ lines in the spectra of stars 11 and 12 could therefore suggest that they are binaries composed of a hot stripped star and a low-mass main sequence companion -- each component not expected to show \HeI\ lines.
 
Because of the limited signal to noise ratio in the optical spectra, we can confirm the presence of lines of ionized nitrogen and carbon only in two of the stars with the best signal, but note possible hints of lines present in other stars in the sample as well. We confirm the presence of \NIII\ lines at 4631, 4634, 4641, and 4642 \AA\ for star 16 and the presence of \CIII\ lines at 4647, 4650, and 4651 \AA\ for stars 15 and 16. There seem to be hints of the same \CIII\ lines in addition to the \CIV\ $\lambda\lambda 5801/12$ doublet in star 11. 
As illustrated in \figref{fig:example_SpT}, lines of ionized nitrogen may be visible in composite spectra of stripped star binaries. We have not detected \NV\ or \NIV\ lines in the stars in the ``Composite'' group. However, with the current signal-to-noise ratio in the spectra and given that these weak features from the stripped star should be further weakened by the main-sequence star, it is not surprising that we have not yet discovered them.

\subsubsection{Class 3: ``B-type'' group}

The optical spectra for the stars in the ``B-type'' group are shown in \figreftwo{fig:optical_spectra_Btype}{fig:optical_spectra_Btype2}. 
These spectra can be considered to be typical for B-type stars. The hydrogen Balmer series is dominating the spectral morphology and plenty of lines of neutral helium are present. 
B-type stars are typically classified using lines of silicon and magnesium \cite{2009ssc..book.....G}, but these are too weak to be possible to identify given the noise level in our spectra. 
\cite{2020A&A...634A..51B} used the helium lines \HeII\ $\lambda 5412$ and \HeI\ $\lambda 6678$ to classify B-stars. When adopting their method, we identify that the stars in the ``B-type'' group should have spectral types of B1-B7, since all have \HeI\ $\lambda 6678$ present but none have \HeII\ $\lambda 5412$.

We note that the width of the \HeI\ features vary between the stars in the ``B-type'' group. While none of the stars show very broad features, it is possible that stellar rotation can have affected the shapes of these narrow lines. 
We do not identify any spectral features of metals apart from the interstellar calcium doublet H and K. 
Finally, it may be worth to note that the spectrum of star number 24 appears somewhat anomalous, with possibly weaker Balmer features compared to the other stars. However, this is speculative given the very low signal-to-noise of the spectrum.

\section{Physical Properties of Observed Stars in the ``Helium star type'' Spectral Class}\label{sec:Physical_properites}

In the main text, we described how we estimated the stellar properties of the stars in Class 1. Here, we discuss several of these steps in more detail. In addition, we present the estimated ranges for the stellar properties in \tabref{tab:estimate_stellar_properties}.

\subsection{Effective temperature}\label{sec:Teff_diagnostic}

The temperature of the atmosphere of stars directly affects the ionization balance of different elements and therefore also the spectral morphology. 
The helium ionization balance has been used to limit the effective temperatures of main sequence O and B-type stars, which typically exhibit \HeII\ and/or \HeI\ lines \cite{2009ssc..book.....G,2020A&A...634A..51B}. 
The nitrogen ionization balance is more sensitive to temperature differences in stars hotter than $\sim 40,000$~K compared to the helium ionization balance. However, although stripped stars are thought to have nitrogen enriched surfaces, the nitrogen lines are weak, which can make them hard to detect (see \figref{fig:optical_spectra_isolated}). Therefore, we choose to use the helium ionization balance as the main temperature determinant, but also describe how the obtained temperature range can be refined if nitrogen lines are identified in the spectrum. 
\newline

\begin{figure}
\centering
\includegraphics[width=0.49\textwidth]{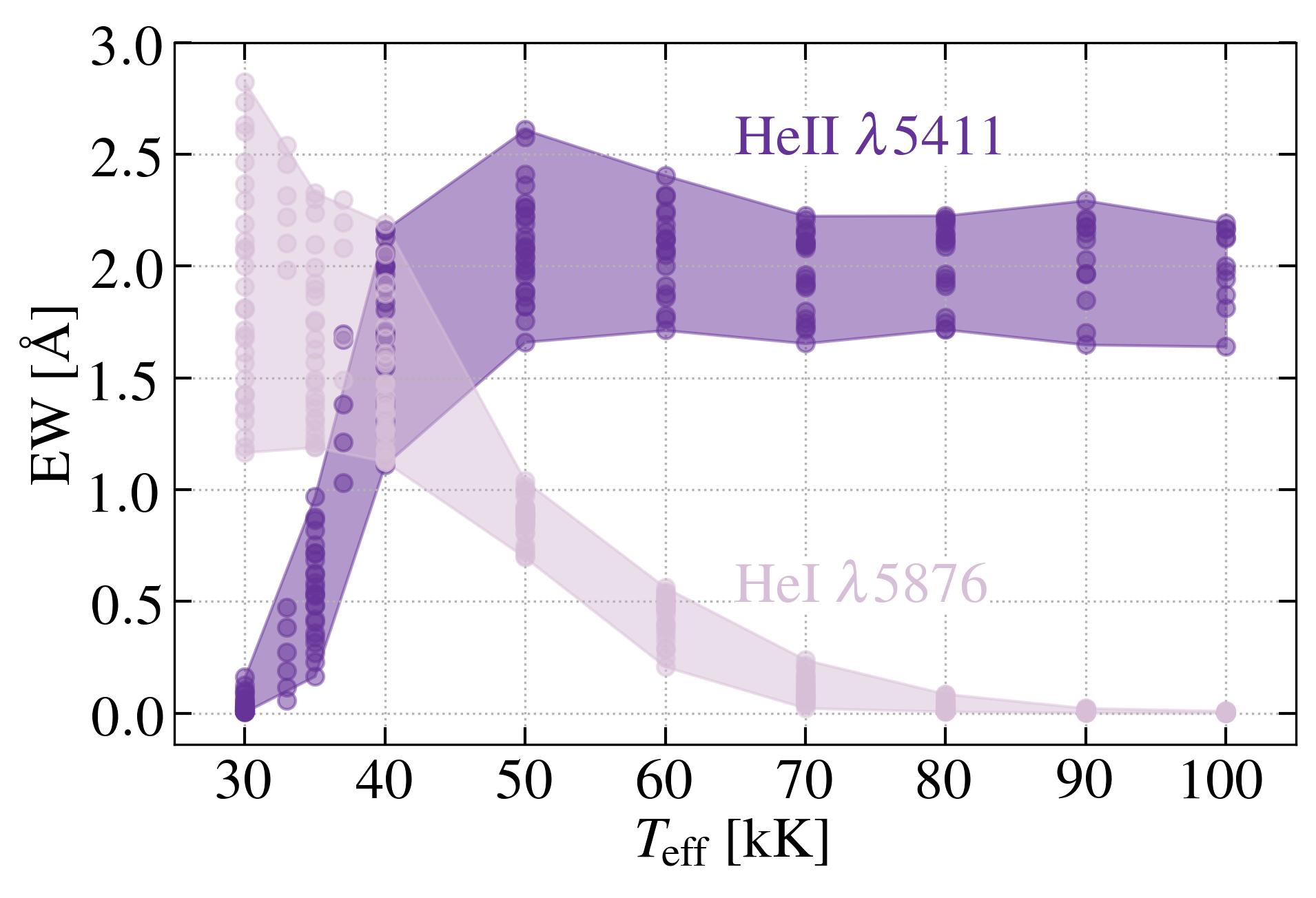}
\includegraphics[width=0.49\textwidth]{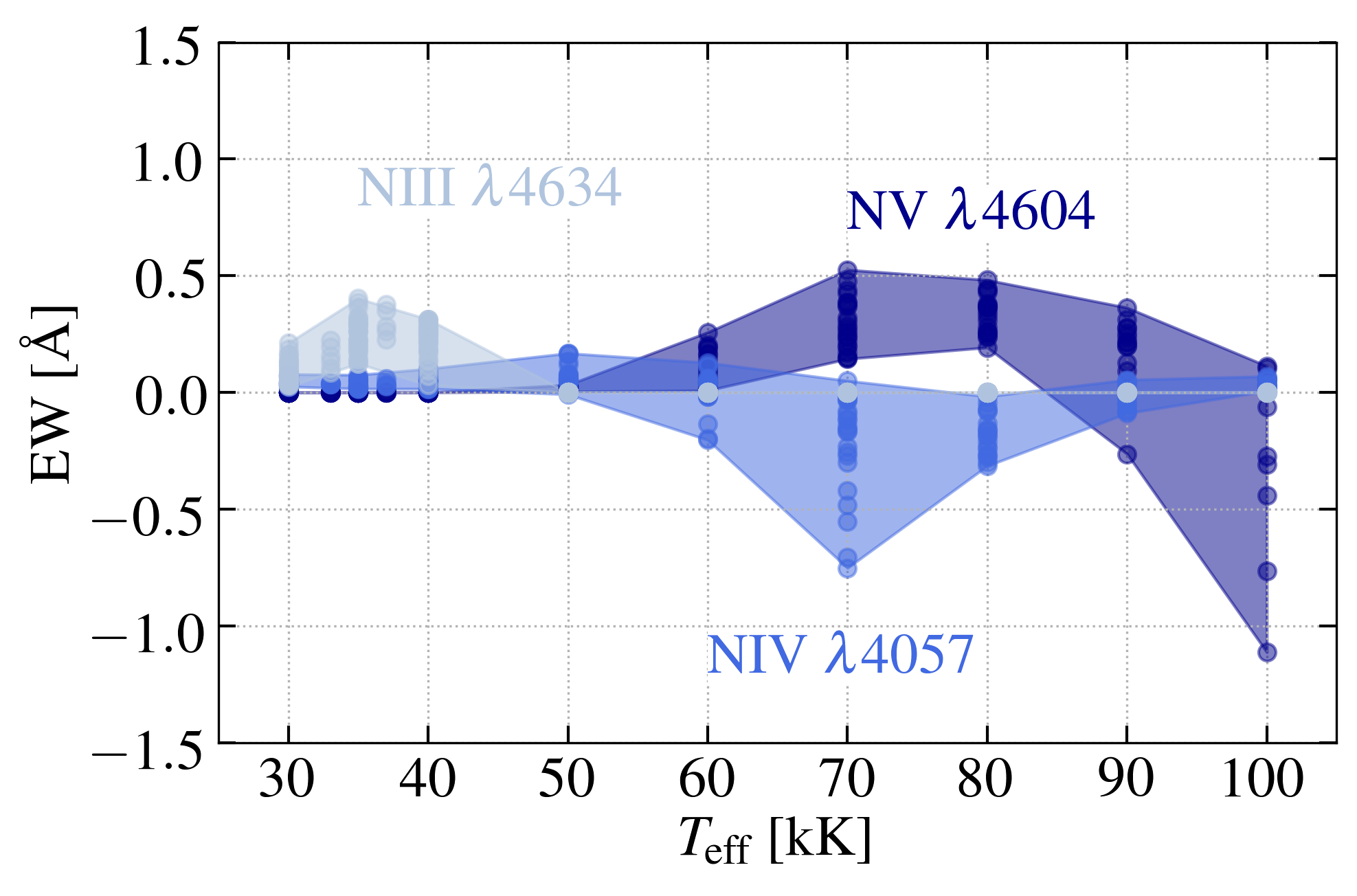}
\caption{The ionization balance of helium (left) and nitrogen (right) can be used to constrain the effective temperature of hot stars. We show equivalent widths of \HeI\ $\lambda 5876$ (light purple) \& \HeII\ $\lambda5411$ (dark purple) and \NIII\ $\lambda 4634$ (light blue), \NIV\ $\lambda4057$ (medium blue) \& \NV\ $\lambda4604$ (dark blue) measured for spectral models for stripped stars and as function of effective temperature.}
\label{fig:ionization_balances}
\end{figure}

\noindent \textbf{Constraints from helium lines} $\quad$
By inspecting the spectral models described in \secref{sec:agnostic_grid}, we found that the \HeI\ line that is present at the highest temperatures is \HeI\ $\lambda$5876, which can be detected for temperatures up to $60,000$~K and in some cases even $70,000$~K. For the \HeII\ lines, \HeII\ $\lambda$4686 is the \HeII\ line present at the lowest temperatures, appearing when $T_{\mathrm{eff}} \gtrsim 30,000$~K, but it is very sensitive to wind mass loss and may be partially filled in by a weak wind in observed spectra. Therefore we search among the other \HeII\ lines and find that \HeII\ $\lambda$5411 is most suitable to use, because it is a strong line, which also is minimally perturbed by other spectral features. For $T_{\rm eff} >50,000$~K, the equivalent width of \HeII\ $\lambda$5411 is not affected by increasing temperature, but the equivalent width of \HeI\ $\lambda$5876 decreases with increasing temperature. This makes the comparison between the equivalent widths of these two lines useful for a temperature estimate. This temperature sensitivity is visualized in the left panel of \figref{fig:ionization_balances}, where we show the equivalent widths of these two lines measured in the spectral models and plotted as function of effective temperature.

In \figref{fig:diagnostic_plots}a, we display the equivalent widths of \HeII\ $\lambda$5411 and \HeI\ $\lambda$5876 measured in the spectral model grid for stripped stars and mark their respective temperatures with colored dots and shaded regions. The separation in temperature is clear for models with temperatures up to about $60,000$~K. This implies that the diagram can be used for rough temperature determinations up to temperatures of about $60,000$~K. Above this threshold, the models overlap in the diagram and a lower limit of $70,000$~K can be obtained. 
Based on this comparison, all stars in Class 1 have effective temperatures of $50,000$~K or higher. 
\HeI\ $\lambda$5876 was not detected for six of the stars (stars number 1, 2, 3, 4, 6, and 8), indicating that their effective temperatures are $\gtrsim 70,000$~K, with the exception of star number 8 which has larger uncertainty and its effective temperature can be constrained to $\gtrsim 60,000$~K. For two stars (stars number 5 and 7) \HeI\ $\lambda$5876 was detected.  
We estimate the effective temperature of star number 5 to $\sim 60,000$~K and of star number 7 to $\sim 60,000$-$70,000$~K. 
The effective temperature estimates are summarized in the second column of \tabref{tab:estimate_stellar_properties}.

In \figref{fig:diagnostic_plots}a, we also mark the location of the O and B-star grids from TLUSTY. As shown, these main sequence star models have smaller equivalent widths than the stripped star models grids and are therefore located in the bottom left corner of the diagnostic diagram. The reason their equivalent widths are smaller is because of their lower surface mass fraction of helium. In fact, this composition dependence is also present in our model grids, with the most helium poor models showing smaller equivalent widths and with increasing surface helium content their equivalent widths also increase. 
\newline

\noindent \textbf{Constraints from nitrogen lines} $\quad$
Lines of doubly (\NIII), triply (\NIV), and quadruply (\NV) ionized nitrogen are expected to provide a more accurate estimate for the effective temperature if they are detected in the stellar spectrum (cf.\ \cite{2012A&A...543A..95R,2017A&A...600A..82G}). The sensitivity is displayed in the right panel of \figref{fig:ionization_balances}, where we use three of the strongest nitrogen lines: \NIII\ $\lambda$4634, \NIV\ $\lambda$4057, and \NV\ $\lambda$4604. The figure shows that \NIII\ $\lambda$4634 is only expected for effective temperatures of $\lesssim 40,000$~K, while \NIV\ $\lambda$4057 is expected in emission when the effective temperature is between $\sim 60,000$ and $\sim 80,000$~K and \NV\ $\lambda$4604 is expected in absorption for effective temperatures between $\sim 60,000$ and $\sim 90,000$~K, however, it can flip into emission if the effective temperature is higher than $\sim 90,000$~K. \NIV\ $\lambda4057$ can also appear in absorption, but it should be weak and if \NIII\ $\lambda 4634$ and \NV\ $\lambda 4604$ are not present, the temperature range can be narrowed down to $\sim 50,000$~K. 

We have identified the presence of nitrogen lines in the spectra of five of the stars in the Helium star type  group. Using the above described relations between line presence and effective temperature, we can constrain the effective temperature of star number 1 to $\gtrsim 90,000$~K (\NV\ $\lambda 4604$ is in emission), star number 2 to $\sim 70,000$-$90,000$~K (\NIV\ $\lambda 4057$ is present in emission), 
star number 5 to $\sim 60,000$-$80,000$~K (\NIV\ $\lambda 4057$ is present in emission) and star number 6 to $\sim 70,000$-$80,000$~K (\NIV\ $\lambda 4057$ is present in emission). 
Informed by the presence of nitrogen lines and the limits from the helium ionization balance, we provide an extra column in \tabref{tab:estimate_stellar_properties} ($T_{\rm eff, N}$) where we provide refined estimates for the effective temperature.
It is likely that the other stars in Class 1 also will show nitrogen lines if higher signal-to-noise ratio of their spectra was obtained. We note, however, that the nitrogen lines are also affected by the nitrogen abundance, whose variation we have not explored in the spectral model grid.

\begin{table}
\centering
\caption{The estimated ranges for the effective temperature, surface gravity, surface helium mass fraction, and surface hydrogen mass fraction for the stars in the Helium star type group. The first column for effective temperature is determined using the helium ionization balance shown in {color{red}\figref{fig:diagnostic_plots}a}, while in the second column we have narrowed the temperature range when the nitrogen ionization balance could be used.}
\label{tab:estimate_stellar_properties}
\begin{tabular}{lccccc}
\toprule\midrule
Star    & $T_{\rm eff}$ [kK] & $T_{\rm eff, N}$ [kK]        & $\log_{10} (g/{\rm cm\; s}^{-2})$ & $X_{\rm He, surf}$ & $X_{\rm H, surf}$ \\
\midrule
\textbf{SMC} \\
1     & $\gtrsim 70$  & $\gtrsim 90$      & $\sim$5.0-5.5 & $\sim 0.69$   & $\sim 0.3$ \\  
2    & $\gtrsim 70$  & $\sim 70-90$      & $\sim$5.2-5.7 & $\sim 0.69$   & $\sim 0.3$ \\  
3    & $\gtrsim 70$  & --                & $\gtrsim$5.2 & $\sim 0.89$   & $\sim 0.1$ \\ 
4    & $\gtrsim 70$  & --                & $\gtrsim$5.5 & $\sim 0.89$   & $\sim 0.1$ \\ \\

\textbf{LMC} \\
5    & $\sim 60$     & $\sim 60-80$      & --            & $\sim 0.98$   & $\sim 0.01$ \\
6     & $\gtrsim 70$  & $\sim 70-80$      & $\sim$5.0-5.7 & $\sim 0.89$   & $\sim 0.1$ \\
7    & $\sim 60-70$     & --                & $\gtrsim$ 5.5 & $\sim 0.69$   & $\sim 0.3$ \\    
8     & $\gtrsim 60$  & --                & $\sim$4.8-5.2 & $\sim 0.89$   & $\sim 0.1$ \\ 
\bottomrule
\end{tabular}
\end{table}

\subsection{Surface gravity}\label{sec:logg_diagnostic}

Surface gravity is well known to affect the width and shape of spectral lines, but it also affects their strengths and equivalent widths. 
For the hydrogen Balmer series and the \HeII\ Pickering series, the equivalent widths of the short-wavelength lines decrease with increasing surface gravity, while the equivalent widths of the long-wavelength lines increase with increasing surface gravity. 
We exploit this effect to estimate the surface gravity of the stars in the ``Helium star type'' group. We choose to use the spectral lines \HeII\ $\lambda$3835/H$\eta$ and \HeII\ $\lambda$4860/H$\beta$ to represent the short- and long-wavelength lines. These two lines are both blends of lines in the hydrogen Balmer series and the \HeII\ Pickering series, making a comparison less dependent on the surface mass fraction of hydrogen or helium. These lines are also minimally affected by other spectral features.

In \figref{fig:diagnostic_plots}b, we show the diagnostic plot for estimating the surface gravity. 
We have chosen to limit the models that are shown to $T_{\rm eff} \geq 50,000$~K to avoid clutter and excessive overlap. 
A monotonic trend is visible: the equivalent width of \HeII\ $\lambda$3835/H$\eta$ decreases with increasing surface gravity, while the equivalent width of \HeII\ $\lambda$4860/H$\beta$ increases. Although several regions overlap in the diagram, it is clear that low surface gravities ($\log _{10} g \sim 4-4.5$, red and orange colors) are relatively well separated in the upper left part from the higher surface gravity models in the bottom right part. 

For comparison, we also show the location of the O-stars in the TLUSTY grid with $T_{\rm eff} \geq 50,000$~K. These suggest that hot main-sequence stars should have higher equivalent width of \HeII\ $\lambda$3835/H$\eta$ compared to our models. As in the case of effective temperature, the difference in location is because of the slight difference in surface composition -- the higher hydrogen abundance in main-sequence stars allow for stronger Balmer features. 
All stars in the sample are consistent with having surface gravities higher than main-sequence stars. 
None of the observed stars overlap with regions where the models have $\log_{10} g =$ 4, 4.3 or 4.5 and none of them overlap with the location of hot main-sequence stars. 
 
By comparing to the models, we can constrain the surface gravity for stars 1, 2, 6, and 8, which all have both lines measured, to $\sim$5.0-5.5, $\sim$5.0-5.7, $\sim$5.0-5.5, and $\sim$4.8-5.2, respectively. \HeII\ $\lambda$3835/H$\eta$ is not detected in the spectra of stars 3, 4, and 7. Therefore, using their upper limits, we estimate their surface gravity to be $\gtrsim$5.2, $\gtrsim $5.5, and $\gtrsim $5.5, respectively. 
Although both lines were detected for star number 5, the measurements do not overlap with any of the expectations from the models and we therefore consider its surface gravity estimate inconclusive. There could be several reasons to why the measured equivalent widths for star number 5 do not overlap with those of the models. We note, for example, the difficulty of determining the continuum level in observed spectra. This could lead to a star appearing as an outlier even though a well-fitting model has been found. Comparing the equivalent widths of just two lines in the spectrum is also an approximate method and careful spectral fitting is necessary for better determining the stellar properties. 

\subsection{Helium enrichment}\label{sec:XHes}

The strengths of helium and hydrogen lines are directly related to the helium and hydrogen content in a stellar atmosphere. Therefore, because every second line in the \HeII\ Pickering series is blended with a hydrogen Balmer line, the presence of hydrogen in a stellar atmosphere can be recognized if every second Pickering line (the ones blended with Balmer lines) is stronger. 
We choose to utilize this expected variation by comparing the equivalent widths of a pure helium line with that of a hydrogen line blended with a helium line. For these, we have chosen to use the pure helium blend \HeI/\HeII\ $\lambda$4026 and the helium/hydrogen blend \HeII\ $\lambda $4100/H$\delta$. These two lines are also minimally affected by other spectral features.

The models with the different surface helium (hydrogen) mass fractions 0.98 (0.01), 0.89 (0.1), 0.69 (0.3), and 0.49 (0.5) in the spectral model grid (\secref{sec:agnostic_grid}) clearly separate when plotting the equivalent widths of \HeI/\HeII\ $\lambda$4026 and \HeII\ $\lambda $4100/H$\delta$ in \figref{fig:diagnostic_plots}c. The figure shows that with increasing helium content, the pure helium line becomes stronger and the hydrogen/helium blend becomes weaker (see also \figref{fig:diagnostic_plots}b). In the panel, we zoom in on the region for the high temperature models ($T_{\rm eff} \gtrsim 50,000$~K). 
Based on this analysis, all of the stars in the ``Helium star type'' group are consistent with having helium enriched surfaces. 
For all of the stars, both of the lines are detected and we infer that the surface mass fraction of helium (hydrogen) for star number 5 is $\sim 0.98$ ($\sim 0.01$), for stars number 3, 4, 6 and 8 it is $\sim 0.89$ ($\sim 0.1$), and for stars number 1, 2, and 7 it is $\sim 0.69$ ($\sim 0.3$). 

Massive main-sequence stars are expected to have surface mass fractions of helium (hydrogen) of $\sim 0.3$ ($\sim 0.7$), which is far from any of our predictions for the stars in Class 1. However, the main-sequence O-star models from TLUSTY overlap with some of our helium-rich models. The TLUSTY models that overlap are models with lower surface gravity: the ones that overlap with the models with surface helium mass fraction of 0.69 have lower $\log g$ than any of our models ($\log g < 4.0$), while the ones overlapping with the models with surface helium mass fraction of 0.49 have primarily $\log g < 4.5$. 
\newline

\section{Other Possible Origins for the Observed Spectroscopic Sample}\label{sec:nature}

\begin{figure}
\centering
\includegraphics[width=\textwidth]{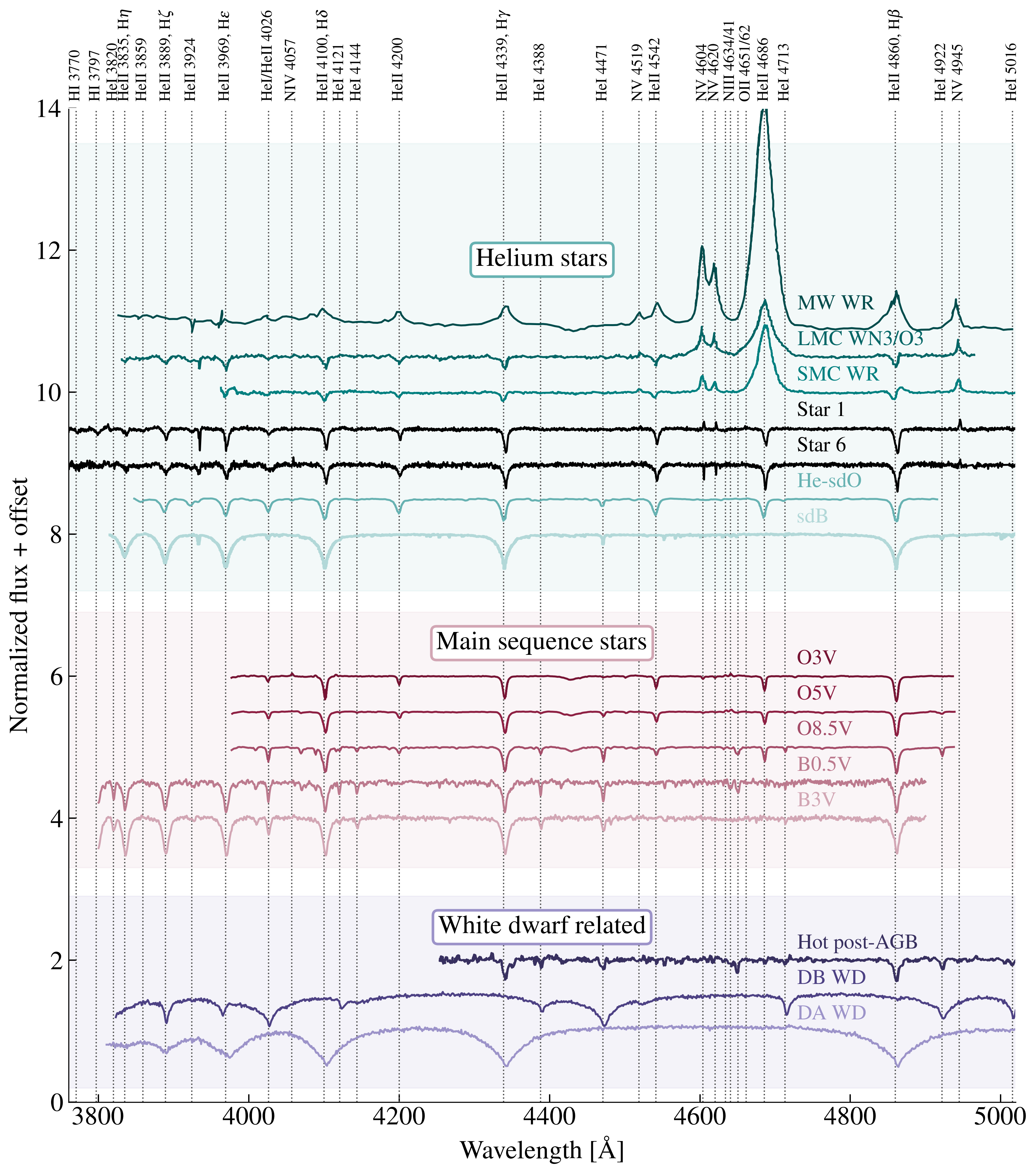}
\caption{Observed spectra of a variety of types of hot stars compared to the spectra of stars 1 and 6 from our spectroscopic sample. In teal shades, we show helium-rich stars. In red shades, we show main-sequence OB stars. In purple shades we show white dwarf related objects. 
From top to bottom, we show 
(1) the Milky Way Wolf-Rayet star WR 152 \cite{1995A&AS..113..459H, 2009ssc..book.....G}, 
(2) the WN3/O3 star LMC 170-2 \cite{2014ApJ...788...83M}, 
(3) the SMC Wolf-Rayet star SMC AB 12 \cite{2015A&A...581A..21H}, 
(4) the SMC star we refer to as star 1 and that is part of the ``Helium-star'' group, 
(5) the LMC star we refer to as star 6 and that is part of the ``Helium-star'' group, 
(6) the Galactic He-sdO star HD~49798 \cite{1998PASP..110.1315G}, 
(7) the Galactic sdB star J2256+0656 \cite{2011A&A...526A..39G}, 
(8) the Galactic O3 star HD 64568 \cite{2011ApJS..193...24S}, 
(9) the Galactic O5 star HD 46150 \cite{2011ApJS..193...24S},
(10) the Galactic O8.5 star HD 46149 \cite{2011ApJS..193...24S},
(11) the Galactic B0.5 star HD 36960 \cite{1990PASP..102..379W}, 
(12) the Galactic B3 star $\eta$ Hyd (also HD 74280) \cite{1990PASP..102..379W},
(13) the hot post-AGB star LSE~45 with spectral type B2I \cite{2003A&A...404..305G}, 
(14) the DB type white dwarf Feige 4 \cite{2013ApJS..204....5K},
(15) the DA type white dwarf J07402413+2029372 \cite{2013ApJS..204....5K}. 
The spectra for white dwarfs (14 and 15) were obtained from \url{https://www.montrealwhitedwarfdatabase.org/home.html} \cite{2017ASPC..509....3D}. 
}
\label{fig:comparison_spectra}
\end{figure}

From the analysis presented in \S~\ref{sec:Physical_properites}, it is clear that the Class 1 stars are hot, compact and helium-rich, while their absolute magnitudes in the CMDs of \figref{fig:CMD} are intermediate between those of known Wolf-Rayet, WN3/O3, and subdwarf stars. Together, these properties are all consistent with expectations for intermediate-mass helium stars, stripped via binary interaction. Here, we investigate other possible origins for these stars. Throughout, we refer to Fig.~\ref{fig:comparison_spectra} where we compare the spectral morphologies observed in our sample to other classes of hot stars. 

\subsection{Over-corrected reddening for main-sequence stars}

In Fig.~\ref{fig:CMD} we adopt a uniform extinction for the LMC/SMC of $A_V = $0.38 and 0.22 mag, respectively (see \S~\ref{sec:UVexcess}). However, the full distribution of $A_V$ values for LMC/SMC stars presented by the MCPS spans 0 to $>$ 1 mag, with implications for our identification of a UV excess. The steepness of the upper MS coupled with the slope of the reddening vector imply that if the the true reddening to any object is higher than our adopted values, they would still possess a UV excess. However, stars with lower reddening could potentially overlap with the MS. 

While the precise CMD locations of the Class 1 and 2 stars will likely shift with more detailed modeling of their line-of-sight extinctions, their spectral morphologies rule out a low-reddening MS star interpretation. In particular, the presence of He\textsc{ii} absorption is only expected in MS stars of type O to B0.5 (Fig.~\ref{fig:comparison_spectra}), which are more luminous than our sample even for our current A$_V$ assumptions. In contrast, it is possible that some of the Class 3 objects are low-reddening B-type MS stars. While one object exhibits binary motion, we have obtained only a single spectrum of most objects in this class. Orbital solutions and UV spectroscopy will help in fully confirming the nature of these stars.
We remind that theory predicts that this spectral morphology should exist for composites consisting of a B-type companion and a lower mass stripped star in the location of the CMD where the Class 3 objects are found (see \S~\ref{sec:prediction}, also \cite{2021AJ....161..248W}).

\subsection{Stars Stripped Via Strong Stellar Winds or Eruptions}
We consider whether our observed sample does contain helium stars, but that were stripped by a different mechanism than binary interaction. Both the physical properties and CMD locations are consistent with expectations for stripped cores that originated from stars with initial masses $\lesssim$20 M$_\odot$. Based on current empirical prescriptions, stars in this mass range do not have sufficient MS or red supergiant mass loss rates to strip their hydrogen envelopes \cite{2021ApJ...922...55B}. In addition, stars below 20 M$_\odot$ are far from their Eddington Limit while on the MS and in the Hertzsprung gap, and are therefore not expected to undergo LBV-like eruptions in the canonical single-star picture.  While this traditional single-star picture for LBV eruptions has been questioned \cite{Smith2015}, known LBVs and LBV-candidates predominantly have luminosities indicative of initial masses $>$20 M$_\odot$ \cite{2019MNRAS.488.1760S}. Finally, while pre-explosion, episodic mass loss may occur in a wide-range of stellar masses/types, it is either predicted to be too weak to expel substantial amounts of material as is the case for wave-driven ejections (cf.\ \cite{2017MNRAS.470.1642F}), or only predicted for significantly more massive stars (helium cores with $>45\Msun$) as is the case for pulsational pair-instability \cite{2017ApJ...836..244W, 2019ApJ...887...53F, 2022arXiv220506386W}.   

Higher mass single stars may be able to strip off their own hydrogen-rich envelopes due to a combination of strong stellar winds and/or eruptions \cite{2014ARA&A..52..487S}. 

However, their exposed helium cores will have high luminosities and are, therefore, expected to have have high mass-loss rates. This should result in strong emission features in their optical spectra, making them appear as Wolf-Rayet stars \cite{2007ARA&A..45..177C}. No such strong wind lines are observed in our sample. Indeed, we see a natural progression in the strength of emission line features between WR stars, the new class of WN3/O3 stars in the LMC \cite{2017ApJ...841...20N}, and our Class 1 objects (see Fig.~\ref{fig:comparison_spectra}) that mimics their relative locations in the CMD. We therefore disfavor an interpretation where either the reddening values or distances were underestimated for the objects in our sample, placing them at much higher luminosities. This, coupled with the confirmed binary motion for Class 1 and Class 2 objects favors binary interaction as the means by which these objects were stripped.

\subsection{Faint and Hot Foreground Stars}

In \S~\ref{sec:kinematics} we found that the radial velocities and proper motions of the stars in our sample were consistent with the 3D kinematics of known OB stars in the Magellanic Clouds. However, this does not rule out foreground contamination for any individual object. In particular, we note that stars 5 and 6 had \emph{Gaia} proper motions slightly offset from the bulk of the LMC, although they also display high excess noise and a poor astrometric goodness-of-fit. Here we examine the possibility that the Class 1 and 2 objects are intrinsically faint foreground objects. For such an object to appear bluewards of the B-type MS it would need to be hot: either a white dwarf (WD) or subdwarf.

The lack of detected parallaxes combined with the large average radial velocities of our sample imply that---if they were foreground objects---they must be located at distances $>$2 kpc and likely within the halo. This rules out WDs which, due to their intrinsic faintness, would need to be located within $\sim$1 kpc to pollute our sample. In addition, the optical spectra of WDs are easily recognizable: they have very high surface gravities ($\log g \gtrsim 7$) that cause their spectral lines to be significantly broader than observed in our spectroscopic sample (Fig.~\ref{fig:comparison_spectra}). 

On the other hand, low-mass stars stripped of their H-rich envelope via interaction with a companion, subdwarfs, could potentially contaminate our sample if located in the halo with distances of $\sim 5-10$ kpc. As they are the same type of objects as intermediate-mass stripped stars, but simply have lower mass, they have sometimes similar spectral morphology (see e.g., \cite{2013A&A...551A..31D, 1990A&A...235..234D, 2003A&A...402..335A}). In particular, the class of He-sdO stars, with their strong \HeII\ absorption bear the closest resemblance to the Class 1 objects (see Fig.~\ref{fig:comparison_spectra}). However, because the helium MS moves to cooler temperatures at lower luminosities (mimicking the MS), very few subdwarfs have effective temperatures higher than $\sim$40~kK \cite{2016PASP..128h2001H}. The 1.5\Msun\ subdwarf in HD~49798 is possibly the subdwarf most similar to intermediate mass stripped stars, with $T_{\rm eff}= 47.5$kK and $\log g \sim 4.25$ \cite{1978A&A....70..653K}. Thus, with significantly hotter temperatures ($>$60-90kK) our Class 1 objects are distinct from typical subdwarfs. 

Alternatively, it has been argued that some He-sdO stars found in the galactic halo \cite{2020A&A...635A.193G} may not be core He-burning stars, but rather later evolutionary phases of lower mass sdB stars or WD merger products \cite{2016PASP..128h2001H}. While such objects may eventually reach hotter temperatures than observed typical subdwarfs, the lifetime of this phase should be short, and such objects should therefore be rare. 
To estimate possible contamination from such systems, we examine a control field at a similar galactic latitude to the Magellanic Clouds. We query the $\emph{Gaia}$ database for all stars located within a 10 degree radius of ($\alpha$,$\delta$) (J2000) $=$ (22:13:28.17, $-$63:06:55.92). We then restrict this sample to stars that have (i) 19.5 mag $<$ m$_G$ $<$ 16.5 mag (ii) no parallax detected at the 3$\sigma$ level, and (iii) proper motions consistent with the distributions found for the Magellanic Clouds. For point (iii), we calculate $\chi^{2}$ in the same manner described in \S~\ref{sec:kinematics} and require that either a star have $\chi^{2} < 11.6$ when compared to the distribution of likely LMC or SMC members or $\chi^{2} < 40$ and \textsc{astrometric\_excess\_noise\_sig} $>$ 4 when compared to the LMC (as was found for stars 5 and 6). In total, $\sim$78,000 stars in the control field pass these cuts. 

Plotting these objects in a \emph{Gaia} $m_G$ versus $BP-RP$ CMD, we find the stars in our spectroscopic sample would be among the bluest objects in the control field (while UV magnitudes are critical to distinguish candidate intermediate mass helium stars from the O/B MS, for low reddening values even optical photometry is sufficient to distinguish from the lower mass MS). In total, we identify only 10 stars in the control field that have no detected parallax, have proper motions similar to those observed for stars in the LMC/SMC, and have \emph{Gaia} $BP-RP$ colors as blue as our Class 1 and 2 objects for a given apparent magnitude. Given that this control field has an area $\gtrsim$25$\times$ larger than the LMC/SMC area covered by the UV data of the SUMaC survey and that this analysis does not assess whether these objects also have radial velocities consistent with the Magellanic Clouds, we predict $<$1 foreground object along the line-of-site to the Clouds with colors, magnitudes, and kinematics similar to that observed in our sample. It is unlikely our spectroscopic sample is dominated by foreground objects.

\subsection{Evolved lower-mass stars}

Low- to intermediate-mass ($\sim 1-8\Msun$) single stars that lose their hydrogen-rich envelopes during the asymptotic giant branch and are on the way to become white dwarfs (post-AGB stars) very briefly reach high temperatures and luminosities \cite{2016A&A...588A..25M}, consistent with what is expected for intermediate mass stripped stars. However, because this evolutionary phase is very short, these stars are expected to be rare. Using the MIST stellar evolution models for LMC metallicity, we find that the total time post-AGB stars spend in the region of the CMD overlapping with our spectroscopic sample (bluewards of the ZAMS and brighter than 19.5 AB mag at the distance of the Magellanic Clouds) ranges from $\sim$20--8500 years. We estimate the number of post-AGB stars that could contaminate our sample ($N_{\rm PAGB}$) based on the number of known AGB stars in the Clouds ($N_{\rm AGB}$), and the relative lifetimes of the AGB and post-AGB phases ($\tau_{\rm AGB}$ and $\tau_{\rm PAGB}$) as:

\begin{equation}\label{eq:HLO2}
   N_{\rm PAGB} = N_{\rm AGB} \int_{M_1}^{M_2} m^\Gamma \tau_{\rm PAGB}\, dm  \left( \int_{M_1}^{M_2} m^\Gamma \tau_{\rm AGB} \, dm \right)^{-1},
\end{equation}

\noindent where $\Gamma$ is the slope of the initial mass function, taken to be $-2.35$ \cite{Salpeter1955}. Taking the sample of AGB stars in the Clouds from \cite{Boyer2011}, and the lifetimes of the MIST stellar models in the AGB phase as described by \cite{Ogrady2020} we expect $\lesssim$1 post-AGB star in the region of the CMDs occupied by our sample.

In addition, because the hydrogen envelope was recently shed, post-AGB stars are expected to be closely surrounded by dusty circumstellar material \cite{2003ARA&A..41..391V}. The central stars have been observed to have very diverse spectral morphologies but, because the hottest phases are the briefest, most have cooled to B-type or later. Common features therefore include Balmer lines, which can appear in emission due to the expelled material (e.g., \cite{2006A&A...458..173S, 2014MNRAS.439.2211K}), and weak or absent \HeII\ lines \cite{2012A&A...543A..11M,2000A&AS..145..269P} (see Fig.~\ref{fig:comparison_spectra}). In contrast, none of the objects in our sample were identified as post-AGB candidates based on the infrared colors \cite{2020A&A...641A.142S}, none are emission-line stars, and all of the Class 1 and 2 objects show \HeII\ lines in their optical spectra. Thus, coupled with the expected rarity of this class, it is unlikely that post-AGB stars are a major contaminant.

\subsection{Accreting white dwarfs}
White dwarfs that accrete material from a hydrogen-rich companion star at high rates ($\gtrsim 10^{-7} \Msunyr$) undergo stable nuclear burning, which results in that they grow in mass, become hotter and more luminous \cite{2007ApJ...663.1269N}. With expected temperatures of $\sim 100,000-1,000,000$ K, these objects should appear blue-wards of the ZAMS (cf.\ \cite{2013ApJ...771...13L}). \cite{2007ApJ...663.1269N} predicted that these rapidly accreting white dwarfs may reach bolometric luminosities of $\log_{10} L/\Lsun \sim 3-5$, increasing with increasing temperature. Although they are bolometrically luminous, because they are so hot, only a small fraction of the light is emitted in optical and ultraviolet wavelengths. 
If the donor is helium-rich, the white dwarf can withstand higher accretion rates before puffing up to giant sizes (up to $5 \times 10^{-6} \Msunyr$, \cite{2019ApJ...878..100W}), but because of the lower energy generation rate from burning helium compared to hydrogen, similar bolometric luminosities may be expected.

The optical spectra of accreting white dwarfs are expected to show emission lines associated with the ongoing accretion and/or accretion disk. An example of such an object could be the emission-line star LMC~N66 \cite{2003A&A...409..969H}, which has strong emission lines of ionized helium and nitrogen. No star in our spectroscopic sample show such strong stellar emission lines or any signs of ongoing accretion.

\subsection{Rotational mixing and chemically homogeneous evolution}
Theoretical models predict that stellar rotation can give rise to interior mixing when the metallicity is low ($Z < Z_{\odot} / 2$), which, if the rotation rate is sufficiently high, can result in the evolution of chemically homogeneous stars \cite{1987A&A...178..159M}. Because CNO-processed material are efficiently brought to the surface and they slowly convert the hydrogen to helium, these stars should become hot, compact and helium-rich \cite{2009A&A...497..243D, 2011A&A...530A.115B}. However, rotational mixing is predicted to only be efficient for stars more massive than $\sim20$\Msun\ \cite{2011A&A...530A.115B}, which after central hydrogen burning results in similar-mass helium stars. These should be significantly brighter than the stars in our spectroscopic sample \cite{2015A&A...581A..15S}. 
Moreover, we note that none of the stars in the spectroscopic sample show signs of high rotation rates (cf.\ e.g., \cite{2013A&A...560A..29R}).

\section{Additional Figures and Tables}

\clearpage

\begin{figure}
\centering
\includegraphics[width=\textwidth]{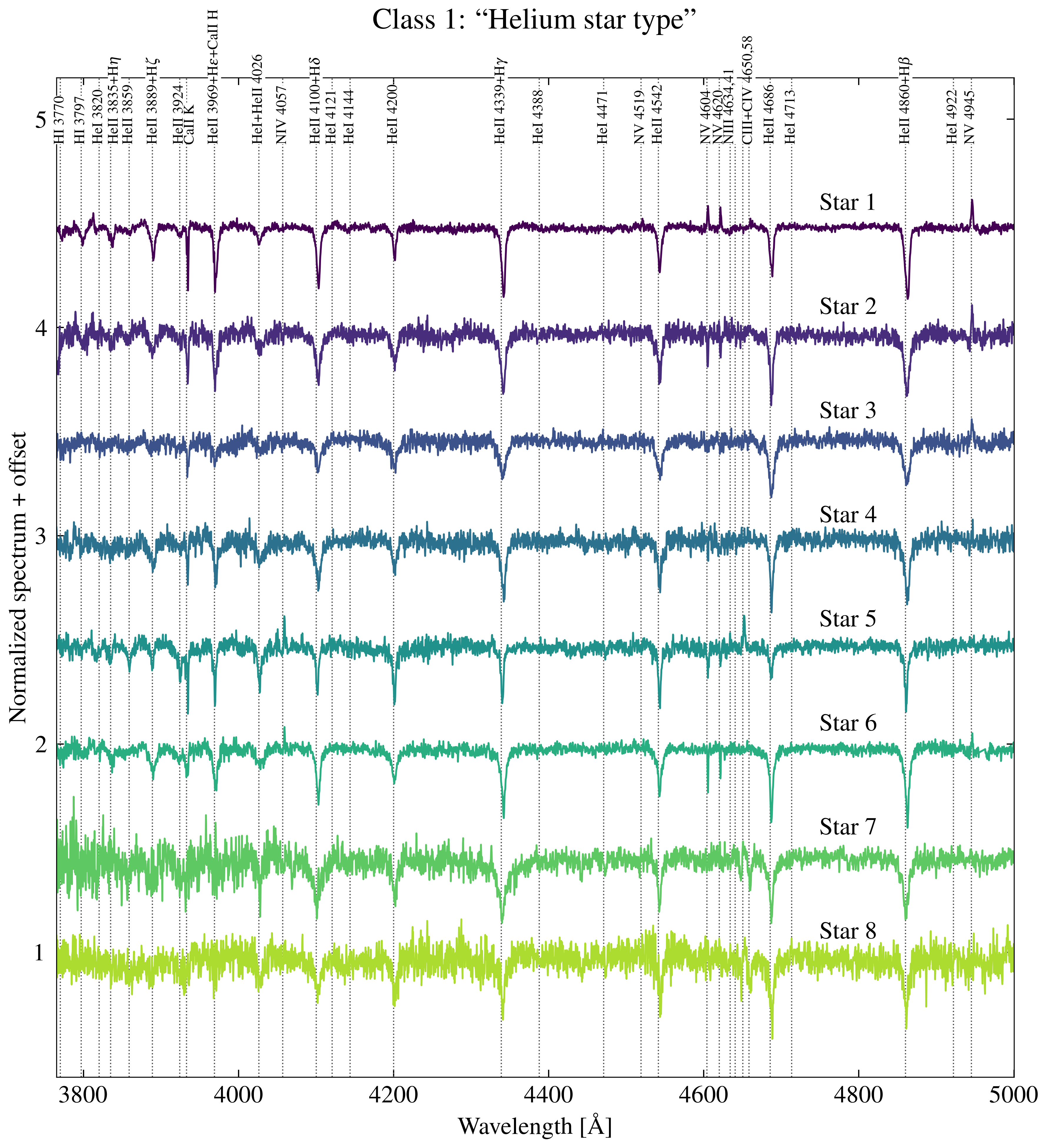}
\caption{The normalized optical spectra of the observed stripped stars classified as ``Helium-star type''. Sorted in luminosity and grouped in LMC and SMC targets. Sky lines and lines belonging to surrounding nebulae are shaded in gray.}
\label{fig:optical_spectra_isolated}
\end{figure}

\begin{figure}
\centering
\includegraphics[width=\textwidth]{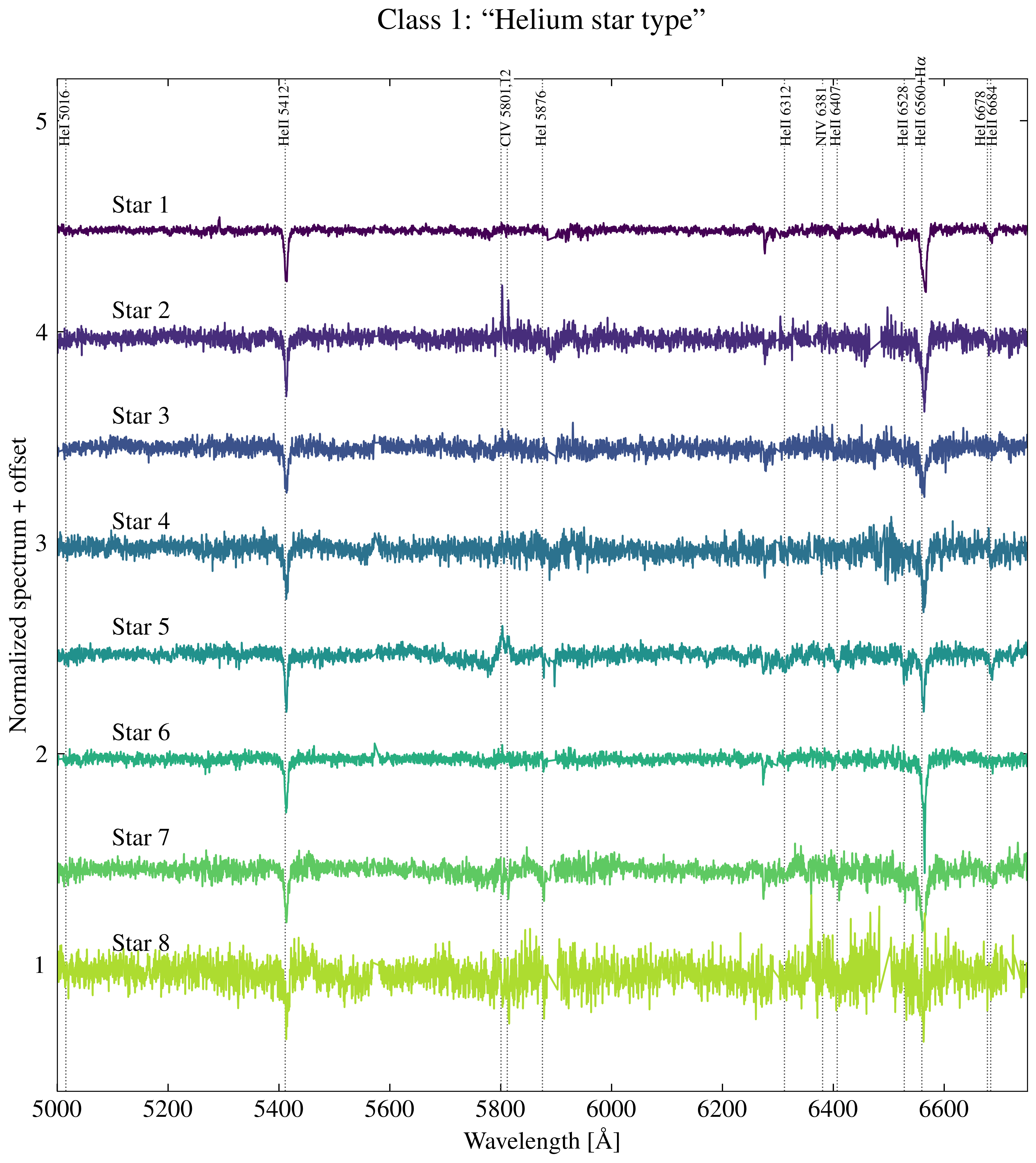}
\caption{Continuation of \figref{fig:optical_spectra_isolated}.}
\label{fig:optical_spectra_isolated2}
\end{figure}

\begin{figure}
\centering
\includegraphics[width=\textwidth]{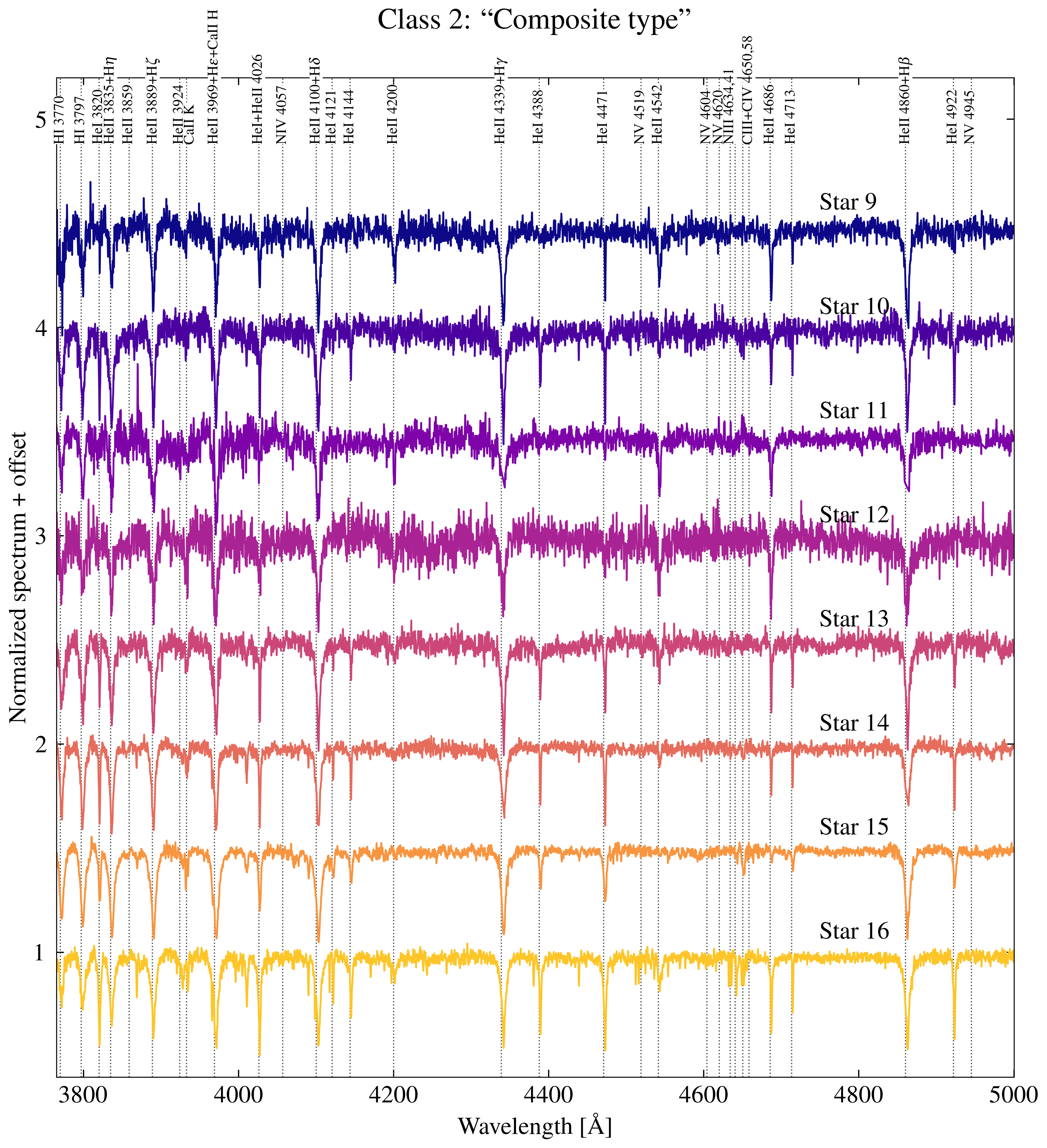}
\caption{Normalized optical spectra for the observed stars in the ``Composite'' group. Prominent short-wavelength Balmer lines, characteristic to B-type stars, are present in combination with pure \HeII\ lines, which are characteristic to a hotter component, such as a stripped star.}
\label{fig:optical_spectra_composite}
\end{figure}

\begin{figure}
\centering
\includegraphics[width=\textwidth]{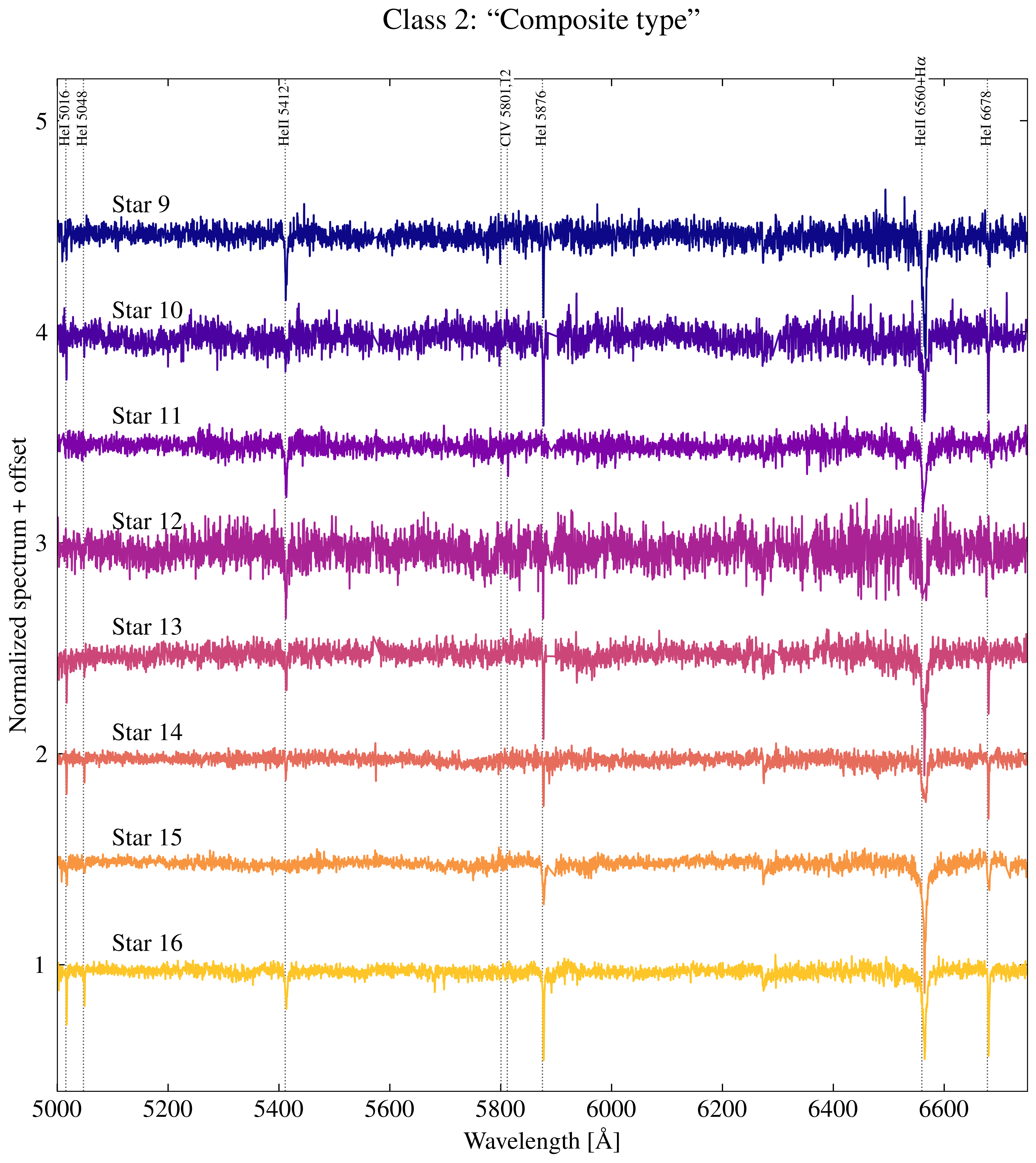}
\caption{Continuation of \figref{fig:optical_spectra_composite}.}
\label{fig:optical_spectra_composite2}
\end{figure}

\begin{figure}
\centering
\includegraphics[width=\textwidth]{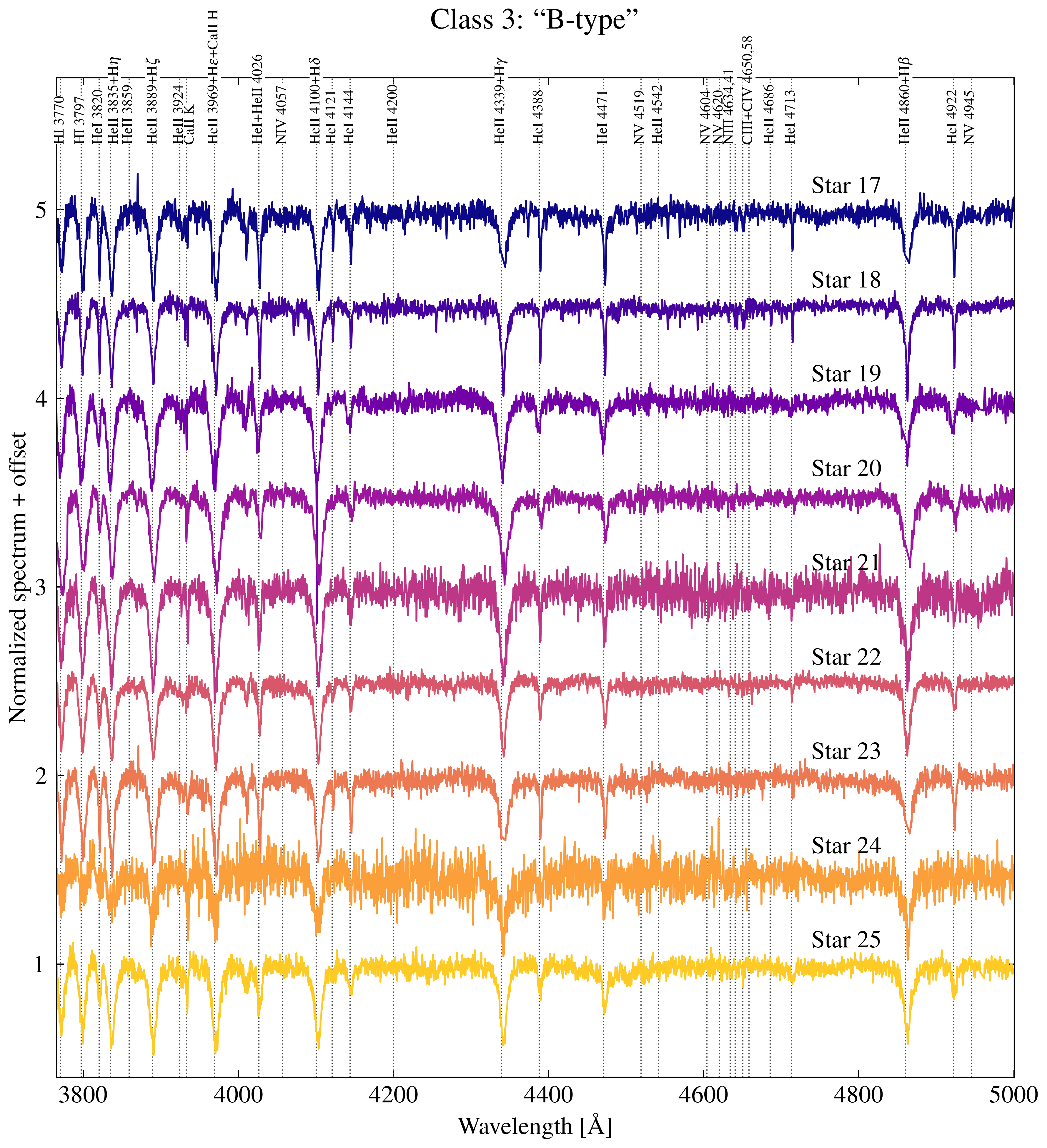}
\caption{Normalized optical spectra for the observed stars in the ``B-type'' group. Prominent short-wavelength Balmer lines, characteristic to B-type stars, are present. Although these stars show UV excess, no sign of the hot component is visible in the optical.}
\label{fig:optical_spectra_Btype}
\end{figure}

\begin{figure}
\centering
\includegraphics[width=\textwidth]{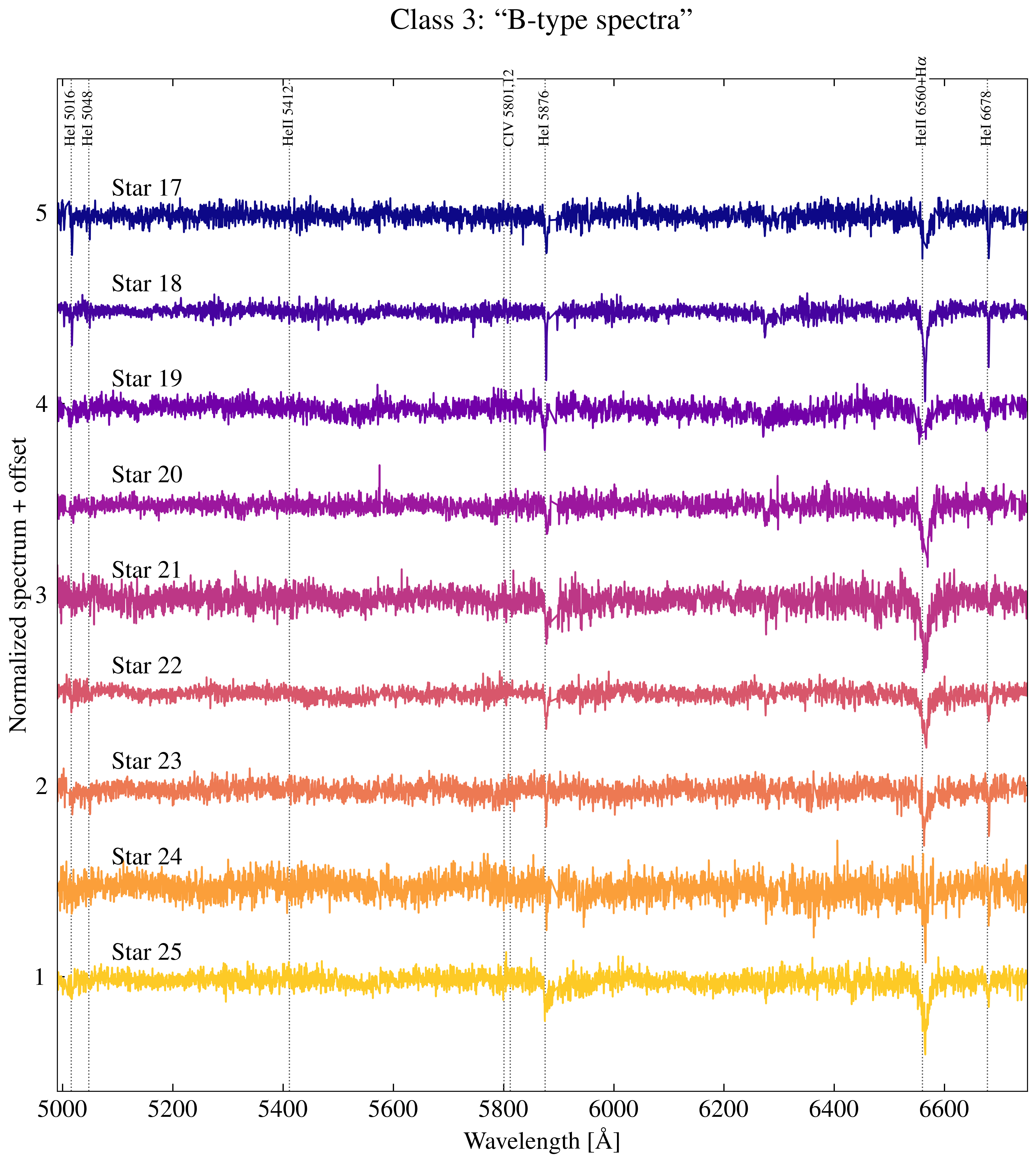}
\caption{Continuation of \figref{fig:optical_spectra_Btype}.}
\label{fig:optical_spectra_Btype2}
\end{figure}

\clearpage


\begin{table}
\begin{center}
\caption{Changes in the spectral models for stripped stars compared to the models presented in \cite{2018A&A...615A..78G}. From left to right we present the initial mass of the progenitor star, the mass of the resulting stripped star, the logarithm of the wind mass loss rate (previous, current), the radial extent the stellar atmosphere is computed for (previous, current), and the number of mesh points in the grid (previous, current).}
\label{tab:updated_stripped_star_models}
\input{Tables/table_updates.tex}
\newline 
{\textbf{Notes.} Subscript "p" refers to previous model parameters and subscript "c" refers to current model parameters.}
\end{center}
\end{table}

\begin{table}
\centering
\caption{Absolute magnitudes of the spectral models of stripped stars computed using binary evolutionary models (see \S~\ref{sec:MESAspec}). The first column gives the mass of the star in \Msun. Columns 2 to 8 give the absolute magnitudes of the models in the Swift filters and Generic Bessell filters in the AB magnitude system, while columns 9 to 15 give the corresponding values in the Vega magnitude system. }
\label{tab:absolute_magnitudes_stripped_stars}
{\small
\input{Tables/table_absmag_stripped}
}
\end{table}

\begin{table}
\centering
\caption{Parameters assumed for the CMFGEN models of main sequence O and B type stars. From left to right we display the initial mass, the current mass, the effective temperature, the bolometric luminosity, the stellar radius, the surface gravity, the surface mass fraction of hydrogen and helium, the wind mass loss rate, the terminal wind speed, the extent of the stellar atmosphere in units of stellar radii assumed for atmosphere calculation, and the number of mesh points. We compute the spectra for the stars at 20\% (top), 60\% (middle) and 90\% (bottom) of the duration through the main-sequence evolution. }
\label{tab:OB_properties}
{\small
\input{Tables/table_prop_allOB}
}
\end{table}

\begin{table}
\centering
\caption{Absolute magnitudes of the spectral models of OB stars computed using binary evolutionary models that are (from top to bottom) 20\%, 60\%, and 90\% through their main-sequence duration (see \S~\ref{sec:MESAMSspec}). The first column gives the mass of the star. Columns 2 to 8 give the absolute magnitudes of the models in the Swift filters and Generic Bessell filters in the AB magnitude system, while columns 9 to 15 give the corresponding values in the Vega magnitude system.}
\label{tab:absolute_magnitudes_OB}
{\small
\include{Tables/table_absmag_allOB}
}
\end{table}


\begin{landscape}
\begin{table}
\centering
\caption{Photometry for the spectroscopic sample (AB magnitudes). Coordinates are given in degrees. $UVBI-$magnitudes are from the MCPS and have simply been converted from Vega to AB magnitudes. $UVW2,UVM2,UVW1-$magnitudes are from this work. For reference, we also provide the number of stars located within a 5 arcsecond radius of each target (n5), the closest star to each target in arcseconds (closest), and the fraction of flux within a 5 arcsecond radius that \emph{Tractor} assigned to each target in all three UV bands (W2-FF, M2-FF, and W1-FF)}
\label{tab:photom}
{\small
\input{Tables/Photometry_V3}
}
\end{table}
\end{landscape}

\begin{table}
\centering
\caption{Equivalent width measurements in \AA ngstr\"{o}ms for the stars in the spectroscopic sample. In cases where a line is not detected, we provide a range of values that encompasses three sigma upper limits for both absorption lines (positive values) and emission lines (negative values).}
\label{tab:ew}
{\small
\input{Tables/Equivalent_widths}
}
\end{table}

\begin{table}
\centering
\caption{Kinematic properties for the stars in the spectroscopic sample. From left to right, we display: the star number, the number of spectra obtained, the measured average radial velocity together with maximum and minimum measured radial velocity, the parallax with associated errors, the proper motion in right ascension and declination with associated errors, the $\chi^2$ values for the 2D proper motion comparison and the $\chi^2$ values for the 3D motion comparison (including radial velocity). Parallax and proper motion values are from \emph{Gaia}DR3 and listed for convenience.}
\label{tab:kinematics}
{\small
\input{Tables/Kinematics_V3}
}
\end{table}

\end{document}

%% file: Tables/table_updates.tex
\begin{tabular}{cccccccccccc} 
\toprule\midrule 
$M_{\mathrm{init}}$ & $M_{\mathrm{strip}}$ & $\log_{10} L$ & $T_{\rm{eff}}$ & $\log_{10} \dot{M}_{\mathrm{wind, p}}$ & $\log_{10} \dot{M}_{\mathrm{wind, c}}$ & RMAX$_{\mathrm{p}}$ & RMAX$_{\mathrm{c}}$ & ND$_{\mathrm{p}}$ & ND$_{\mathrm{c}}$ \\ 
 
 [$M_{\odot}$] & [$M_{\odot}$] & [$L_{\odot}$] & [kK] & [$M_{\odot}$ yr$^{-1}$] & [$M_{\odot}$ yr$^{-1}$] & [$R_{\star}$] & [$R_{\star}$] &  &  \\ 
\midrule 
6.66 & 1.77 & 3.3 & 54.0 & -8.14 & -10.0 & 70 & 60 & 40 & 100\\ 
7.37 & 2.04 & 3.5 & 57.0 & -7.84 & -8.84 & 101 & 70 & 40 & 80\\ 
8.15 & 2.34 & 3.7 & 60.4 & -7.6 & -9.0 & 101 & 70 & 40 & 80\\ 
9.00 & 2.69 & 3.9 & 63.7 & -7.38 & -9.0 & 101 & 35 & 40 & 80\\ 
9.96 & 3.10 & 4.0 & 67.5 & -7.16 & -9.0 & 101 & 71 & 40 & 90\\ 
11.01 & 3.58 & 4.2 & 71.2 & -6.94 & -9.0 & 506 & 25 & 40 & 80\\ 
12.17 & 4.13 & 4.4 & 75.1 & -6.73 & -9.0 & 507 & 50 & 40 & 80\\ 
13.45 & 4.75 & 4.5 & 78.8 & -6.53 & -8.0 & 507 & 80 & 40 & 80\\ 
14.87 & 5.47 & 4.7 & 82.7 & -6.33 & -8.0 & 507 & 516 & 40 & 130\\ 
16.44 & 6.28 & 4.8 & 86.7 & -6.15 & -8.0 & 507 & 101 & 40 & 140\\ 
18.17 & 7.14 & 4.9 & 84.8 & -6.03 & -8.0 & 509 & 130 & 80 & 80\\ 
\bottomrule
\end{tabular}

%% file: Tables/table_absmag_stripped.tex
\begin{tabular}{|l|ccccccc|ccccccc|}
 \toprule\midrule 
 $M_{\mathrm{strip}}$ & \multicolumn{7}{c}{AB magnitude system} & \multicolumn{7}{c}{Vega magnitude system}\\ 
  {[$M_{\odot}$]} & UVW2 & UVM2 & UVW1 & U & B & V & I & UVW2 & UVM2 & UVW1 & U & B & V & I\\ 
\midrule 
 0.37 & 4.45 & 4.55 & 4.63 & 4.94 & 4.86 & 5.15 & 5.80 & 2.75 & 2.86 & 3.21 & 4.18 & 4.95 & 5.14 & 5.36\\ 
0.42 & 3.75 & 3.88 & 3.99 & 4.42 & 4.44 & 4.76 & 5.45 & 2.05 & 2.20 & 2.57 & 3.66 & 4.52 & 4.75 & 5.00\\ 
0.47 & 3.18 & 3.33 & 3.45 & 3.95 & 4.03 & 4.38 & 5.09 & 1.48 & 1.65 & 2.03 & 3.19 & 4.12 & 4.37 & 4.64\\ 
0.54 & 2.62 & 2.78 & 2.92 & 3.47 & 3.61 & 3.98 & 4.71 & 0.92 & 1.10 & 1.50 & 2.71 & 3.70 & 3.97 & 4.26\\ 
0.62 & 2.06 & 2.23 & 2.37 & 2.96 & 3.15 & 3.54 & 4.28 & 0.36 & 0.55 & 0.95 & 2.20 & 3.24 & 3.52 & 3.83\\ 
0.70 & 1.63 & 1.80 & 1.94 & 2.55 & 2.78 & 3.18 & 3.92 & -0.07 & 0.11 & 0.52 & 1.79 & 2.86 & 3.16 & 3.47\\ 
0.80 & 1.21 & 1.36 & 1.50 & 2.13 & 2.42 & 2.82 & 3.58 & -0.49 & -0.32 & 0.08 & 1.37 & 2.50 & 2.81 & 3.14\\ 
0.92 & 0.83 & 0.98 & 1.12 & 1.75 & 2.05 & 2.46 & 3.22 & -0.86 & -0.70 & -0.30 & 0.99 & 2.14 & 2.45 & 2.78\\ 
1.05 & 0.50 & 0.66 & 0.79 & 1.44 & 1.74 & 2.15 & 2.91 & -1.20 & -1.03 & -0.63 & 0.68 & 1.82 & 2.14 & 2.47\\ 
1.20 & 0.15 & 0.31 & 0.45 & 1.09 & 1.41 & 1.83 & 2.59 & -1.54 & -1.37 & -0.97 & 0.33 & 1.49 & 1.81 & 2.14\\ 
1.38 & -0.15 & 0.01 & 0.14 & 0.79 & 1.12 & 1.54 & 2.30 & -1.85 & -1.68 & -1.28 & 0.03 & 1.20 & 1.52 & 1.85\\ 
1.54 & -0.34 & -0.17 & -0.04 & 0.61 & 0.95 & 1.38 & 2.14 & -2.04 & -1.86 & -1.46 & -0.15 & 1.03 & 1.36 & 1.69\\ 
1.77 & -0.66 & -0.49 & -0.35 & 0.31 & 0.65 & 1.08 & 1.85 & -2.35 & -2.17 & -1.77 & -0.45 & 0.74 & 1.07 & 1.40\\ 
2.04 & -0.96 & -0.79 & -0.65 & 0.02 & 0.36 & 0.80 & 1.57 & -2.66 & -2.47 & -2.07 & -0.74 & 0.45 & 0.78 & 1.12\\ 
2.34 & -1.24 & -1.06 & -0.92 & -0.24 & 0.11 & 0.55 & 1.32 & -2.94 & -2.74 & -2.34 & -1.00 & 0.19 & 0.53 & 0.87\\ 
2.69 & -1.53 & -1.35 & -1.21 & -0.52 & -0.17 & 0.27 & 1.05 & -3.23 & -3.04 & -2.63 & -1.28 & -0.08 & 0.26 & 0.60\\ 
3.10 & -1.79 & -1.61 & -1.46 & -0.76 & -0.40 & 0.04 & 0.82 & -3.49 & -3.29 & -2.88 & -1.52 & -0.32 & 0.02 & 0.37\\ 
3.58 & -2.07 & -1.87 & -1.73 & -1.02 & -0.66 & -0.21 & 0.57 & -3.76 & -3.56 & -3.15 & -1.78 & -0.57 & -0.23 & 0.12\\ 
4.13 & -2.32 & -2.12 & -1.97 & -1.26 & -0.89 & -0.44 & 0.34 & -4.02 & -3.81 & -3.39 & -2.02 & -0.81 & -0.46 & -0.11\\ 
4.75 & -2.57 & -2.37 & -2.22 & -1.49 & -1.12 & -0.67 & 0.11 & -4.26 & -4.05 & -3.64 & -2.25 & -1.04 & -0.69 & -0.33\\ 
5.47 & -2.79 & -2.58 & -2.43 & -1.70 & -1.33 & -0.88 & -0.09 & -4.49 & -4.27 & -3.85 & -2.46 & -1.25 & -0.89 & -0.53\\ 
6.28 & -3.00 & -2.79 & -2.64 & -1.90 & -1.52 & -1.07 & -0.28 & -4.70 & -4.47 & -4.06 & -2.66 & -1.44 & -1.08 & -0.72\\ 
7.14 & -3.37 & -3.16 & -3.01 & -2.27 & -1.90 & -1.44 & -0.65 & -5.06 & -4.84 & -4.43 & -3.03 & -1.81 & -1.46 & -1.10\\ 
\bottomrule 
\end{tabular}
 

%% file: Tables/table_prop_allOB.tex
\begin{tabular}{cccccccccccc}
 \toprule\midrule 
 $M_{\rm init}$ & $M_{\rm current}$ & $T_{\rm eff}$ & $L_{\rm bol}$ & $R_{\rm eff}$ & $\log _{10} g_{\rm eff}$ & $X_{\rm H, surf}$ & $X_{\rm He, surf}$ & $\dot{M}_{w}$ & $v_{\infty}$ & RMAX & ND\\ 
 \midrule 
\multicolumn{10}{l}{\textbf{20\% through the main-sequence evolution}}\\
\midrule
 2.21 & 2.20 &   11210 &      35 & 1.59 & 4.38 & 0.74 & 0.25 & 1.00e-10 & 1895 &   50 &   66\\ 
2.44 & 2.44 &   11990 &      52 & 1.68 & 4.37 & 0.74 & 0.25 & 1.00e-10 & 1937 &  131 &   66\\ 
2.70 & 2.69 &   12820 &      76 & 1.77 & 4.37 & 0.74 & 0.25 & 1.00e-10 & 1984 &  100 &   66\\ 
2.99 & 2.98 &   13670 &     110 & 1.87 & 4.37 & 0.74 & 0.25 & 2.43e-11 & 2029 &  131 &   66\\ 
3.30 & 3.29 &   14550 &     159 & 1.98 & 4.36 & 0.74 & 0.25 & 2.43e-11 & 2073 &  131 &   66\\ 
3.65 & 3.64 &   15460 &     228 & 2.10 & 4.35 & 0.85 & 0.15 & 2.43e-11 & 2116 &  131 &   66\\ 
4.04 & 4.02 &   16410 &     325 & 2.23 & 4.34 & 0.83 & 0.17 & 1.14e-10 & 2161 &  131 &   66\\ 
4.46 & 4.45 &   17400 &     460 & 2.36 & 4.34 & 0.84 & 0.15 & 2.54e-11 & 2209 &  131 &   66\\ 
4.93 & 4.91 &   18410 &     655 & 2.52 & 4.33 & 0.82 & 0.18 & 4.60e-11 & 2250 &  131 &   66\\ 
5.45 & 5.43 &   19470 &     922 & 2.67 & 4.32 & 0.80 & 0.20 & 3.67e-11 & 2297 &  130 &   66\\ 
6.03 & 6.00 &   20560 &    1288 & 2.83 & 4.31 & 0.74 & 0.25 & 1.14e-10 & 2346 &  130 &   66\\ 
6.66 & 6.64 &   21680 &    1797 & 3.00 & 4.30 & 0.74 & 0.25 & 2.21e-10 & 2393 &  130 &   66\\ 
7.37 & 7.34 &   22830 &    2497 & 3.19 & 4.29 & 0.74 & 0.25 & 4.23e-10 & 2440 &  130 &   66\\ 
8.15 & 8.11 &   24010 &    3447 & 3.39 & 4.29 & 0.74 & 0.25 & 7.98e-10 & 2489 &  130 &   66\\ 
9.00 & 8.97 &   25240 &    4724 & 3.59 & 4.28 & 0.74 & 0.25 & 2.10e-10 & 2544 &   66 &   66\\ 
9.96 & 9.92 &   26470 &    6479 & 3.82 & 4.27 & 0.74 & 0.25 & 2.10e-10 & 2592 &  130 &   66\\ 
11.01 & 10.96 &   27740 &    8785 & 4.05 & 4.26 & 0.74 & 0.25 & 4.46e-10 & 2647 &  130 &   66\\ 
12.17 & 12.12 &   29080 &   11870 & 4.29 & 4.26 & 0.74 & 0.25 & 9.17e-10 & 2701 &  128 &   66\\ 
13.45 & 13.39 &   30390 &   15960 & 4.55 & 4.25 & 0.74 & 0.25 & 1.82e-09 & 2756 &  118 &   66\\ 
14.87 & 14.80 &   31710 &   21270 & 4.83 & 4.24 & 0.74 & 0.25 & 3.45e-09 & 2814 &  221 &   65\\ 
16.44 & 16.36 &   33040 &   28180 & 5.12 & 4.23 & 0.74 & 0.25 & 6.34e-09 & 2873 &  114 &   90\\ 
18.17 & 18.08 &   34360 &   37040 & 5.43 & 4.23 & 0.74 & 0.25 & 1.12e-08 & 2934 &  101 &   65\\ 
\midrule
\multicolumn{10}{l}{\textbf{60\% through the main-sequence evolution}}\\
\midrule
 2.44 & 2.44 &   11270 &      71 & 2.22 & 4.13 & 0.74 & 0.25 & 1.00e-10 & 1686 &   50 &   66\\ 
2.70 & 2.69 &   12110 &     104 & 2.32 & 4.14 & 0.90 & 0.10 & 1.00e-10 & 1734 &   51 &   86\\ 
2.99 & 2.98 &   12950 &     151 & 2.45 & 4.13 & 0.86 & 0.14 & 1.00e-10 & 1776 &   51 &   86\\ 
3.30 & 3.29 &   13820 &     219 & 2.58 & 4.13 & 0.85 & 0.14 & 1.56e-12 & 1817 &   51 &   66\\ 
3.65 & 3.64 &   14740 &     315 & 2.72 & 4.13 & 0.83 & 0.16 & 3.30e-12 & 1861 &   25 &   66\\ 
4.04 & 4.02 &   15660 &     454 & 2.89 & 4.12 & 0.82 & 0.18 & 1.00e-10 & 1898 &   51 &   66\\ 
4.46 & 4.44 &   16660 &     643 & 3.04 & 4.12 & 0.80 & 0.19 & 1.95e-11 & 1946 &   51 &   66\\ 
4.93 & 4.91 &   17670 &     911 & 3.22 & 4.11 & 0.79 & 0.20 & 4.25e-11 & 1989 &   51 &   66\\ 
5.45 & 5.43 &   18700 &    1289 & 3.42 & 4.11 & 0.78 & 0.22 & 9.21e-11 & 2029 &   51 &   66\\ 
6.03 & 6.00 &   19760 &    1807 & 3.62 & 4.10 & 0.74 & 0.25 & 2.32e-10 & 2072 &   51 &   66\\ 
6.66 & 6.63 &   20850 &    2527 & 3.85 & 4.09 & 0.74 & 0.25 & 4.51e-10 & 2113 &   51 &   66\\ 
7.37 & 7.33 &   21970 &    3512 & 4.09 & 4.08 & 0.74 & 0.25 & 8.65e-10 & 2155 &   51 &   66\\ 
8.15 & 8.12 &   21830 &    4851 & 4.87 & 3.97 & 0.74 & 0.25 & 1.64e-09 & 2198 &   51 &   66\\ 
9.00 & 8.96 &   24310 &    6717 & 4.62 & 4.06 & 0.74 & 0.25 & 3.10e-09 & 2243 &   51 &   66\\ 
9.96 & 9.91 &   25510 &    9236 & 4.92 & 4.05 & 0.74 & 0.25 & 3.81e-10 & 2287 &   51 &   66\\ 
11.01 & 10.96 &   26720 &   12560 & 5.22 & 4.04 & 0.74 & 0.25 & 8.27e-10 & 2333 &   51 &   66\\ 
12.17 & 12.11 &   27960 &   16940 & 5.54 & 4.03 & 0.74 & 0.25 & 1.72e-09 & 2381 &   51 &   66\\ 
\midrule
\multicolumn{10}{l}{\textbf{90\% through the main-sequence evolution}}\\
\midrule
 2.99 & 2.98 &   11100 &     215 & 3.96 & 3.71 & 0.74 & 0.25 & 1.00e-10 & 1395 &   51 &   66\\ 
3.30 & 3.29 &   11890 &     315 & 4.18 & 3.71 & 0.79 & 0.21 & 4.28e-12 & 1429 &   25 &   66\\ 
3.65 & 3.64 &   12730 &     459 & 4.40 & 3.71 & 0.78 & 0.22 & 1.03e-11 & 1464 &   51 &   66\\ 
4.04 & 4.02 &   13610 &     662 & 4.62 & 3.71 & 0.77 & 0.23 & 2.27e-11 & 1502 &   51 &   66\\ 
4.46 & 4.44 &   14530 &     949 & 4.86 & 3.71 & 0.76 & 0.23 & 1.00e-10 & 1541 &   51 &   66\\ 
4.93 & 4.91 &   15410 &    1361 & 5.17 & 3.70 & 0.76 & 0.24 & 1.11e-10 & 1570 &   51 &   66\\ 
5.45 & 5.43 &   16380 &    1924 & 5.44 & 3.70 & 0.75 & 0.24 & 2.30e-10 & 1609 &   51 &   66\\ 
6.03 & 5.99 &   17320 &    2716 & 5.79 & 3.69 & 0.74 & 0.25 & 4.97e-10 & 1640 &   51 &   66\\ 
6.66 & 6.62 &   18310 &    3801 & 6.12 & 3.69 & 0.74 & 0.25 & 9.71e-10 & 1676 &   51 &   66\\ 
7.37 & 7.31 &   19270 &    5311 & 6.54 & 3.67 & 0.74 & 0.25 & 1.88e-09 & 1704 &   51 &   66\\ 
8.15 & 8.09 &   19090 &    7330 & 7.82 & 3.56 & 0.74 & 0.25 & 3.55e-09 & 1738 &   51 &   66\\ 
\bottomrule 
\end{tabular}

%% file: Tables/table_absmag_allOB.tex
\begin{tabular}{|l|ccccccc|ccccccc|}
 \toprule\midrule 
 $M_{\mathrm{init}}$ & \multicolumn{7}{c}{AB magnitude system} & \multicolumn{7}{c}{Vega magnitude system}\\

{\footnotesize [$M_{\odot}$]} &  UVW2 & UVM2 & UVW1 & U & B & V & I & UVW2 & UVM2 & UVW1 & U & B & V & I\\ 
 
\midrule 
\multicolumn{10}{l}{\textbf{20\% through the main-sequence evolution}}\\
\midrule
 2.21 & 2.64 & 2.59 & 2.50 & 2.07 & 1.32 & 1.47 & 1.96 & 0.95 & 0.90 & 1.08 & 1.31 & 1.40 & 1.45 & 1.51\\ 
2.44 & 2.16 & 2.12 & 2.06 & 1.72 & 1.05 & 1.22 & 1.74 & 0.46 & 0.44 & 0.64 & 0.96 & 1.13 & 1.21 & 1.29\\ 
2.70 & 1.68 & 1.65 & 1.61 & 1.36 & 0.78 & 0.98 & 1.51 & -0.02 & -0.04 & 0.19 & 0.60 & 0.87 & 0.96 & 1.06\\ 
2.99 & 1.25 & 1.23 & 1.21 & 1.05 & 0.53 & 0.75 & 1.30 & -0.45 & -0.45 & -0.21 & 0.29 & 0.62 & 0.73 & 0.85\\ 
3.30 & 0.83 & 0.82 & 0.82 & 0.73 & 0.28 & 0.51 & 1.07 & -0.87 & -0.86 & -0.60 & -0.03 & 0.36 & 0.49 & 0.63\\ 
3.65 & 0.42 & 0.44 & 0.45 & 0.42 & 0.04 & 0.28 & 0.86 & -1.28 & -1.25 & -0.97 & -0.34 & 0.12 & 0.27 & 0.41\\ 
4.04 & 0.01 & 0.04 & 0.06 & 0.10 & -0.22 & 0.04 & 0.63 & -1.68 & -1.64 & -1.36 & -0.66 & -0.13 & 0.03 & 0.19\\ 
4.46 & -0.35 & -0.31 & -0.27 & -0.17 & -0.45 & -0.17 & 0.43 & -2.05 & -1.99 & -1.69 & -0.93 & -0.36 & -0.19 & -0.01\\ 
4.93 & -0.75 & -0.70 & -0.65 & -0.49 & -0.71 & -0.43 & 0.20 & -2.44 & -2.38 & -2.07 & -1.25 & -0.63 & -0.44 & -0.25\\ 
5.45 & -1.11 & -1.06 & -1.00 & -0.80 & -0.96 & -0.66 & -0.03 & -2.81 & -2.74 & -2.43 & -1.56 & -0.88 & -0.68 & -0.48\\ 
6.03 & -1.47 & -1.44 & -1.40 & -1.15 & -1.25 & -0.93 & -0.28 & -3.17 & -3.13 & -2.82 & -1.91 & -1.17 & -0.95 & -0.72\\ 
6.66 & -1.83 & -1.80 & -1.75 & -1.46 & -1.51 & -1.18 & -0.51 & -3.53 & -3.49 & -3.17 & -2.22 & -1.43 & -1.20 & -0.96\\ 
7.37 & -2.18 & -2.15 & -2.10 & -1.77 & -1.77 & -1.43 & -0.75 & -3.88 & -3.84 & -3.52 & -2.53 & -1.69 & -1.45 & -1.20\\ 
8.15 & -2.52 & -2.50 & -2.44 & -2.08 & -2.03 & -1.68 & -0.99 & -4.22 & -4.18 & -3.86 & -2.84 & -1.95 & -1.70 & -1.44\\ 
9.00 & -2.85 & -2.82 & -2.76 & -2.37 & -2.28 & -1.92 & -1.22 & -4.55 & -4.51 & -4.18 & -3.13 & -2.19 & -1.94 & -1.67\\ 
9.96 & -3.17 & -3.13 & -3.05 & -2.62 & -2.51 & -2.15 & -1.44 & -4.87 & -4.81 & -4.47 & -3.38 & -2.42 & -2.16 & -1.89\\ 
11.01 & -3.49 & -3.44 & -3.35 & -2.88 & -2.74 & -2.37 & -1.65 & -5.19 & -5.12 & -4.77 & -3.64 & -2.65 & -2.38 & -2.10\\ 
12.17 & -3.81 & -3.75 & -3.64 & -3.13 & -2.96 & -2.58 & -1.86 & -5.51 & -5.43 & -5.06 & -3.89 & -2.88 & -2.60 & -2.30\\ 
13.45 & -4.13 & -4.05 & -3.93 & -3.39 & -3.20 & -2.81 & -2.07 & -5.82 & -5.74 & -5.35 & -4.15 & -3.12 & -2.83 & -2.52\\ 
14.87 & -4.40 & -4.33 & -4.19 & -3.63 & -3.42 & -3.03 & -2.28 & -6.10 & -6.01 & -5.61 & -4.39 & -3.34 & -3.04 & -2.73\\ 
16.44 & -4.68 & -4.60 & -4.46 & -3.89 & -3.66 & -3.26 & -2.51 & -6.38 & -6.29 & -5.88 & -4.65 & -3.58 & -3.28 & -2.96\\ 
18.17 & -4.92 & -4.83 & -4.69 & -4.11 & -3.87 & -3.46 & -2.71 & -6.62 & -6.52 & -6.11 & -4.87 & -3.78 & -3.48 & -3.16\\ 
\midrule
\multicolumn{10}{l}{\textbf{60\% through the main-sequence evolution}}\\
\midrule
 2.44 & 1.90 & 1.84 & 1.75 & 1.31 & 0.57 & 0.73 & 1.23 & 0.20 & 0.16 & 0.33 & 0.55 & 0.65 & 0.72 & 0.78\\ 
2.70 & 1.37 & 1.33 & 1.27 & 0.93 & 0.30 & 0.49 & 1.00 & -0.33 & -0.35 & -0.15 & 0.17 & 0.39 & 0.48 & 0.56\\ 
2.99 & 0.89 & 0.86 & 0.83 & 0.58 & 0.04 & 0.25 & 0.78 & -0.81 & -0.82 & -0.59 & -0.18 & 0.13 & 0.23 & 0.33\\ 
3.30 & 0.48 & 0.47 & 0.45 & 0.29 & -0.20 & 0.03 & 0.57 & -1.22 & -1.21 & -0.97 & -0.47 & -0.11 & 0.01 & 0.13\\ 
3.65 & 0.03 & 0.03 & 0.03 & -0.05 & -0.46 & -0.22 & 0.35 & -1.66 & -1.65 & -1.39 & -0.81 & -0.37 & -0.23 & -0.10\\ 
4.04 & -0.37 & -0.36 & -0.35 & -0.36 & -0.71 & -0.45 & 0.12 & -2.07 & -2.04 & -1.77 & -1.12 & -0.62 & -0.47 & -0.32\\ 
4.46 & -0.74 & -0.72 & -0.70 & -0.65 & -0.94 & -0.67 & -0.08 & -2.44 & -2.41 & -2.12 & -1.41 & -0.86 & -0.69 & -0.52\\ 
4.93 & -1.13 & -1.10 & -1.07 & -0.96 & -1.20 & -0.91 & -0.30 & -2.82 & -2.79 & -2.49 & -1.72 & -1.11 & -0.93 & -0.75\\ 
5.45 & -1.51 & -1.49 & -1.45 & -1.28 & -1.46 & -1.16 & -0.53 & -3.21 & -3.17 & -2.87 & -2.04 & -1.38 & -1.18 & -0.98\\ 
6.03 & -1.87 & -1.86 & -1.82 & -1.61 & -1.73 & -1.42 & -0.77 & -3.57 & -3.54 & -3.24 & -2.37 & -1.65 & -1.43 & -1.22\\ 
6.66 & -2.24 & -2.22 & -2.18 & -1.92 & -1.99 & -1.67 & -1.01 & -3.93 & -3.91 & -3.60 & -2.68 & -1.91 & -1.68 & -1.45\\ 
7.37 & -2.59 & -2.57 & -2.53 & -2.23 & -2.25 & -1.92 & -1.24 & -4.28 & -4.26 & -3.95 & -2.99 & -2.17 & -1.93 & -1.69\\ 
8.15 & -2.94 & -2.93 & -2.88 & -2.59 & -2.62 & -2.28 & -1.61 & -4.64 & -4.61 & -4.30 & -3.35 & -2.53 & -2.30 & -2.06\\ 
9.00 & -3.27 & -3.25 & -3.20 & -2.84 & -2.76 & -2.41 & -1.72 & -4.97 & -4.94 & -4.62 & -3.60 & -2.68 & -2.43 & -2.17\\ 
9.96 & -3.59 & -3.57 & -3.50 & -3.10 & -3.00 & -2.64 & -1.94 & -5.29 & -5.25 & -4.92 & -3.86 & -2.91 & -2.66 & -2.39\\ 
11.01 & -3.92 & -3.88 & -3.80 & -3.36 & -3.23 & -2.87 & -2.16 & -5.62 & -5.57 & -5.22 & -4.12 & -3.14 & -2.88 & -2.61\\ 
12.17 & -4.24 & -4.19 & -4.09 & -3.61 & -3.46 & -3.08 & -2.36 & -5.93 & -5.87 & -5.51 & -4.37 & -3.37 & -3.10 & -2.81\\ 
\midrule
\multicolumn{10}{l}{\textbf{90\% through the main-sequence evolution}}\\
\midrule
 2.99 & 0.74 & 0.67 & 0.58 & 0.07 & -0.67 & -0.49 & -0.00 & -0.96 & -1.01 & -0.84 & -0.69 & -0.58 & -0.51 & -0.45\\ 
3.30 & 0.20 & 0.15 & 0.08 & -0.31 & -0.95 & -0.75 & -0.24 & -1.50 & -1.53 & -1.34 & -1.07 & -0.86 & -0.76 & -0.69\\ 
3.65 & -0.27 & -0.31 & -0.35 & -0.64 & -1.20 & -0.98 & -0.46 & -1.97 & -1.99 & -1.77 & -1.40 & -1.12 & -1.00 & -0.91\\ 
4.04 & -0.73 & -0.75 & -0.78 & -0.98 & -1.45 & -1.22 & -0.68 & -2.42 & -2.43 & -2.20 & -1.74 & -1.37 & -1.24 & -1.12\\ 
4.46 & -1.15 & -1.17 & -1.18 & -1.30 & -1.70 & -1.45 & -0.89 & -2.85 & -2.85 & -2.60 & -2.06 & -1.62 & -1.47 & -1.34\\ 
4.93 & -1.55 & -1.57 & -1.57 & -1.62 & -1.96 & -1.70 & -1.12 & -3.25 & -3.25 & -2.99 & -2.38 & -1.88 & -1.71 & -1.57\\ 
5.45 & -1.94 & -1.95 & -1.94 & -1.92 & -2.21 & -1.93 & -1.33 & -3.64 & -3.63 & -3.36 & -2.68 & -2.12 & -1.94 & -1.78\\ 
6.03 & -2.33 & -2.34 & -2.32 & -2.24 & -2.47 & -2.18 & -1.57 & -4.03 & -4.02 & -3.74 & -3.00 & -2.38 & -2.19 & -2.01\\ 
6.66 & -2.70 & -2.71 & -2.69 & -2.55 & -2.72 & -2.42 & -1.79 & -4.40 & -4.39 & -4.11 & -3.31 & -2.64 & -2.43 & -2.24\\ 
7.37 & -3.07 & -3.08 & -3.05 & -2.87 & -2.99 & -2.67 & -2.03 & -4.77 & -4.76 & -4.47 & -3.63 & -2.90 & -2.69 & -2.48\\ 
8.15 & -3.42 & -3.43 & -3.40 & -3.23 & -3.36 & -3.04 & -2.40 & -5.12 & -5.11 & -4.83 & -3.99 & -3.27 & -3.06 & -2.85\\ 
\bottomrule 
\end{tabular}

%% file: Tables/Photometry_V3.tex
\begin{tabular}{lcc cc cc cc cc c c c c}
\toprule\midrule
Star & RA & DEC & n5 & closest &UVW2 & W2-FF & UVM2 & M2-FF & UVW1 & W1-FF & U & B & V & I \\ 
\midrule
1	&  &	&	0	&	5.33	&	16.18	$\pm$	0.05	&	1.00	&	16.26	$\pm$	0.07	&	1.00	&	16.40	$\pm$	0.07	&	0.99	&	16.72	$\pm$	0.04	&	17.14	$\pm$	0.03	&	17.45	$\pm$	0.04	&	18.22	$\pm$	0.05	\\
2	&  &	&	1	&	4.00	&	17.94	$\pm$	0.09	&	0.92	&	17.98	$\pm$	0.10	&	0.91	&	18.05	$\pm$	0.17	&	0.92	&	18.29	$\pm$	0.05	&	18.63	$\pm$	0.04	&	18.93	$\pm$	0.06	&	19.63	$\pm$	0.07	\\
3	&  &	&	2	&	3.77	&	17.86	$\pm$	0.07	&	0.95	&	18.00	$\pm$	0.14	&	0.97	&	18.17	$\pm$	0.10	&	0.93	&	18.44	$\pm$	0.04	&	18.86	$\pm$	0.03	&	19.22	$\pm$	0.04	&	19.77	$\pm$	0.08	\\
4	&  &	&	3	&	2.46	&	17.82	$\pm$	0.11	&	0.62	&	17.88	$\pm$	0.11	&	0.61	&	18.01	$\pm$	0.13	&	0.59	&	18.46	$\pm$	0.10	&	18.86	$\pm$	0.03	&	19.22	$\pm$	0.06	&	19.99	$\pm$	0.08	\\
5	&  &	&	1	&	4.47	&	17.73	$\pm$	0.08	&	0.66	&	17.80	$\pm$	0.11	&	0.63	&	17.81	$\pm$	0.11	&	0.56	&	17.97	$\pm$	0.06	&	18.03	$\pm$	0.04	&	18.33	$\pm$	0.05	&	19.03	$\pm$	0.06	\\
6	&  &	&	3	&	3.54	&	17.31	$\pm$	0.10	&	0.82	&	17.56	$\pm$	0.10	&	0.76	&	17.62	$\pm$	0.09	&	0.73	&	18.02	$\pm$	0.07	&	18.30	$\pm$	0.04	&	18.57	$\pm$	0.06	&	19.30	$\pm$	0.06	\\
7	&  &	&	1	&	3.56	&	17.86	$\pm$	0.09	&	0.95	&	17.97	$\pm$	0.10	&	0.92	&	18.06	$\pm$	0.10	&	0.92	&	18.44	$\pm$	0.07	&	18.54	$\pm$	0.08	&	18.68	$\pm$	0.19	&	...			\\
8	&  &	&	0	&	6.74	&	18.16	$\pm$	0.07	&	1.00	&	18.27	$\pm$	0.09	&	1.00	&	18.40	$\pm$	0.08	&	1.00	&	18.83	$\pm$	0.07	&	19.10	$\pm$	0.05	&	19.48	$\pm$	0.09	&	20.31	$\pm$	0.14	\\
9	&  &	&	1	&	2.18	&	17.87	$\pm$	0.11	&	0.89	&	17.93	$\pm$	0.12	&	0.92	&	18.00	$\pm$	0.08	&	0.90	&	18.43	$\pm$	0.05	&	18.49	$\pm$	0.06	&	18.68	$\pm$	0.11	&	19.63	$\pm$	0.13	\\
10	&  &	&	0	&	5.24	&	18.01	$\pm$	0.09	&	0.94	&	18.02	$\pm$	0.10	&	0.93	&	18.25	$\pm$	0.10	&	0.91	&	18.47	$\pm$	0.32	&	18.74	$\pm$	0.15	&	18.96	$\pm$	0.10	&	19.77	$\pm$	0.08	\\
11	&  &	&	0	&	4.99	&	17.47	$\pm$	0.09	&	1.00	&	17.57	$\pm$	0.09	&	1.00	&	17.68	$\pm$	0.08	&	1.00	&	18.14	$\pm$	0.05	&	18.11	$\pm$	0.03	&	18.43	$\pm$	0.09	&	19.13	$\pm$	0.10	\\
12	&  &	&	1	&	4.54	&	18.46	$\pm$	0.08	&	0.83	&	18.42	$\pm$	0.09	&	0.82	&	18.59	$\pm$	0.10	&	0.75	&	18.70	$\pm$	0.07	&	19.12	$\pm$	0.09	&	19.26	$\pm$	0.18	&	20.21	$\pm$	0.22	\\
13	&  &	&     1	&	2.58	&	18.07	$\pm$	0.07	&	0.90	&	18.17	$\pm$	0.12	&	0.92	&	18.21	$\pm$	0.12	&	0.90	&	18.30	$\pm$	0.04	&	18.68	$\pm$	0.08	&	18.81	$\pm$	0.20	&	19.42	$\pm$	0.05	\\
14	&  &	&      1	&	4.78	&	17.09	$\pm$	0.05	&	1.00	&	17.19	$\pm$	0.05	&	0.99	&	17.28	$\pm$	0.06	&	0.99	&	17.61	$\pm$	0.04	&	17.67	$\pm$	0.04	&	17.96	$\pm$	0.03	&	18.85	$\pm$	0.04	\\
15	&  &	&	1	&	1.48	&	16.12	$\pm$	0.27	&	0.23	&	16.02	$\pm$	0.39	&	0.24	&	16.24	$\pm$	0.24	&	0.19	&	16.69	$\pm$	0.07	&	16.55	$\pm$	0.06	&	16.59	$\pm$	0.10	&	17.45	$\pm$	0.12	\\
16	&  &	&	1	&	2.42	&	18.08	$\pm$	0.15	&	0.92	&	18.13	$\pm$	0.10	&	0.93	&	18.20	$\pm$	0.11	&	0.91	&	18.13	$\pm$	0.07	&	18.50	$\pm$	0.07	&	18.77	$\pm$	0.13	&	19.94	$\pm$	0.15	\\
17	&  &	&	1	&	2.36	&	16.54	$\pm$	0.14	&	0.23	&	16.52	$\pm$	0.22	&	0.23	&	16.79	$\pm$	0.14	&	0.19	&	16.77	$\pm$	0.05	&	16.89	$\pm$	0.04	&	17.05	$\pm$	0.04	&	17.86	$\pm$	0.06	\\
18	&  &	&	1	&	4.44	&	15.14	$\pm$	0.05	&	0.99	&	15.18	$\pm$	0.05	&	0.99	&	15.28	$\pm$	0.05	&	0.99	&	15.74	$\pm$	0.03	&	15.72	$\pm$	0.04	&	16.13	$\pm$	0.09	&	16.75	$\pm$	0.05	\\
19	&  &	&	3	&	3.16	&	17.33	$\pm$	0.40	&	0.08	&	16.81	$\pm$	0.28	&	0.11	&	16.82	$\pm$	0.24	&	0.13	&	17.26	$\pm$	0.05	&	17.13	$\pm$	0.04	&	17.40	$\pm$	0.14	&	17.62	$\pm$	0.08	\\
20	&  &	&	2	&	2.26	&	17.67	$\pm$	0.31	&	0.18	&	17.76	$\pm$	0.13	&	0.16	&	17.73	$\pm$	0.12	&	0.17	&	18.39	$\pm$	0.05	&	17.99	$\pm$	0.07	&	18.19	$\pm$	0.10	&	18.90	$\pm$	0.08	\\
21	&  &	&	1	&	1.80	&	17.71	$\pm$	0.17	&	0.49	&	17.70	$\pm$	0.16	&	0.47	&	17.67	$\pm$	0.22	&	0.47	&	18.14	$\pm$	0.05	&	18.06	$\pm$	0.04	&	18.26	$\pm$	0.05	&	18.83	$\pm$	0.05	\\
22	&  &	&	2	&	1.76	&	16.20	$\pm$	0.12	&	0.50	&	16.00	$\pm$	0.11	&	0.52	&	16.11	$\pm$	0.24	&	0.57	&	16.65	$\pm$	0.04	&	16.71	$\pm$	0.04	&	17.11	$\pm$	0.06	&	17.72	$\pm$	0.04	\\
23	&  &	&	2	&	3.55	&	16.99	$\pm$	0.06	&	0.94	&	17.05	$\pm$	0.07	&	0.90	&	17.15	$\pm$	0.08	&	0.90	&	17.17	$\pm$	0.03	&	17.31	$\pm$	0.03	&	17.84	$\pm$	0.10	&	18.27	$\pm$	0.04	\\
24	&  &	&	2	&	2.75	&	17.82	$\pm$	0.26	&	0.09	&	18.25	$\pm$	0.29	&	0.06	&	18.58	$\pm$	0.39	&	0.04	&	18.63	$\pm$	0.08	&	19.06	$\pm$	0.08	&	19.05	$\pm$	0.12	&	19.83	$\pm$	0.14	\\
25	&  &	&	2	&	2.14	&	16.93	$\pm$	0.21	&	0.19	&	16.92	$\pm$	0.33	&	0.19	&	17.02	$\pm$	0.18	&	0.19	&	17.35	$\pm$	0.09	&	17.38	$\pm$	0.03	&	17.68	$\pm$	0.05	&	18.31	$\pm$	0.06	\\
\bottomrule
\end{tabular}

%% file: Tables/Equivalent_widths.tex
\begin{tabular}{lcccccccc}
\toprule\midrule 
Star & Class & He\textsc{ii} $\lambda 3835$+H$\eta$ & He\textsc{i}+He\textsc{ii} $\lambda 4026$ & He\textsc{ii} $\lambda 4100$+H$\delta$ & He\textsc{ii} $\lambda4339$+H$\gamma$ & He\textsc{ii} $\lambda4860$+H$\beta$ & He\textsc{ii} $\lambda5411$ & He\textsc{i} $\lambda5876$\\ 
\midrule 
1 & 1 & $0.46\pm 0.02 $ & $0.58\pm 0.04 $ & $1.75\pm 0.02 $ & $2.18\pm 0.04 $ & $2.62\pm 0.05 $ & $1.90\pm 0.04 $ & $-0.09$-$ 0.06 $\\ 
2 & 1 & $0.42\pm 0.08 $ & $0.96\pm 0.07 $ & $2.22\pm 0.13 $ & $2.60\pm 0.08 $ & $2.87\pm 0.10 $ & $1.83\pm 0.09 $ & $-0.17$-$ 0.19 $\\ 
3 & 1 & $-0.03$-$ 0.08 $ & $0.73\pm 0.05 $ & $1.40\pm 0.08 $ & $2.08\pm 0.02 $ & $2.43\pm 0.10 $ & $1.89\pm 0.09 $ & $-0.07$-$ 0.18 $\\ 
4 & 1 & $-0.07$-$ 0.33 $ & $1.14\pm 0.10 $ & $1.92\pm 0.11 $ & $2.39\pm 0.06 $ & $2.75\pm 0.13 $ & $2.31\pm 0.23 $ & $-0.16$-$ 0.26 $\\ 
5 & 1 & $0.27\pm 0.05 $ & $1.15\pm 0.06 $ & $1.04\pm 0.05 $ & $1.29\pm 0.09 $ & $1.81\pm 0.08 $ & $1.72\pm 0.08 $ & $0.21\pm 0.06 $\\ 
6 & 1 & $0.40\pm 0.08 $ & $0.97\pm 0.05 $ & $1.73\pm 0.04 $ & $2.68\pm 0.09 $ & $2.71\pm 0.05 $ & $2.08\pm 0.07 $ & $-0.11$-$ 0.12 $\\ 
7 & 1 & $-0.15$-$ 0.19 $ & $1.06\pm 0.10 $ & $2.24\pm 0.40 $ & $4.49\pm 0.20 $ & $2.78\pm 0.14 $ & $2.26\pm 0.17 $ & $0.37\pm 0.09 $\\ 
8 & 1 & $0.24\pm 0.04 $ & $1.00\pm 0.10 $ & $1.61\pm 0.07 $ & $2.33\pm 0.13 $ & $2.17\pm 0.06 $ & $1.93\pm 0.09 $ & $-0.11$-$ 0.52 $\\ 
\midrule 
9 & 2 & $1.39\pm 0.10 $ & -- & -- & -- & -- & $1.46\pm 0.13 $ & --\\ 
10 & 2 & $3.35\pm 0.23 $ & -- & -- & -- & -- & $0.62\pm 0.17 $ & --\\ 
11 & 2 & $1.48\pm 0.15 $ & -- & -- & -- & -- & $1.42\pm 0.14 $ & --\\ 
12 & 2 & $1.84\pm 0.21 $ & -- & -- & -- & -- & $1.25\pm 0.15 $ & --\\ 
13 & 2 & $2.56\pm 0.10 $ & -- & -- & -- & -- & $0.71\pm 0.11 $ & --\\ 
14 & 2 & $2.50\pm 0.05 $ & -- & -- & -- & -- & $0.24\pm 0.02 $ & --\\ 
15 & 2 & $3.87\pm 0.51 $ & -- & -- & -- & -- & $-0.07$-$ 0.14 $ & --\\ 
16 & 2 & $2.39\pm 0.07 $ & -- & -- & -- & -- & $0.94\pm 0.06 $ & --\\ 
\midrule 
17 & 3 & $3.57\pm 0.10 $ & -- & -- & -- & -- & $-0.22$-$ 0.16 $ & --\\ 
18 & 3 & $3.22\pm 0.11 $ & -- & -- & -- & -- & $-0.10$-$ 0.11 $ & --\\ 
19 & 3 & $4.93\pm 0.36 $ & -- & -- & -- & -- & $-0.14$-$ 0.18 $ & --\\ 
20 & 3 & $4.58\pm 0.15 $ & -- & -- & -- & -- & $-0.13$-$ 0.14 $ & --\\ 
21 & 3 & $5.26\pm 0.08 $ & -- & -- & -- & -- & $-0.20$-$ 0.20 $ & --\\ 
22 & 3 & $3.71\pm 0.26 $ & -- & -- & -- & -- & $-0.15$-$ 0.08 $ & --\\ 
23 & 3 & $4.17\pm 0.08 $ & -- & -- & -- & -- & $-0.13$-$ 0.15 $ & --\\ 
24 & 3 & $2.41\pm 0.34 $ & -- & -- & -- & -- & $-0.14$-$ 0.28 $ & --\\ 
25 & 3 & $4.49\pm 0.16 $ & -- & -- & -- & -- & $-0.17$-$ 0.17 $ & --\\ 
\bottomrule
\end{tabular}

%% file: Tables/Kinematics_V3.tex
\begin{tabular}{lccc ccc ccc cc}
\toprule\midrule 
 Star & Galaxy& Class & N$_{\rm spec}$ & $v_{\rm avg}$ &$v_{\rm min}$ & $v_{\rm max}$&  $\pi$ & $\mu_{\alpha}$ & $\mu_{\delta}$ & $\chi^2_{\mathrm{2D}}$ & $\chi^2_{\mathrm{3D}}$\\ 
&&& [km s$^{-1}$] & [km s$^{-1}$] & [km s$^{-1}$] & [mas] & [mas/yr] & [mas/yr] & & \\ 
\midrule 
1	& SMC & 1 & 	18	&	151	&	127	&	174	&	 $-0.094\pm0.074$ 	&	 $0.700\pm0.090$ 	&	 $-1.287\pm0.079$ 	&	0.1	&	0.2\\ 
2	& SMC &1 &	11	&	156	&	102	&	193	&	 $0.067\pm0.188$ 	&	 $-0.150\pm0.264$ 	&	 $-0.636\pm0.218$ 	&	10.4	&	3.3\\ 
3	& SMC &1 &	30	&	206	&	73	&	278	&	 $-0.519\pm0.215$ 	&	 $0.261\pm0.267$ 	&	 $-0.986\pm0.201$ 	&	2.5	&	1.2\\ 
4	& SMC &1 &	7	&	202	&	153	&	221	&	 $0.177\pm0.209$ 	&	 $1.077\pm0.300$ 	&	 $-0.793\pm0.252$ 	&	3.3	&	1.2\\ 
5	& LMC &1 &	14	&	280	&	250	&	307	&	 $0.294\pm0.155$ 	&	 $2.070\pm0.202$ 	&	 $-1.431\pm0.216$ 	&	15.7	&	16.9\\ 
6	& LMC &1 &	18	&	270	&	241	&	301	&	 $0.144\pm0.195$ 	&	 $3.717\pm0.219$ 	&	 $1.579\pm0.257$ 	&	40.2	&	25.4\\ 
7	& LMC &1 &	8	&	269	&	243	&	281	&	 $-0.104\pm0.236$ 	&	 $2.122\pm0.346$ 	&	 $-0.172\pm0.359$ 	&	1.2	&	2.6\\ 
8	& LMC &1 &	6	&	296	&	284	&	304	&	 $0.290\pm0.239$ 	&	 $1.148\pm0.278$ 	&	 $0.807\pm0.343$ 	&	3.9	&	2.6\\ 
\midrule
9	& LMC &2 &	7	&	307	&	301	&	312	&	 $0.200\pm0.155$ 	&	 $1.872\pm0.173$ 	&	 $0.401\pm0.212$ 	&	0.1	&	1.9\\ 
10	& LMC &2 &	1	&	251	&	251	&	251	&	 $-0.016\pm0.180$ 	&	 $2.127\pm0.243$ 	&	 $-0.207\pm0.249$ 	&	1.7	&	2.6\\ 
11	& LMC &2 &	4	&	215	&	203	&	221	&	 $-0.005\pm0.134$ 	&	 $1.481\pm0.183$ 	&	 $-0.864\pm0.178$ 	&	9.5	&	13.6\\ 
12	& LMC &2 &	2	&	299	&	289	&	309	&	 $0.109\pm0.231$ 	&	 $1.033\pm0.295$ 	&	 $-0.790\pm0.327$ 	&	9.4	&	8.1\\ 
13	& LMC &2 &	2	&	284	&	284	&	284	&	 $-0.452\pm0.190$ 	&	 $0.917\pm0.240$ 	&	 $1.188\pm0.245$ 	&	9.5	&	5.6\\ 
14	& LMC &2 &	7	&	235	&	231	&	241	&	 $0.106\pm0.092$ 	&	 $1.913\pm0.130$ 	&	 $0.134\pm0.122$ 	&	0.2	&	2.2\\ 
15	& LMC &2 &	2	&	280	&	278	&	282	&	 $-0.121\pm0.077$ 	&	 $2.319\pm0.126$ 	&	 $0.456\pm0.086$ 	&	3	&	2.2\\ 
16	& LMC &2 &	11	&	276	&	220	&	336	&	 $-0.114\pm0.131$ 	&	 $1.527\pm0.230$ 	&	 $0.758\pm0.164$ 	&	1.8	&	0.8\\ 
\midrule
17	& LMC &3 &	1	&	282	&	282	&	282	&	 $-0.195\pm0.076$ 	&	 $1.566\pm0.104$ 	&	 $0.368\pm0.108$ 	&	0.9	&	0.3\\ 
18	& LMC &3 &	1	&	276	&	276	&	276	&	 $0.009\pm0.038$ 	&	 $1.681\pm0.047$ 	&	 $0.270\pm0.050$ 	&	0.3	&	0.2\\ 
19	& LMC &3 &	1	&	269	&	269	&	269	&	 $0.221\pm0.069$ 	&	 $1.950\pm0.100$ 	&	 $0.196\pm0.093$ 	&	0.2	&	0.5\\ 
20	& LMC &3 &	5	&	291	&	282	&	296	&	 $0.095\pm0.153$ 	&	 $1.406\pm0.188$ 	&	 $0.084\pm0.203$ 	&	2.1	&	1.3\\ 
21	& SMC &3 &	1	&	142	&	142	&	142	&	 $0.315\pm0.130$ 	&	 $0.677\pm0.156$ 	&	 $-1.470\pm0.164$ 	&	1.2	&	1.3\\ 
22	& SMC &3 &	1	&	196	&	196	&	196	&	 $-0.106\pm0.064$ 	&	 $0.895\pm0.089$ 	&	 $-1.091\pm0.078$ 	&	1.4	&	0.5\\ 
23	& SMC &3 &	1	&	154	&	154	&	154	&	 $-0.121\pm0.074$ 	&	 $0.814\pm0.094$ 	&	 $-1.110\pm0.086$ 	&	0.7	&	0.6\\ 
24	& SMC &3 &	1	&	140	&	140	&	140	&	 $0.719\pm0.212$ 	&	 $0.798\pm0.261$ 	&	 $-2.141\pm0.280$ 	&	8.2	&	3.1\\ 
25	& SMC &3 &	1	&	204	&	204	&	204	&	 $0.080\pm0.084$ 	&	 $0.891\pm0.112$ 	&	 $-1.395\pm0.108$ 	&	1	&	0.8\\ 
\bottomrule
\end{tabular}

%% file: CompleteManuscriptV2_doublespace.bbl
\begin{thebibliography}{100}

\bibitem{2012Sci...337..444S}
H.~{Sana}, {\it et~al.\/}, {\it Science\/} {\bf 337}, 444 (2012).

\bibitem{2017ApJS..230...15M}
M.~{Moe}, R.~{Di Stefano}, {\it \apjs\/} {\bf 230}, 15 (2017).

\bibitem{Drout2011}
M.~R. {Drout}, {\it et~al.\/}, {\it \apj\/} {\bf 741}, 97 (2011).

\bibitem{Lyman2016}
J.~D. {Lyman}, {\it et~al.\/}, {\it \mnras\/} {\bf 457}, 328 (2016).

\bibitem{Smith2011}
N.~{Smith}, W.~{Li}, A.~V. {Filippenko}, R.~{Chornock}, {\it \mnras\/} {\bf
  412}, 1522 (2011).

\bibitem{Eldridge2013}
J.~J. {Eldridge}, M.~{Fraser}, S.~J. {Smartt}, J.~R. {Maund}, R.~M. {Crockett},
  {\it \mnras\/} {\bf 436}, 774 (2013).

\bibitem{2017ApJ...846..170T}
T.~M. {Tauris}, {\it et~al.\/}, {\it \apj\/} {\bf 846}, 170 (2017).

\bibitem{Stanway2016}
E.~R. {Stanway}, J.~J. {Eldridge}, G.~D. {Becker}, {\it \mnras\/} {\bf 456},
  485 (2016).

\bibitem{2020A&A...636A..47S}
A.~{Saxena}, {\it et~al.\/}, {\it \aap\/} {\bf 636}, A47 (2020).

\bibitem{2020A&A...634A.134G}
Y.~{G{\"o}tberg}, {\it et~al.\/}, {\it \aap\/} {\bf 634}, A134 (2020).

\bibitem{2016PASP..128h2001H}
U.~{Heber}, {\it \pasp\/} {\bf 128}, 082001 (2016).

\bibitem{2007ARA&A..45..177C}
P.~A. {Crowther}, {\it \araa\/} {\bf 45}, 177 (2007).

\bibitem{2021AJ....161..248W}
L.~{Wang}, {\it et~al.\/}, {\it \aj\/} {\bf 161}, 248 (2021).

\bibitem{2017MNRAS.464.2066S}
M.~M. {Shara}, {\it et~al.\/}, {\it \mnras\/} {\bf 464}, 2066 (2017).

\bibitem{1992Natur.355..703V}
M.~H. {van Kerkwijk}, {\it et~al.\/}, {\it \nat\/} {\bf 355}, 703 (1992).

\bibitem{2008A&A...485..245G}
J.~H. {Groh}, A.~S. {Oliveira}, J.~E. {Steiner}, {\it \aap\/} {\bf 485}, 245
  (2008).

\bibitem{2018A&A...615A..78G}
Y.~{G{\"o}tberg}, {\it et~al.\/}, {\it \aap\/} {\bf 615}, A78 (2018).

\bibitem{MM}
Materials and methods are available as supplementary materials.

\bibitem{Siegel2015}
M.~H. {Siegel}, C.~A. {Gronwall}, L.~M.~Z. {Hagen}, E.~A. {Hoversten}, {\it
  arXiv e-prints\/} p. arXiv:1504.02369 (2015).

\bibitem{Hagen2017}
L.~M.~Z. {Hagen}, {\it et~al.\/}, {\it \mnras\/} {\bf 466}, 4540 (2017).

\bibitem{Lang2016b}
D.~{Lang}, D.~W. {Hogg}, D.~J. {Schlegel}, {\it \aj\/} {\bf 151}, 36 (2016).

\bibitem{Zaritsky2002}
D.~{Zaritsky}, J.~{Harris}, I.~B. {Thompson}, E.~K. {Grebel}, P.~{Massey}, {\it
  \aj\/} {\bf 123}, 855 (2002).

\bibitem{Zaritsky2004}
D.~{Zaritsky}, J.~{Harris}, I.~B. {Thompson}, E.~K. {Grebel}, {\it \aj\/} {\bf
  128}, 1606 (2004).

\bibitem{1998ApJ...496..407H}
D.~J. {Hillier}, D.~L. {Miller}, {\it \apj\/} {\bf 496}, 407 (1998).

\bibitem{2003ApJS..146..417L}
T.~{Lanz}, I.~{Hubeny}, {\it \apjs\/} {\bf 146}, 417 (2003).

\bibitem{2007ApJS..169...83L}
T.~{Lanz}, I.~{Hubeny}, {\it \apjs\/} {\bf 169}, 83 (2007).

\bibitem{2002AJ....123.2754W}
N.~R. {Walborn}, {\it et~al.\/}, {\it \aj\/} {\bf 123}, 2754 (2002).

\bibitem{Neugent2017}
K.~F. {Neugent}, P.~{Massey}, D.~J. {Hillier}, N.~{Morrell}, {\it \apj\/} {\bf
  841}, 20 (2017).

\bibitem{2013A&A...551A..31D}
J.~S. {Drilling}, C.~S. {Jeffery}, U.~{Heber}, S.~{Moehler}, R.~{Napiwotzki},
  {\it \aap\/} {\bf 551}, A31 (2013).

\bibitem{2020ApJ...890...51E}
T.~{Ertl}, S.~E. {Woosley}, T.~{Sukhbold}, H.~T. {Janka}, {\it \apj\/} {\bf
  890}, 51 (2020).

\bibitem{2017ApJ...840...10Y}
S.-C. {Yoon}, L.~{Dessart}, A.~{Clocchiatti}, {\it \apj\/} {\bf 840}, 10
  (2017).

\bibitem{2021ApJ...922...55B}
E.~R. {Beasor}, B.~{Davies}, N.~{Smith}, {\it \apj\/} {\bf 922}, 55 (2021).

\bibitem{2003ARA&A..41..391V}
H.~{van Winckel}, {\it \araa\/} {\bf 41}, 391 (2003).

\bibitem{2003A&A...409..969H}
W.~R. {Hamann}, M.~{Pe{\~n}a}, G.~{Gr{\"a}fener}, M.~T. {Ruiz}, {\it \aap\/}
  {\bf 409}, 969 (2003).

\bibitem{2011A&A...530A.115B}
I.~{Brott}, {\it et~al.\/}, {\it \aap\/} {\bf 530}, A115 (2011).

\bibitem{2011ApJS..192....3P}
B.~{Paxton}, {\it et~al.\/}, {\it \apjs\/} {\bf 192}, 3 (2011).

\bibitem{2013ApJS..208....4P}
B.~{Paxton}, {\it et~al.\/}, {\it \apjs\/} {\bf 208}, 4 (2013).

\bibitem{2015ApJS..220...15P}
B.~{Paxton}, {\it et~al.\/}, {\it \apjs\/} {\bf 220}, 15 (2015).

\bibitem{2017A&A...607L...8V}
J.~S. {Vink}, {\it \aap\/} {\bf 607}, L8 (2017).

\bibitem{2019MNRAS.486.4451G}
A.~{Gilkis}, J.~S. {Vink}, J.~J. {Eldridge}, C.~A. {Tout}, {\it \mnras\/} {\bf
  486}, 4451 (2019).

\bibitem{2015A&A...580A..27M}
E.~{Moravveji}, C.~{Aerts}, P.~I. {P{\'a}pics}, S.~A. {Triana}, B.~{Vandoren},
  {\it \aap\/} {\bf 580}, A27 (2015).

\bibitem{1990A&A...231..116H}
D.~J. {Hillier}, {\it \aap\/} {\bf 231}, 116 (1990).

\bibitem{1999isw..book.....L}
H.~J.~G.~L.~M. {Lamers}, J.~P. {Cassinelli}, {\it {Introduction to Stellar
  Winds}\/} (1999).

\bibitem{2000A&A...360..227N}
T.~{Nugis}, H.~J.~G.~L.~M. {Lamers}, {\it \aap\/} {\bf 360}, 227 (2000).

\bibitem{2020A&A...634A..79S}
T.~{Shenar}, A.~{Gilkis}, J.~S. {Vink}, H.~{Sana}, A.~A.~C. {Sand er}, {\it
  \aap\/} {\bf 634}, A79 (2020).

\bibitem{2014A&A...570A..38B}
J.~M. {Bestenlehner}, {\it et~al.\/}, {\it \aap\/} {\bf 570}, A38 (2014).

\bibitem{2011A&A...536A..58R}
J.~G. {Rivero Gonz{\'a}lez}, J.~{Puls}, F.~{Najarro}, {\it \aap\/} {\bf 536},
  A58 (2011).

\bibitem{2012A&A...537A..79R}
J.~G. {Rivero Gonz{\'a}lez}, J.~{Puls}, F.~{Najarro}, I.~{Brott}, {\it \aap\/}
  {\bf 537}, A79 (2012).

\bibitem{2014A&A...564A..30G}
J.~H. {Groh}, G.~{Meynet}, S.~{Ekstr{\"o}m}, C.~{Georgy}, {\it \aap\/} {\bf
  564}, A30 (2014).

\bibitem{2001A&A...369..574V}
J.~S. {Vink}, A.~{de Koter}, H.~J.~G.~L.~M. {Lamers}, {\it \aap\/} {\bf 369},
  574 (2001).

\bibitem{1995ApJ...455..269L}
H.~J.~G.~L.~M. {Lamers}, T.~P. {Snow}, D.~M. {Lindholm}, {\it \apj\/} {\bf
  455}, 269 (1995).

\bibitem{1992IAUS..149..225K}
R.~L. {Kurucz}, {\it The Stellar Populations of Galaxies\/}, B.~{Barbuy},
  A.~{Renzini}, eds. (1992), vol. 149 of {\it IAU Symposium\/}, p. 225.

\bibitem{2019A&A...629A.134G}
Y.~{G{\"o}tberg}, S.~E. {de Mink}, J.~H. {Groh}, C.~{Leitherer}, C.~{Norman},
  {\it \aap\/} {\bf 629}, A134 (2019).

\bibitem{2021ApJ...908...67S}
Y.~{Shao}, X.-D. {Li}, {\it \apj\/} {\bf 908}, 67 (2021).

\bibitem{2017ApJ...842..125Z}
E.~{Zapartas}, {\it et~al.\/}, {\it \apj\/} {\bf 842}, 125 (2017).

\bibitem{1994A&A...290..119P}
O.~R. {Pols}, {\it \aap\/} {\bf 290}, 119 (1994).

\bibitem{2019A&A...624A..66R}
M.~{Renzo}, {\it et~al.\/}, {\it \aap\/} {\bf 624}, A66 (2019).

\bibitem{1998ApJ...493..440G}
D.~R. {Gies}, {\it et~al.\/}, {\it \apj\/} {\bf 493}, 440 (1998).

\bibitem{2017ApJ...843...60W}
L.~{Wang}, D.~R. {Gies}, G.~J. {Peters}, {\it \apj\/} {\bf 843}, 60 (2017).

\bibitem{2018ApJ...853..156W}
L.~{Wang}, D.~R. {Gies}, G.~J. {Peters}, {\it \apj\/} {\bf 853}, 156 (2018).

\bibitem{2016ApJ...823..102C}
J.~{Choi}, {\it et~al.\/}, {\it \apj\/} {\bf 823}, 102 (2016).

\bibitem{2009ssc..book.....G}
R.~O. {Gray}, J.~{Corbally}, Christopher, {\it {Stellar Spectral
  Classification}\/} (2009).

\bibitem{Brosch1999}
N.~{Brosch}, M.~{Shara}, J.~{MacKenty}, D.~{Zurek}, B.~{McLean}, {\it \aj\/}
  {\bf 117}, 206 (1999).

\bibitem{Simons2014}
R.~{Simons}, D.~{Thilker}, L.~{Bianchi}, T.~{Wyder}, {\it Advances in Space
  Research\/} {\bf 53}, 939 (2014).

\bibitem{Monet2003}
D.~G. {Monet}, {\it et~al.\/}, {\it \aj\/} {\bf 125}, 984 (2003).

\bibitem{Lang2010}
D.~{Lang}, D.~W. {Hogg}, K.~{Mierle}, M.~{Blanton}, S.~{Roweis}, {\it \aj\/}
  {\bf 139}, 1782 (2010).

\bibitem{Lang2016}
D.~{Lang}, D.~W. {Hogg}, D.~{Mykytyn}, {\it Astrophysics Source Code Library,
  record ascl:1604.008\/}  (2016).

\bibitem{photutils}
L.~Bradley, {\it et~al.\/}, astropy/photutils: 1.0.0 (2020).

\bibitem{astropy2018}
{Astropy Collaboration}, {\it et~al.\/}, {\it \aj\/} {\bf 156}, 123 (2018).

\bibitem{2008SPIE.7014E..54M}
J.~L. {Marshall}, {\it et~al.\/}, {\it Ground-based and Airborne
  Instrumentation for Astronomy II\/}, I.~S. {McLean}, M.~M. {Casali}, eds.
  (2008), vol. 7014 of {\it Society of Photo-Optical Instrumentation Engineers
  (SPIE) Conference Series\/}, p. 701454.

\bibitem{Kelson2000}
D.~D. {Kelson}, G.~D. {Illingworth}, P.~G. {van Dokkum}, M.~{Franx}, {\it
  \apj\/} {\bf 531}, 159 (2000).

\bibitem{Kelson2003}
D.~D. {Kelson}, {\it \pasp\/} {\bf 115}, 688 (2003).

\bibitem{pyraf2012}
{Science Software Branch at STScI}, {\it Astrophysics Source Code Library,
  record ascl:1207.011\/}  (2012).

\bibitem{Tonry1979}
J.~{Tonry}, M.~{Davis}, {\it \aj\/} {\bf 84}, 1511 (1979).

\bibitem{Pietrzynski.G.2013.LMCDistance}
G.~{Pietrzy{\'n}ski}, {\it et~al.\/}, {\it \nat\/} {\bf 495}, 76 (2013).

\bibitem{Hilditch.R.2005.SMCDistance}
R.~W. {Hilditch}, I.~D. {Howarth}, T.~J. {Harries}, {\it \mnras\/} {\bf 357},
  304 (2005).

\bibitem{Gordon2003}
K.~D. {Gordon}, G.~C. {Clayton}, K.~A. {Misselt}, A.~U. {Landolt}, M.~J.
  {Wolff}, {\it \apj\/} {\bf 594}, 279 (2003).

\bibitem{Evans2008}
C.~J. {Evans}, I.~D. {Howarth}, {\it \mnras\/} {\bf 386}, 826 (2008).

\bibitem{Evans2015}
C.~J. {Evans}, {\it et~al.\/}, {\it \aap\/} {\bf 574}, A13 (2015).

\bibitem{Evans2015b}
C.~J. {Evans}, J.~T. {van Loon}, R.~{Hainich}, M.~{Bailey}, {\it \aap\/} {\bf
  584}, A5 (2015).

\bibitem{Gaia.Collaboration.2020.EDR3}
{Gaia Collaboration}, {\it et~al.\/}, {\it arXiv e-prints\/} p.
  arXiv:2012.01533 (2020).

\bibitem{Bailer-Jones2021}
C.~A.~L. {Bailer-Jones}, J.~{Rybizki}, M.~{Fouesneau}, M.~{Demleitner},
  R.~{Andrae}, {\it VizieR Online Data Catalog\/} p. I/352 (2021).

\bibitem{Gaia2018}
{Gaia Collaboration}, {\it et~al.\/}, {\it \aap\/} {\bf 616}, A12 (2018).

\bibitem{Ogrady2020}
A.~J.~G. {O'Grady}, {\it et~al.\/}, {\it \apj\/} {\bf 901}, 135 (2020).

\bibitem{Aadland2018}
E.~{Aadland}, P.~{Massey}, K.~F. {Neugent}, M.~R. {Drout}, {\it \aj\/} {\bf
  156}, 294 (2018).

\bibitem{Ogrady2022}
{O'Grady et~al.}, {\it submitted to ApJ\/}  (2022).

\bibitem{2014ApJ...788...83M}
P.~{Massey}, K.~F. {Neugent}, N.~{Morrell}, D.~J. {Hillier}, {\it \apj\/} {\bf
  788}, 83 (2014).

\bibitem{2012A&A...545A..95M}
F.~{Martins}, D.~J. {Hillier}, {\it \aap\/} {\bf 545}, A95 (2012).

\bibitem{2009PhDT.......273H}
H.~A. {Hirsch}, {Hot subluminous stars: On the Search for Chemical Signatures
  of their Genesis}, Ph.D. thesis, Friedrich-Alexander University
  Erlangen-N{\"u}rnberg (2009).

\bibitem{2020A&A...634A..51B}
J.~{Bodensteiner}, {\it et~al.\/}, {\it \aap\/} {\bf 634}, A51 (2020).

\bibitem{2012A&A...543A..95R}
J.~G. {Rivero Gonz{\'a}lez}, J.~{Puls}, P.~{Massey}, F.~{Najarro}, {\it \aap\/}
  {\bf 543}, A95 (2012).

\bibitem{2017A&A...600A..82G}
N.~J. {Grin}, {\it et~al.\/}, {\it \aap\/} {\bf 600}, A82 (2017).

\bibitem{1995A&AS..113..459H}
W.-R. {Hamann}, L.~{Koesterke}, U.~{Wessolowski}, {\it \aaps\/} {\bf 113}, 459
  (1995).

\bibitem{2015A&A...581A..21H}
R.~{Hainich}, {\it et~al.\/}, {\it \aap\/} {\bf 581}, A21 (2015).

\bibitem{1998PASP..110.1315G}
R.~F. {Garrison}, {\it \pasp\/} {\bf 110}, 1315 (1998).

\bibitem{2011A&A...526A..39G}
S.~{Geier}, {\it et~al.\/}, {\it \aap\/} {\bf 526}, A39 (2011).

\bibitem{2011ApJS..193...24S}
A.~{Sota}, {\it et~al.\/}, {\it \apjs\/} {\bf 193}, 24 (2011).

\bibitem{1990PASP..102..379W}
N.~R. {Walborn}, E.~L. {Fitzpatrick}, {\it \pasp\/} {\bf 102}, 379 (1990).

\bibitem{2003A&A...404..305G}
G.~{Gauba}, M.~{Parthasarathy}, B.~{Kumar}, R.~K.~S. {Yadav}, R.~{Sagar}, {\it
  \aap\/} {\bf 404}, 305 (2003).

\bibitem{2013ApJS..204....5K}
S.~J. {Kleinman}, {\it et~al.\/}, {\it \apjs\/} {\bf 204}, 5 (2013).

\bibitem{2017ASPC..509....3D}
P.~{Dufour}, {\it et~al.\/}, {\it 20th European White Dwarf Workshop\/}, P.~E.
  {Tremblay}, B.~{Gaensicke}, T.~{Marsh}, eds. (2017), vol. 509 of {\it
  Astronomical Society of the Pacific Conference Series\/}, p.~3.

\bibitem{Smith2015}
N.~{Smith}, R.~{Tombleson}, {\it \mnras\/} {\bf 447}, 598 (2015).

\bibitem{2019MNRAS.488.1760S}
N.~{Smith}, {\it et~al.\/}, {\it \mnras\/} {\bf 488}, 1760 (2019).

\bibitem{2017MNRAS.470.1642F}
J.~{Fuller}, {\it \mnras\/} {\bf 470}, 1642 (2017).

\bibitem{2017ApJ...836..244W}
S.~E. {Woosley}, {\it \apj\/} {\bf 836}, 244 (2017).

\bibitem{2019ApJ...887...53F}
R.~{Farmer}, M.~{Renzo}, S.~E. {de Mink}, P.~{Marchant}, S.~{Justham}, {\it
  \apj\/} {\bf 887}, 53 (2019).

\bibitem{2022arXiv220506386W}
S.~E. {Woosley}, N.~{Smith}, {\it arXiv e-prints\/} p. arXiv:2205.06386 (2022).

\bibitem{2014ARA&A..52..487S}
N.~{Smith}, {\it \araa\/} {\bf 52}, 487 (2014).

\bibitem{2017ApJ...841...20N}
K.~F. {Neugent}, P.~{Massey}, D.~J. {Hillier}, N.~{Morrell}, {\it \apj\/} {\bf
  841}, 20 (2017).

\bibitem{1990A&A...235..234D}
S.~{Dreizler}, U.~{Heber}, K.~{Werner}, S.~{Moehler}, K.~S. {de Boer}, {\it
  \aap\/} {\bf 235}, 234 (1990).

\bibitem{2003A&A...402..335A}
A.~{Ahmad}, C.~S. {Jeffery}, {\it \aap\/} {\bf 402}, 335 (2003).

\bibitem{1978A&A....70..653K}
R.~P. {Kudritzki}, K.~P. {Simon}, {\it \aap\/} {\bf 70}, 653 (1978).

\bibitem{2020A&A...635A.193G}
S.~{Geier}, {\it \aap\/} {\bf 635}, A193 (2020).

\bibitem{2016A&A...588A..25M}
M.~M. {Miller Bertolami}, {\it \aap\/} {\bf 588}, A25 (2016).

\bibitem{Salpeter1955}
E.~E. {Salpeter}, {\it \apj\/} {\bf 121}, 161 (1955).

\bibitem{Boyer2011}
M.~L. {Boyer}, {\it et~al.\/}, {\it \aj\/} {\bf 142}, 103 (2011).

\bibitem{2006A&A...458..173S}
O.~{Su{\'a}rez}, {\it et~al.\/}, {\it \aap\/} {\bf 458}, 173 (2006).

\bibitem{2014MNRAS.439.2211K}
D.~{Kamath}, P.~R. {Wood}, H.~{Van Winckel}, {\it \mnras\/} {\bf 439}, 2211
  (2014).

\bibitem{2012A&A...543A..11M}
D.~R.~C. {Mello}, S.~{Daflon}, C.~B. {Pereira}, I.~{Hubeny}, {\it \aap\/} {\bf
  543}, A11 (2012).

\bibitem{2000A&AS..145..269P}
M.~{Parthasarathy}, J.~{Vijapurkar}, J.~S. {Drilling}, {\it \aaps\/} {\bf 145},
  269 (2000).

\bibitem{2020A&A...641A.142S}
R.~{Szczerba}, {\it et~al.\/}, {\it \aap\/} {\bf 641}, A142 (2020).

\bibitem{2007ApJ...663.1269N}
K.~{Nomoto}, H.~{Saio}, M.~{Kato}, I.~{Hachisu}, {\it \apj\/} {\bf 663}, 1269
  (2007).

\bibitem{2013ApJ...771...13L}
K.~{Lepo}, M.~{van Kerkwijk}, {\it \apj\/} {\bf 771}, 13 (2013).

\bibitem{2019ApJ...878..100W}
T.~L.~S. {Wong}, J.~{Schwab}, {\it \apj\/} {\bf 878}, 100 (2019).

\bibitem{1987A&A...178..159M}
A.~{Maeder}, {\it \aap\/} {\bf 178}, 159 (1987).

\bibitem{2009A&A...497..243D}
S.~E. {de Mink}, {\it et~al.\/}, {\it \aap\/} {\bf 497}, 243 (2009).

\bibitem{2015A&A...581A..15S}
D.~{Sz{\'e}csi}, {\it et~al.\/}, {\it \aap\/} {\bf 581}, A15 (2015).

\bibitem{2013A&A...560A..29R}
O.~H. {Ram{\'{\i}}rez-Agudelo}, {\it et~al.\/}, {\it \aap\/} {\bf 560}, A29
  (2013).

\end{thebibliography}
